\documentclass[]{elsart}
\usepackage{graphics}
\usepackage{times}
\usepackage{latexsym}
\usepackage{amssymb}

\newcommand{\sla}{\!\!\!\!/ \,}

\def\beq{\begin{equation}}
\def\eeq{\end{equation}}
\def\bea{\begin{eqnarray}}
\def\eea{\end{eqnarray}}

\begin{document}
%\draft
\begin{frontmatter}

\title{Direct Photons from Relativistic Heavy-Ion Collisions}

\author{Thomas Peitzmann} 
\address{University of M{\"u}nster, 48149 M{\"u}nster, Germany}
\ead{peitzmann@ikp.uni-muenster.de}
\author{Markus H. Thoma}
%\thanks[label1]{Heisenberg Fellow} 
\address{Max-Planck-Institut f\"ur extraterrestrische Physik, 
Giessenbachstra{\ss}e, 85748 Garching, Germany} 
\ead{thoma@mpe.mpg.de}
\begin{abstract}
Direct photons have been proposed as a promising signature for the
quark-gluon plasma (QGP) formation in relativistic heavy-ion 
collisions. Recently WA98 presented the first data on direct photons
in $Pb$+$Pb$-collisions at SPS. At the same time RHIC started with
its experimental program. The discovery of the QGP in these 
experiments relies on a comparison of data with theoretical
predictions for QGP signals. In the case of direct photons
new results for the production rates of thermal photons from the QGP 
and a hot hadron gas as well as for prompt photons from initial
hard parton scatterings have been proposed recently. Based on these
rates a variety of different hydrodynamic models, describing the 
space-time evolution of the fireball, have been adopted for 
calculating the direct photon spectra. The results have been compared to the
WA98 data and predictions for RHIC and LHC have been made. So far
the conclusions of the various models are controversial.

The aim of the present review is to provide a comprehensive
and up-to-date survey and status report on the experimental and 
theoretical aspects of direct photons in relativistic heavy-ion 
collisions. 
\end{abstract}

\begin{keyword}
relativistic heavy-ion collisions \sep direct photons

\PACS 25.75.-q \sep 25.20.Lj
\end{keyword}

\end{frontmatter}

\newpage

\tableofcontents
\section{Introduction}\label{sec:intro}

The major motivation to study relativistic heavy-ion collisions is
the search for the quark-gluon plasma (QGP), a potential new state of 
matter where
colored quarks and gluons are no longer confined into hadrons and
chiral symmetry is restored. 
The phase transition to quark matter has been predicted first for
the interior of neutron stars \cite{Itoh70,Collins75} and afterwards
in high-energy nucleus-nucleus collisions \cite{Chapline78,Chin78,Anishetty80}.
Subsequently it has been studied in great detail in lattice QCD 
\cite{Karsch01}. The quark-gluon plasma phase could provide insight in
the important non-perturbative features that usually govern hadronic
physics.

A wealth of knowledge has been accumulated by the early experiments
especially at the CERN SPS accelerator (see e.g. \cite{QM1999,QM2001}).
Many of the properties of
these collisions have been studied and interesting observations have
been made concerning non-trivial behavior of the strongly interacting
matter, most notably the suppression of J/$\psi$ production beyond the
expectation from normal nuclear effects, the enhancement of
strangeness production, modifications of the dilepton spectrum and
direct photon production in excess of known extrapolations from
particle physics. Some of these observations were actually predicted
to happen in relation to the phase transition to a QGP,
and one possible conclusion, guided by Ockham's razor\footnote{``Law of
Parsimony'' by William of Ockham, 14th century}, is to see the
experimental hints as evidence, though ``circumstantial'', of the new
phase \cite{Heinz00}.

However, a real understanding of the related physical concepts is extremely
difficult. Not only are most of the involved processes soft, and
thereby in the domain of large coupling constants where perturbation
theory breaks down, but the system itself is a multi-particle system,
which is already a challenge in situations where the underlying
interaction is much weaker. Although one might hope that in large
enough nuclei the system might be governed at least partially by laws
of thermodynamics, and thus be treatable, the conditions are
complicated further by the need
to control the residual non-equilibrium aspects.

To study such a complicated system one wishes for a probe that is not
equally complicated in itself. The production of hadrons is of course
governed by the strong interaction and therefore adds to the
complication.  One possible way out might be the study of hard
processes where QCD, the theory of strong interaction, enters the
perturbative regime and is calculable. The other avenue involves a
particle that suffers only electromagnetic interaction: Photons ---
both real and virtual --- should be an ideal 
probe.\footnote{For previous reviews on this topic see Refs. 
\protect\cite{Alam96,Alam01}.}  As we
will discuss in the present report, while photon production may be less
difficult to treat than some other processes in hadronic physics, an
adequate treatment in heavy-ion collisions turns out to be far from
trivial.
Experimentally, high energy direct photon measurement has always been 
considered
a challenge. This is true already in particle physics and even more in
the environment of heavy-ion collisions. Nevertheless a lot of
progress has been made and a large amount of experimental data is
available, though mostly from particle physics. Direct photon measurements
in heavy-ion collisions are expected to come into real fruition with
the advent of colliders like RHIC and LHC.

In the present report we attempt to provide a comprehensive review of
the theoretical and experimental aspects of the study of direct
photon production in heavy-ion collisions. We will also touch
photon production in proton-proton collisions as far as we consider
it relevant to our main subject. Because of the large amount of work
existing, we will most likely not be able to do justice to all of it,
and we would like to apologize for any omission or mistreatment of
related publications.

The structure of the present report will be as follows: In the next Section
we will discuss the theoretical status of the photon production from
$pp$ to $AA$ collisions. In particular, we will consider the calculation
of the rates from the QGP, from the hot hadron gas, and from
initial hard collisions. Furthermore, we review some basics of the
hydrodynamical description for deriving photon spectra in heavy-ion collisions.
In Section 3 experimental concepts for measuring direct photons and 
results from $pp$ and $pA$ collisions as well as from $^{16}O$-, 
$^{32}S$- and $^{208}Pb$-induced reactions are reviewed. 
In Section 4 these results are compared
to theoretical calculations, and predictions for RHIC and LHC are presented.
The following summary will conclude this review. Appendix A and B
provide some technical details for calculating the photon production
rate from the QGP.

\section{Theoretical Status}\label{sec:ts}

The theoretical prediction and calculation of the photon emission, i.e. 
yields and spectra, from a thermal system has a long tradition, culminating
in the discovery of quantum physics \cite{Planck00}. In astrophysics the
detection of electromagnetic radiation from the hot surfaces of stars
and other objects, even from the entire universe (Cosmic Microwave Background),
provides the most essential information, such as temperature, 
size, chemical composition etc. In particular deviations from the
pure black-body spectrum are of utmost interest, e.g. to learn about
the composition, evolution, and structure formation in the universe from the
Cosmic Microwave Background \cite{Wight92}. 

The photon emission from the nuclear fireball, created in a relativistic 
heavy-ion collision, differs from the one of macroscopic stellar objects
in the following respect. Whereas the photons in the latter case are 
thermalized when they leave the surface, the mean-free path of the
photons produced in nucleus-nucleus collisions is large compared to the
size of the fireball. Hence, the photons do not interact after their
production and leave the fireball undisturbed. As a consequence they carry 
information about the stage of the fireball at the time of their creation. 
The photon
spectrum, containing photons from all stages, allows therefore to study
the entire evolution of the fireball. Direct photons, together with dileptons
and to some extent hard probes like jet quenching, are therefore a unique 
diagnostic tool for the different phases and the equation of state (EOS) 
of the ultradense
matter produced in high-energy nuclear collisions. Photon production in 
high-energy nuclear and particle physics 
provides information on the momentum distributions of the emitting 
particles. In particle physics this may be used to extract 
information on structure functions. In thermalized systems, expected 
in nuclear collisions, it should yield information on the thermal 
distributions.

To draw conclusions about the state of the matter in the fireball,
created in relativistic heavy-ion collisions, it is necessary to 
compare the experimental data for direct photons with theoretical 
calculations. The ideal theoretical description would be a comprehensive
treatment of the entire space-time evolution of the fireball from the
first contact of the cold nuclei to the freeze-out and subsequent decay
of hadrons, e.g. in a dynamical lattice QCD approach. At the same time all 
participating particle species and their interactions should be included.
Due to the complexity of the problem, e.g. the consistent treatment of
hadronization and the non-perturbative nature of the strong interaction, 
such a systematic investigation is presently
only wishful thinking. Alternatively, 
the different stages of the fireball (initial stage, pre-equilibrium
QGP, thermal QGP, mixed phase\footnote{The existence of a mixed phase
as a consequence of a first order phase transition is questionable since 
recent lattice calculations prefer a cross over \cite{Karsch01b}.}
and hadronization, hot hadron gas,
freeze-out and hadronic decays) are treated separately. Furthermore,
one computes first the production rates of the photons from the different
stages, e.g. at a given temperature. Then these rates are convoluted
with the space-time evolution of the fireball using mostly hydrodynamical 
models. In this way, estimates of the photon spectra are obtained, which
can be compared to experimental results.

In the present chapter we will discuss in detail the status and problems of
calculating production rates of direct photons from a thermal QGP and hadron 
gas as well as from hard scatterings in the initial non-equilibrium stage.
In addition, the various hydrodynamical approaches and their applications 
to photon spectra will be critically reviewed. 

\subsection{Photon Production Rates}\label{subsec:ppr}

In this Section the calculation of the production rates of direct photons
with experimentally relevant energies $E\gg T$
from a thermal QGP, a hot hadron gas (HHG)
and of prompt photons from the initial
phase will be considered. Since direct photons have been proposed as a
promising signature of the QGP formation in relativistic heavy-ion collisions
\cite{Shuryak78,Kajantie81,Halzen82,Kajantie83,Sinha83,Hwa85,Staadt86,Neubert89}, 
emphasis is put on the photon production from the QGP and the calculation
of this rate will be discussed first in detail.

Particle production rates can be computed from the amplitudes of the basic
processes for the particle production, convoluted with
the distribution functions of the par\-ti\-ci\-pating particles \cite{Weldon83}.
For example, the production rate of a particle $A$ with energy $E$ 
follows from
\bea
&&\Gamma _{\rm prod}(E)=\frac{1}{2E} \int  \frac{d^3p_1}{(2\pi )^32E_1} ...
\frac{d^3p_m}{(2\pi )^32E_m}\> \frac{d^3p'_1}{(2\pi )^32E'_1} ...
\frac{d^3p'_n}{(2\pi )^32E'_n}\nonumber \\ 
&& (2\pi)^4\> \delta (P-\sum_{i=1}^m P_i+\sum_{i=1}^n P'_i)\>
\sum_{i,j} |\mathcal{M}|^2\> f_1 ... f_m\> (1\pm f'_1) ... (1\pm f'_n).
\label{prodrate}
\eea
Here $\mathcal{M}$ is the matrix element of the basic process for the 
production of particle $A$, where $m$ particles participate in the 
initial and $n$ particles (denoted by a prime) in the final channel.
$\sum_{i,j}$ indicates the sum over all states of the 
particles in the initial and final states except of the particle $A$, and
$P$, $P_i$, and $P'_j$ are the 4-momenta of the particles. $f_i$ denotes
the distribution functions of the incoming particles and $f'_j$ of the 
outgoing ones (except of $A$). For outgoing bosons, the plus sign holds, 
corresponding to Bose-enhancement, whereas for fermions the minus sign,
corresponding to Pauli-blocking. In an equilibrated system, such as
the QGP or the HHG, the distribution
functions are given by Bose-Einstein or Fermi-Dirac distributions, 
respectively. In high-energy particle physics, such as the production
of prompt photons in $pp$ collisions, the parton structure functions 
are taken.

\subsubsection{Thermal Rates from the QGP}\label{subsubsec:trq}

\begin{figure}[hbt]
       \centerline{\resizebox{12cm}{!}{\includegraphics{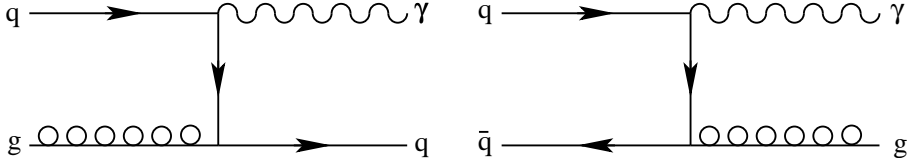}}}
       \caption{Lowest order contributions to photon production from the QGP:
Compton scattering (left) and quark-antiquark annihilation (right).}
       \protect\label{fig2.1}
\end{figure}

A QGP emits photons as every thermal source does. The microscopic process 
is the photon radiation from quarks having an electric charge. Due to 
energy-momentum conservation, these quarks have to interact with the thermal
particles of the QGP in order to emit a photon. Hence, an ideal, 
non-interacting QGP cannot be seen. However, there will always be 
(strong and electromagnetic) interactions in the QGP, such as 
quark-antiquark annihilation.
However, due to energy-momentum conservation the direct annihilation 
of quarks and anti-quarks into real photons is also 
not possible but only into 
virtual photons which can decay into lepton pairs. The production of dileptons 
is another promising signature for the QGP \cite{Gale01}, which, however, 
is not
the topic of the present review. To lowest order perturbation theory, real
photons are produced from the annihilation of a quark-antiquark pair
into a photon and a gluon ($q\bar q\rightarrow g\gamma$)
and by absorption of a gluon by a quark emitting
a photon ($qg\rightarrow q\gamma$), 
similar to Compton scattering in QED (see Fig. \ref{fig2.1}). 
A higher order process 
for the photon production is, for example, bremsstrahlung, where a quark 
radiates a photon by scattering off a gluon or another quark in the QGP.

\vspace*{0.7cm}

The photon production rate can be computed from the matrix elements
of these basic processes by convoluting them with the distribution 
functions of the part\-ici\-pating partons according to Eq. (\ref{prodrate}). 
In the case of processes
with two partons in the initial and one in the final channel, such as 
annihilation and Compton scattering discussed above, the differential
photon production rate is given by \cite{Kapusta91}
\bea
\hspace*{-1cm}\frac{dN}{d^4xd^3p}&=&\frac{1}{(2\pi)^32E}\> 
\int \frac{d^3p_1}{(2\pi)^32E_1} 
\frac{d^3p_2}{(2\pi)^32E_2} \frac{d^3p_3}{(2\pi)^32E_3}\> 
n_1(E_1)n_2(E_2)[1\pm n_3(E_3)]\nonumber \\
&& \sum_{i} \langle |\mathcal{M}|^2\rangle \>
(2\pi)^4\> \delta(P_1+P_2-P_3-P).
\label{rate1}
\eea
Here $P_1$ and $P_2$ are the 4-momenta of the incoming partons, 
$P_3$ of the outgoing parton, and $P$ of the produced photon. Throughout
the paper we use the notation $P=(p_0, {\bf p})$ and $p=|{\bf p}|$,
which is convenient in thermal field theory. For on-shell particles,
the energy is denoted by $p_0=E$. In equilibrium, the distribution functions 
$n_i(E_i)$ are given by the Bose-Einstein distribution, 
$n_B(E_i)=1/[\exp(E_i/T)-1]$, for gluons and by the Fermi-Dirac distribution,
$n_F(E_i)=1/[\exp(E_i/T)+1]$, for quarks, respectively. The factor
$[1\pm n_3(E_3)]$ describes Pauli-blocking (minus sign) in the case of a
final-state quark or Bose-enhancement (plus sign) in the case of a 
final-state gluon. The factor $\langle |\mathcal{M}|^2\rangle$
is the matrix element of the basic process averaged over the initial states
and summed over the final states. The $\sum_{i}$ indicates the sum over
the initial parton states. The delta function, as usual, ensures
energy-momentum conservation. The formula Eq. (\ref{rate1}) can be extended easily
to higher order processes, by integrating over the momenta of the
additional external partons, taking into account also 
their distribution functions. The differential
rate, defined above, determines the number of emitted photons 
with momentum ${\bf p}$ within the interval $[{\bf p}, {\bf p}+d^3p]$ and 
energy $E=p$ from the space-time volume 
$d^4x$. The total rate follows from integrating over the photon momentum.
The observable spectrum is obtained by integrating over the space-time volume,
by using for instance a hydrodynamical model, describing the space-time
evolution of the expanding QGP. The total photon yield
results from an integration of the spectrum over the photon momentum.

\begin{figure}[hbt]
       \centerline{\resizebox{10cm}{!}{\includegraphics{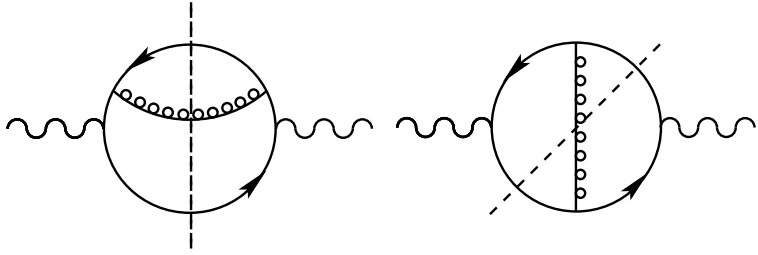}}}
       \caption{2-loop polarization tensor. The dashed lines indicate  
cuts through the diagrams, corresponding to the processes in Fig. \ref{fig2.1}.}
       \protect\label{fig2.2}
\end{figure}

An alternative definition of the differential photon production rate
is based on the polarization tensor or photon self-energy. According to 
cutting rules extended from vacuum quantum field theory 
to finite temperature \cite{Weldon83,Kobes85,Kobes86}, 
the differential rate can be related to
the imaginary part of the polarization tensor $\Pi_{\mu \nu}(P)$ on its 
mass shell ($p_0=E=p$)
\cite{Gale91}
\beq 
\frac{dN}{d^4xd^3p}=-\frac{1}{(2\pi)^3}\>\frac{1}{E}\> \frac{1}{{\exp (E/T)-1}}\>
{\rm Im}\, {\Pi_\mu}^\mu (E).
\label{rate2}
\eeq
This expression is exact to first order in the
electromagnetic coupling $\alpha$ and to all orders in the strong coupling
constant\footnote{The strong coupling constant at finite temperature
depends on the temperature (effective, temperature-dependent running coupling
constant) \cite{Karsch88}. However, for most applications in the following 
we will
use a mean value of $\alpha_s=0.2$ - 0.5, which is typical for temperatures
reachable in relativistic heavy-ion collisions.}
$\alpha_s=g^2/4\pi$. Therefore, it contains in contrast to the definition
Eq. (\ref{rate1}), which holds only for $2\rightarrow 2$ reactions, also
higher order processes like bremsstrahlung 
if the photon self-energy is chosen accordingly.
The lowest order annihilation and Compton processes correspond to a 
polarization tensor containing one quark loop and one internal gluon line
as shown in Fig. \ref{fig2.2}. Cutting these diagrams reproduces the processes 
of Fig. \ref{fig2.1}
in an illustrative way. 

Now we will discuss the various attempts for calculating the production
rate of energetic photons ($E\gg T$) from an equilibrated QGP.

\medskip

{\it Pre-HTL rate:} Before the invention of the Hard-Thermal-Loop (HTL)
improved perturbation theory (see below), the QGP photon rates have been 
calculated using the perturbative matrix elements for the processes of
Fig. \ref{fig2.1} together with Eq. (\ref{rate1}) \cite{Shuryak78,Kajantie81,Hwa85,Neubert89}.
In Ref.\cite{Hwa85}, even bremsstrahlung has been considered in this way. 
The derivation
of the differential production rate of energetic photons
($E\gg T$), produced by the processes of Fig. \ref{fig2.1},
is presented in Appendix A. In the case of two thermalized 
quark flavors with bare mass $m_0$ it is given by 
\cite{Kapusta91}
\beq
\frac{dN}{d^4xd^3p}=\frac{5}{18\pi^2}\> \alpha \> \alpha_s \> \frac{T^2}{E}\> 
e^{-E/T}\> \ln \frac{ET}{m_0^2}, 
\label{pre_htl}
\eeq
where $ET\gg m_0^2$ has been assumed.

It was noted that there is a logarithmic infrared (IR) sensitivity, 
i.e. the rate diverges logarithmically if the mass of the exchanged 
quark in Fig. \ref{fig2.1} tends to zero. Therefore, 
Kajantie and Ruuskanen argued \cite{Kajantie83} that the bare quark mass
should be replaced by an effective thermal quark mass. This means that
even the production of energetic photons is sensitive to in-medium effects 
of the QGP, since the exchange of soft quarks plays an important role in the
production mechanism. A systematic treatment of in-medium effects 
is provided by the HTL resummation technique, discussed below. 
Kajantie and Ruuskanen \cite{Kajantie83} used an effective, 
temperature-dependent quark mass 
calculated from the quark self-energy in the high temperature limit
as discussed in Appendix B.
The result is \cite{Klimov82,Weldon82}: $m_q^2=g^2T^2/6$. 
For $g=1.5$ - $2.5$ corresponding to realistic values 
$\alpha_s\simeq 0.2$ - $0.5$, 
one gets
$m_q=0.6$ - $1.0$ $T$. For typical temperatures of the QGP, e.g. $T=200$
MeV, the effective quark mass is much larger than the bare mass of
up and down quarks ($m_{u,d}\simeq 5$ - $10$ MeV) and of the same order as the
bare strange quark mass. Hence, neglecting in-medium effects, i.e.
adopting the bare instead of the effective
quark mass in Eq. (\ref{pre_htl}), leads to an overestimation of the rate.
In the weak coupling limit, in which perturbation 
theory holds, the logarithm in Eq. (\ref{pre_htl}) has to be replaced now
by $\ln (E/\alpha_s T)$, neglecting a constant of the order of 1. 
As we will see below, using the
HTL technique, this result is the leading logarithm approximation for
the rate.

\medskip

{\it 1-loop HTL rate:} Using only bare propagators (and vertices)
as in Fig.~\ref{fig2.1} or Fig.~\ref{fig2.2} 
for gauge theories 
(QED, QCD) at finite temperature, problems such as IR divergent
and gauge dependent results are encountered. A famous example is the so-called 
plasmon puzzle: the damping rate of a gluon with a long wavelength
or small momentum in a QGP, called plasmon, has been calculated in
different gauges and different results have been found. In particular,
in covariant gauges a negative result was obtained, indicating a puzzling
instability of the QGP in perturbation theory \cite{Parikh89}. Braaten and
Pisarski \cite{Braaten90} argued that naive perturbation theory, using 
only bare propagators (and vertices), is incomplete at finite temperature. 
Higher-order
diagrams, containing infinitely many loops, can contribute to the same 
order in the coupling constant. These diagrams can be taken into account 
by resumming a certain class of diagrams, the hard thermal loops (HTLs). 
These diagrams are 
1-loop diagrams (self-energies and vertex corrections), 
where the loop momentum is hard, i.e. of the order
of the temperature or larger. This approximation agrees with the 
high-temperature limit of these diagrams, which has been computed already
some time ago in the case of the gluon and quark self-energy 
\cite{Klimov82,Weldon82,Weldon82a}. Resumming these self-energies within
a Dyson-Schwinger equation leads to effective gluon and quark propagators,
which describe the propagation of collective gluon and quark modes in
the QGP. These effective propagators (and similar effective vertices) 
have to be used if the momentum of the propagator is soft, i.e. of the 
order $gT$. Otherwise a bare propagator is sufficient. In this way,
gauge invariant results for physical quantities are obtained and their 
IR behavior is improved. In the case of the plasmon damping rate,
Braaten and Pisarski derived a positive, gauge independent result
by using HTL-resummed gluon propagators and vertices \cite{Braaten90a}. 
It is important to 
note, that the HTL-resummation technique relies on the weak coupling limit
assumption, $g\ll 1$, which allows the separation of the soft scale $gT$ 
and the hard scale $T$. The HTL-resummed perturbation theory is 
exemplified in Appendix B, where the photon production rate is calculated
in this way. For a review of the HTL-method and its application see
\cite{LeBellac96,Thoma95,Thoma00,Blaizot01}.

\begin{figure}[hbt]
       \centerline{\resizebox{6cm}{!}{\includegraphics{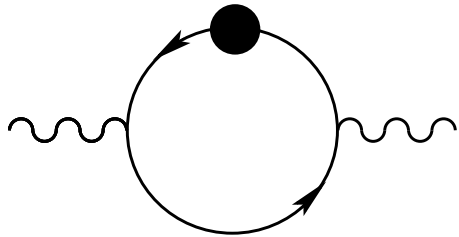}}}
       \caption{Photon self-energy containing a HTL-resummed quark propagator
indicated by the filled circle. The energy and momentum of the resummed
quark propagator are soft, i.e. smaller than the separation scale $q_c$, whereas the 
other quark momentum is hard due to energy-momentum conservation in the case of hard 
photons.}
       \protect\label{fig2.3}
\end{figure}

In the case of massless quarks, the hard photon production rate from the QGP 
is logarithmically IR divergent due to the exchange of
a massless quark, as discussed above. Therefore, the bare quark propagator
has to be replaced by a HTL-resummed one, defined in Fig. \ref{figB.1} of
Appendix B.
According to the rules of the
Braaten-Pisarski method, this has to be done for soft quark momenta.
Therefore, we decompose the rate in a soft and a hard contribution,
introducing a separation scale for the quark momentum $gT\ll q_c \ll T$
\cite{Braaten91}. For the soft contribution, we start from Eq. (\ref{rate2}) and
use the diagram shown in Fig. \ref{fig2.3} as polarization tensor. 
This 1-loop diagram
has a non-vanishing imaginary part since the effective quark propagator
contains an infinite number of loops (see Fig. \ref{figB.1}). Cutting
this diagram through the filled circle reproduces the diagrams shown in
Fig. \ref{fig2.1}, where the bare quark propagator is replaced by a
resummed one. It is not 
necessary to dress both propagators or to use an effective quark-photon
vertex\footnote{The energetic photon resolves the quark-photon vertex 
rendering a vertex correction unnecessary.} since only one internal quark 
line can be soft due to energy-momentum conservation in the case of hard photons. 
The hard contribution
follows from the pre-HTL result, replacing the bare quark mass by
the separation scale $q_c$ (see Appendix A). 
The details of these calculations are presented 
in Appendix B. Adding up the soft and the hard contribution, the 
separation scale cancels. In this way Kapusta et al. \cite{Kapusta91}
and independently Baier et al. \cite{Baier92} found
\beq
\frac{dN}{d^4xd^3p}\biggl |_{\rm 1-loop}=a\> \alpha \alpha _s\> 
e^{-E/T}\> \frac{T^2}{E}\> \ln \frac{0.2317E}{\alpha _sT}, 
\label{1-loop}
\eeq
where $a=0.0281$ for $N_F=2$ thermalized quark flavors and $a=0.0338$ for 
$N_F=3$, respectively. The result has been extended to finite baryon density
by generalizing the HTL-resummation technique to finite quark chemical 
potential \cite{Vija95}. 
It is interesting to note that for finite $\mu $
one has to give up the Boltzmann approximation for the initial
parton distributions in the hard contribution (see Appendix B). 
Otherwise there
is no cancellation of the separation scale after adding the hard and 
the soft part. Therefore, the photon production rate at finite $\mu$
can be determined only numerically.
For $|\mu/T|<1$, the factor $T^2$ in Eq. (\ref{1-loop})
has to be replaced to a good accuracy simply by $T^2+\mu^2/\pi^2$ 
\cite{Traxler95}. 

The 1-loop HTL photon production rate has also been calculated for
a chemically non-equilibrated QGP 
\cite{Shuryak93,Kaempfer94,Strickland94,Traxler96,Baier97},
as discussed at the end of Section \ref{subsubsec:hyd}. 

\medskip

{\it 2-loop HTL rate:} Naively one expects that higher order diagrams
such as brems\-strahl\-ung will contribute only to order $\alpha \, 
\alpha_s^2$.
However, recently Aurenche et al. \cite{Aurenche98} showed that
the 2-loop HTL contribution to the hard photon production rate is 
of order $\alpha \, \alpha_s$, i.e. contributes to Eq. (\ref{1-loop}) beyond the
leading logarithm approximation. In the following, we will only sketch the 
arguments without presenting the calculation in detail.

\begin{figure}[hbt]
       \centerline{\resizebox{12cm}{!}{\includegraphics{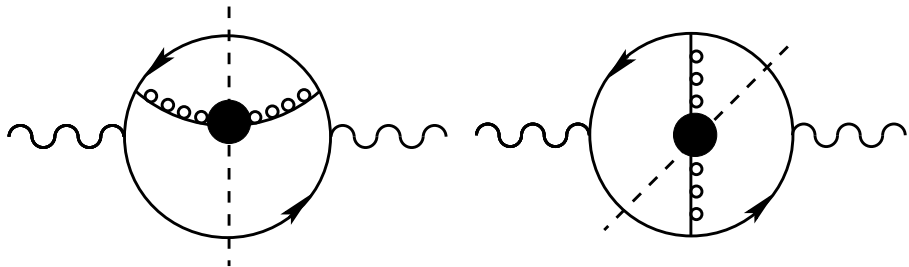}}}
       \caption{2-loop HTL polarization tensor containing a HTL resummed 
gluon propagator indicated by a filled circle. The dashed lines indicate  
cuts through the diagrams, corresponding to the processes in Fig. \ref{fig2.5}.}
       \protect\label{fig2.4}
\end{figure}

The 1-loop HTL contribution of Fig. \ref{fig2.3} to the hard 
photon production rate corresponds to the exchange of a soft, collective
quark. The logarithmic IR singularity in the case of massless bare quarks
is cut off by medium effects (in-medium quark ``mass'') of the order
$gT$. The complete second order HTL 
rate follows from adding the 1-loop HTL contribution
for soft quarks and the 2-loop diagram of Fig. \ref{fig2.2}, where the 
intermediate quark is hard. Note that in Fig. \ref{fig2.2} we assumed that
the gluon is also hard, i.e., it is a thermal particle with an average energy 
of the order $T$. However, if this gluon is soft, there will be a Bose 
enhancement factor $n_B(k_0\sim gT)\simeq T/k_0 \sim 1/g$. Hence, this 
contribution might be important. According to the HTL resummation
method, we therefore have to dress the gluon in 
Fig. \ref{fig2.2}, i.e., to use a HTL-resummed gluon propagator as in 
Fig. \ref{fig2.4}.

\begin{figure}
\centerline{\resizebox{12cm}{!}{\includegraphics{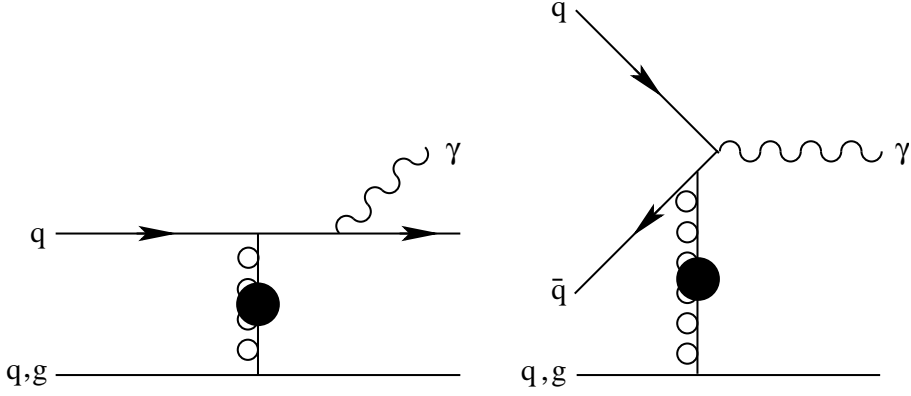}}}
%\vspace*{-2.5cm}
\caption{Photon production processes corresponding to the 2-loop HTL
contribution: bremsstrahlung (left) and annihilation with scattering (right).
The filled circles indicate HTL resummed gluon propagators. The lower line 
indicates either a quark or a gluon.}
\protect\label{fig2.5}
\end{figure}

One contribution to the imaginary part of these diagrams comes from cutting 
through the filled circle of the effective gluon propagator, i.e.
from the imaginary part of the gluon self-energy of the effective gluon
propagator corresponding to
Landau damping of the time-like gluon (see Appendix B). 
Since the HTL gluon self-energy
contains hard quark and gluon loops, physical processes contained in
the imaginary part of Fig. \ref{fig2.4} 
are brems\-strahl\-ung and annihilation with
scattering as shown in Fig. \ref{fig2.5}.

Naively one expects that these diagrams lead to a rate that is reduced
by a factor of $g^2$   
compared to the 1-loop HTL rate Eq. (\ref{1-loop}) due to the
additional vertex. However, caused by a strong collinear IR singularity
it turns out that ${\rm Im}\, {\Pi_\mu}^\mu \sim e^2 g^4$ has to be 
multiplied by a factor $T^2/m^2_\infty$. Here $m^2_\infty=2m_q^2=g^2T^2/3$
is the asymptotic thermal quark mass which cuts off the IR singularity
in the diagrams of Fig. \ref{fig2.5}\footnote{Although 
the IR singularity in Fig. \ref{fig2.5}
is related to the exchange of the gluon, it vanishes due to a thermal
quark mass \cite{Aurenche97}. The asymptotic quark mass enters the
calculation if a resummed instead of a bare quark propagator is used in
Fig. \ref{fig2.4}.}. Hence, the contribution to the photon rate from
Fig. \ref{fig2.4} is of the same order $e^2\, g^2$ as the one from 
Fig. \ref{fig2.2}. This is a typical
problem of perturbative field theory at finite temperature, where due to 
medium effects higher-order diagrams can contribute to lower order in the 
coupling constant. For further examples see e.g. Ref.\cite{Thoma95}. 

\begin{figure}[hbt]
       \centerline{\resizebox{10cm}{!}{\includegraphics{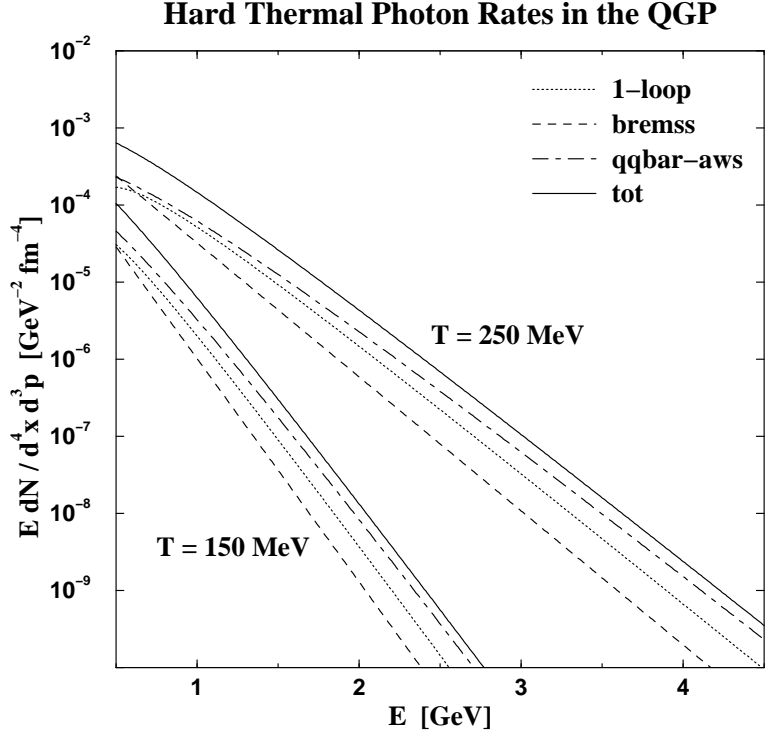}}}
       \caption{Comparison of the 1-loop HTL rate (dotted line), the
bremsstrahlung rate (dashed line), the annihilation-with-scattering
rate (dot-dashed line), and the total rate (solid lines) at $T=150$ MeV
and $T=200$ MeV \cite{Steffen01}.}
       \protect\label{fig2.6}
\end{figure}

Now we present the final result of the tedious 2-loop HTL calculation of 
the production rate of energetic photons ($E\gg T$) \cite{Aurenche98}. 
In the case of bremsstrahlung, it reads
\beq
\frac{dN}{d^4xd^3p}\biggl |_{\rm brems}=b\> \alpha \alpha _s\> e^{-E/T}\>  
\frac{T^2}{E}\>,
\label{brems}
\eeq
where $b=0.0219$ for $N_F=2$ and $b=0.0281$ for $N_F=3$, respectively.
The annihilation with scattering (aws)
in Fig. \ref{fig2.5} leads to 
\beq
\frac{dN}{d^4xd^3p}\biggl |_{\rm aws}=c\> \alpha \alpha _s\> e^{-E/T}\> T,
\label{aws}
\eeq 
where $c=0.0105$ for $N_F=2$ and $c=0.0135$ for $N_F=3$, 
respectively\footnote{In Ref.\cite{Aurenche98} a numerical error led to an 
overestimation of the rate by a factor of 4 \cite{Steffen01}.}. The constants
$b$ and $c$ had to be computed numerically. Comparing Eqs. (\ref{brems}) and
(\ref{aws}) with Eq. (\ref{1-loop}), we observe that the 2-loop HTL rate is
of the same order as the 1-loop HTL rate apart from a factor 
$\ln (1/\alpha_s)$, which comes from the thermal quark mass playing the
role of an IR cutoff in the 1-loop HTL contribution. Moreover, the
annihilation-with-scattering process is due to phase space proportional 
to $T$ instead of $T^2/E$ as in the case of the Compton scattering,
annihilation without scattering, and bremsstrahlung. Hence, that contribution 
dominates at large photon energies. In Fig. \ref{fig2.6} the various 
contributions to the rate are compared at two different 
temperatures, $T=150$ MeV 
and $T=200$ MeV \cite{Steffen01}, 
where a temperature dependent coupling constant
$\alpha_s(T)=6\pi/[(33-2N_F)\ln(8T/T_c)]$ with $T_c=170$ MeV has been
adopted \cite{Karsch88}. Although the extrapolation of the HTL-results 
obtained in the limit $g\ll 1$ to realistic values of the coupling constant
($\alpha_s \simeq 0.3$) is doubtful, one sees the relative importance of the 
individual contributions. In particular one observes the dominant role
of the annihilation-with-scattering contribution above $E=1$ GeV. 

The 2-loop HTL rate has also been generalized to chemical non-equilibrium 
\cite{Mustafa00,Dutta01} but not to a finite chemical potential (finite baryon 
density) so far.

\medskip

\begin{figure}[hbt]
       \centerline{\resizebox{6cm}{!}{\includegraphics{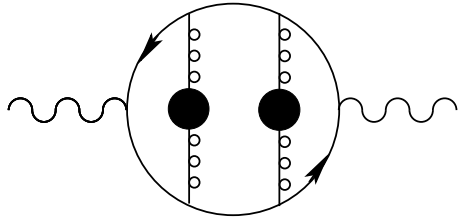}}}
       \caption{Example for a 3-loop HTL polarization tensor.}
       \protect\label{fig2.7}
\end{figure}

{\it Higher-order contributions:} Since 2-loop HTL contributions are as 
important as 1-loop HTL contributions, what about higher-loop diagrams?
Aurenche et al. \cite{Aurenche00} have also investigated this question 
looking at 3-loop HTL diagrams like the one in Fig. \ref{fig2.7}. 
Using power counting
one can show that the 3-loop diagram is proportional to the 2-loop diagram
times a factor $g^2T/\mu$, where $\mu $ is the IR cutoff for the
additional exchanged gluon. In the case of a transverse gluon this cutoff
is provided by the non-perturbative magnetic mass of the order $g^2T$.
Hence, the 3-loop contribution is of the same order as the 2-loop.
This argument is essential the same that has been used by Linde \cite{Linde80}
for showing the break-down of perturbation theory for QCD at finite 
temperature. However, the power counting argument is too restrictive since
there are cancellations of IR singularities between different cuts
of the diagram according to the Kinoshita-Lee-Nauenberg theorem 
\cite{Kinoshita62,Lee64}. Indeed, the sum over the different cuts generates
a kinematical cutoff. However, this cutoff becomes smaller than the
non-perturbative magnetic cutoff if the virtuality of the photon is small. 
In particular, for real photons the rate is always sensitive to the
magnetic cutoff. Hence, the production rate of real photons cannot be
evaluated within perturbation theory. Infinitely many higher order diagrams 
contribute to the same order, $\alpha \, \alpha_s$, as the 2-loop HTL 
diagram. For dileptons with an invariant mass larger than $g^2T$,
on the other hand, the Kinoshita-Lee-Nauenberg cutoff becomes relevant and 
their rate can be accessed perturbatively in the weak coupling limit. 

\begin{figure}[hbt]
       \centerline{\resizebox{6cm}{!}{\includegraphics{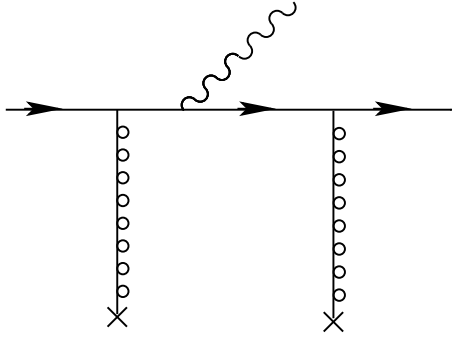}}}
       \caption{Multi-scattering bremsstrahlung affected by the LPM-effect.}
       \protect\label{fig2.8}
\end{figure}

Although there are additional contributions of the same order to the rate 
compared to the 1- and 2-loop HTL contributions, the 1- and 2-loop HTL
rate cannot be used as a lower limit for the photon production rate, 
for there are also destructive interferences in the higher-order 
contributions. They lead to a process known as the Landau-Pomeranchuk-Migdal
(LPM) effect which results in a suppression of the photon emission. Loosely
spoken, a photon will not be emitted if there is not enough time for its
production before the radiating quark will be scattered off another particle.
The production time can be estimated from the uncertainty principle while
the time between two successive collisions follows from the mean-free
path of the quark in the QGP. As an example the bremsstrahlung from
a quark between two scatterings is shown in Fig. \ref{fig2.8}. 
The LPM effect has been
predicted in QED by Landau and Pomeranchuk \cite{Landau53} and Migdal
\cite{Migdal56} a long time ago and recently been confirmed experimentally
at SLAC in the suppression of bremsstrahlung in thick targets
\cite{Anthony95,Anthony97}. Generalized to non-abelian gauge theories, 
it also plays an important role in the energy loss of energetic partons 
in the QGP and the associated jet-quenching \cite{Baier96}. Assuming
for simplification a constant, energy-independent damping rate or width 
for the quark, Aurenche et al. \cite{Aurenche00a} estimated the LPM-effect
in the photon production from the QGP. They showed that for  
bremsstrahlung only low-energy photons, typically with energies below
100 MeV are strongly affected (see also \cite{Cleymans93}), 
whereas in the annihilation-with-scattering
case surprisingly only high energy photons ($E>10$ GeV) are strongly
suppressed. In the interesting energy regime of a few GeV the influence of 
the LPM-effect seems not to be very important. A verification of this 
statement, however, requires a thorough consideration of the LPM-effect for 
the photon production,
going beyond the simplified calculation of Aurenche et al. \cite{Aurenche00a}.
\footnote{Recently, Arnold, Moore, and Yaffe claimed that a rigorous
treatment of the LPM-effect by summing ladder diagrams leads to an
infrared finite result which is sensitive only to the scale $gT$
\cite{Arnold01}. They found that for $\alpha_s=0.2$ and $2.5 \leq E/T \leq 10$
the complete leading order rate agrees within a factor of 2 with the 1-loop
HTL result (\ref{1-loop}) \cite{Arnold01a}.}

Considering a possible suppression of the photon production from the QGP
by the LPM-effect and a possible enhancement by other higher-order 
contributions, the sum of the 1- and 2-loop HTL rate has been used as an
educated guess. Moreover, one has to keep in mind that these rates have been
derived under the unrealistic assumption of $g\ll 1$, which renders their
applicability even more dubious.
Since non-perturbative methods such as lattice QCD do
not allow the calculation of dynamical quantities, e.g. particle production 
rates, at the moment, this estimate appears to be the state of the art.
It might be possible in the future that lattice calculations will be
capable to extract non-perturbative information also for production rates
using the maximum entropy method \cite{Karsch01}. Such information
would be of utmost importance, not only for the photon production but 
basically for all signatures of the QGP formation. 

\subsubsection{Thermal Rates from the Hadron Gas}\label{subsubsec:trh}

In order to calculate the photon spectrum and yield from the fireball
in relativistic heavy-ion collisions, one has to know also the photon
production rate from the HHG since photons will also be 
emitted from this thermal phase following the QGP. Furthermore, the prediction 
of the photon production from the HHG is necessary if one wants to use the 
photon spectrum as a signature for the QGP. For this purpose, one has to 
compare the photon spectrum with and without phase transition, i.e., 
in a hydrodynamical model one has to consider equations of state (EOS) 
describing, on the one hand, a QGP, mixed, and HHG phase and, on the
other hand, a pure HHG phase.

The microscopic description of the thermal photon emission from the HHG
is based on the interactions between hadrons in the heat bath. 
Due to vector meson dominance (VMD), vector 
mesons ($\rho$, $a_1$) play an important role for the photon production.
Furthermore, in particular pions and etas decay into
photons. However, since these processes take place predominantly
after freeze-out,
these decay photons are subtracted from the experimentally observed
spectrum as a huge background (signal to background ratio about 20\%)
for obtaining the direct photon spectrum.
Hence, we will not consider hadronic decays into photons after freeze-out
in the following.

In contrast to the QGP, which can be treated within QCD, one has to adopt 
effective theories for the hadron interactions. Effective theories contain
a certain number of hadron species, whose interactions are determined by 
symmetry and simplicity arguments. The first calculation of the photon
production from the HHG has been performed by Kapusta, Lichard, and Seibert, 
\cite{Kapusta91}. They considered a baryon-free HHG (zero chemical potential)
consisting out of pions, which are the most abundant hadronic constituents
due to their small mass, and rhos, which are important for photon emission
because of VMD.
They started from an effective Lagrangian describing the interaction
between charged pions, coupled to photons, and neutral
rhos
\beq
{\mathcal L}=|D_\mu \Phi|^2-m_\pi^2\> |\Phi |^2-\frac{1}{4}\> \rho_{\mu \nu}
\rho^{\mu \nu} +\frac{1}{2}\> m_\rho^2\> \rho_\mu \rho^\mu -\frac{1}{4}
\> F_{\mu \nu}F^{\mu \nu}.
\label{eff-lagr}
\eeq
Here $D_\mu=\partial _\mu-ieA_\mu -ig_\rho \rho_\mu$ is the covariant
derivative, $\Phi $ is the complex pion field, and $\rho_\mu$ is the rho field.
$\rho_{\mu \nu}=\partial_\mu \rho_\nu -\partial_\nu \rho_\mu$ is the 
field-strength tensor of the rho field and $F_{\mu \nu}=\partial_\mu A_\nu
-\partial_\nu A_\mu$ the one of the electromagnetic field. The pion-rho
coupling constant $g_\rho$ is determined from the decay rate of the process
$\rho 
\rightarrow \pi \pi$, yielding $g_\rho^2/(4\pi )=2.9$. 

\begin{figure}[hbt]
       \centerline{\resizebox{14cm}{!}{\includegraphics{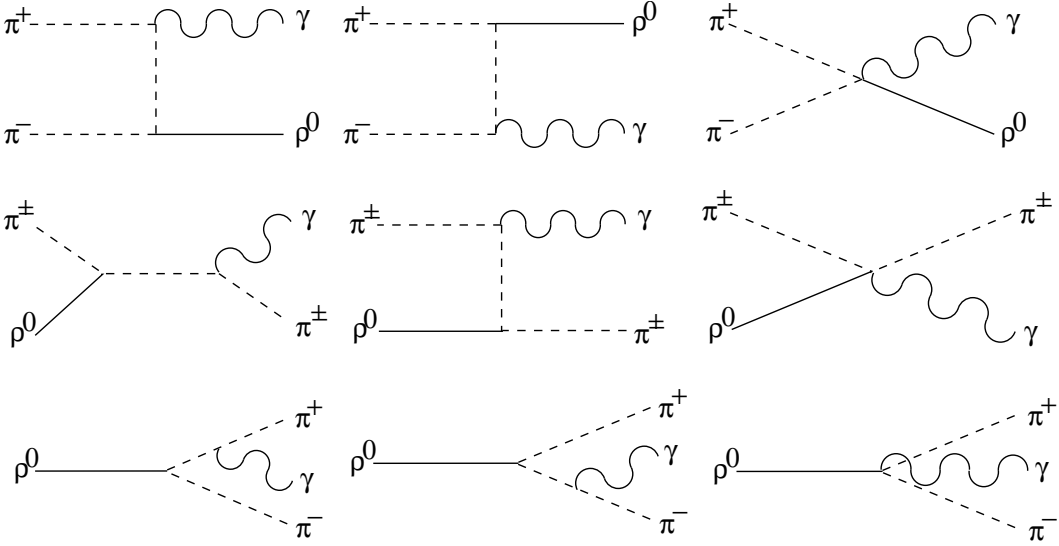}}}
       \caption{Photon production from pions and rhos.}
       \protect\label{fig2.9}
\end{figure}

The lowest order processes from this effective theory are pion annihilation,
$\pi^+ \pi^- \rightarrow \rho \gamma$, ``Compton scattering''
$\pi^\pm \rho \rightarrow \pi^\pm \gamma$, and $\rho$-decay, $\rho \rightarrow
\pi^+ \pi^- \gamma$, as shown in Fig.~\ref{fig2.9}. 
Kapusta, Lichard, and Seibert
have also considered the processes $\pi^+ \pi^- \rightarrow \eta \gamma $,
$\pi^\pm \eta \rightarrow \pi^\pm \gamma$, $\pi^+ \pi^- \rightarrow \gamma
\gamma$, and $\omega \rightarrow \pi^0 \gamma$. Apart from the last all
these processes are suppressed compared to the ones of Fig. \ref{fig2.9} 
by at least
an order of magnitude. The decay $\omega \rightarrow \pi^0 \gamma$ dominates 
over the rho meson decay above a photon energy of $E=0.5$ GeV. However, 
the contribution from the $\omega$-decay to the photon production, 
following from an extrapolation from $pp$ collisions, has 
also been subtracted from the experimental data \cite{Aggarwal:2000:long}.
Hence, the $\omega$-decay contribution is taken into account only partly in 
the spectra presented by WA98.

The matrix elements of the processes shown
in Fig. \ref{fig2.9} and of the other processes, discussed above, have been 
listed, e.g. in Ref.\cite{Sarkar98,Roy99}. 
Folding them with the hadron distribution functions, similar 
as in Eq. (\ref{rate1}), the photon production rate corresponding to these
processes from a HHG has been evaluated numerically. Note that
many of the involved mesons are rather short-living, such as the rho meson. 
Therefore, one should
use modified distributions for unstable particles \cite{Weldon93}.
However, it has been shown that the influence of a finite width of the 
rho meson has a negligible effect on the photon production rate
\cite{Sarkar98}.

Parametrizing the
numerical results, the following closed expressions have been given
for the various rates following from Fig. \ref{fig2.9} \cite{Nadeau92}
\bea
E\frac{dN}{d^4xd^3p}\biggl |_{\pi \pi \rightarrow \rho \gamma} &=& 0.0717\>
T^{1.866}\> \exp (-0.7315/T+1.45/\sqrt{E}-E/T),\nonumber \\
E\frac{dN}{d^4xd^3p}\biggl |_{\pi \rho \rightarrow \pi \gamma} &=& 
T^{2.4}\> \exp [-1/(2TE)^{3/4}-E/T],\nonumber \\
E\frac{dN}{d^4xd^3p}\biggl |_{\rho \rightarrow \pi \pi \gamma} &=& 0.0785\>
T^{4.283}\> E^{-2.976+0.1977/T}\> \exp (-E/T).
\label{nadeau}
\eea
Here the temperature $T$ and photon energy $E$ are to be given in GeV, and
the invariant rate 
then has dimensions of fm$^{-4}$ GeV$^{-2}$. These expressions
are accurate compared to the numerical results
to at least 20\% in the range 100 MeV $<T<$ 200 MeV and
0.2 GeV $<E<$ 3 GeV.   

Comparing the HHG rate
at a temperature of $T=200$ MeV to the 1-loop HTL rate Eq. (\ref{1-loop}),
it was found that both rates have a very similar shape and magnitude.
Hence, Kapusta, Lichard, and Seibert concluded that ``the hadron gas shines just
as brightly as the quark-gluon plasma'' \cite{Kapusta91}. This coincidence
between the rates of the two different phases has also been related to
the ``quark-hadron duality'' \cite{Rapp00,Gallmeister00}. However, as we have
discussed already above, the QGP photon rate is enhanced by 
2-loop HTL corrections and the influence of higher-order corrections is 
unknown. Also the HHG photon rate is changed by including further processes
and particles, in particular the $a_1$ vector meson, as we will discuss below.
Therefore the agreement of both rates might be a mere coincidence.
We will come back to this point below. 

After this first calculation of the photon emission from the HHG, Xiong, 
Shuryak and Brown \cite{Xiong92} found that the process $\pi \rho
\rightarrow \pi \gamma$ is significantly enhanced if an intermediate
$a_1$ resonance state is taken into account. A parametrization
of the numerical result for this contribution reads 
\beq
\hspace*{-1cm}E\frac{dN}{d^4xd^3p}
\biggl |_{\pi \rho \rightarrow a_1\rightarrow \pi \gamma} 
[{\rm{fm}}^{-4}{\rm{GeV}}^{-2}]
= 2.4\> T^{2.15}\> \exp [-1/(1.35TE)^{0.77}-E/T].
\label{xiong}
\eeq
Although the $a_1$ contribution in Ref.\cite{Xiong92} has been 
overestimated\footnote{Xiong, Shuryak and Brown
\cite{Xiong92} proposed
an effective Lagrangian for the interaction between the $a_1$-, the
$\rho$-, and the $\pi$-mesons. The coupling constant was determined from the
decay width of the $a_1$. However, it was overestimated since the full width
instead of the partial width was assumed for each isospin channel in the
photon production via the $a_1$-resonance. This error led to an 
overestimation of the rate by a factor of 4 \cite{Haglin00}.} 
by a factor of 4, the total 
photon rate is enhanced by about a factor of 2 due to this contribution.

The role of the $a_1$ meson on the photon production has been studied
further starting from effective chiral Lagrangians \cite{Song93,Kim96}.
In this way other processes, in which the
$a_1$ participates, and interference effects 
have been included. This leads to a further enhancement
of the rates. However, the final result depends crucially on the specific
form of the Lagrangian and the choice of its parameters, which cannot be 
fixed unambiguously \cite{Gale01,Song93}. Therefore, the final photon rate from
the HHG can easily vary by a factor of about 3 depending on the assumptions
of the effective theory used \cite{Gale01,Song93}. An alternative, more
model independent approach, based on constraints from data (electro
production, tau decay, radiative pion decay, 2-photon fusion) and
general arguments (broken chiral symmetry, current conservation, unitarity)
\cite{Steele96,Steele97,Lee98} indicates a somewhat larger photon production
compared to most estimates from using effective chiral Lagrangians.

\begin{figure}[hbt]
       \centerline{\resizebox{10cm}{!}{\includegraphics{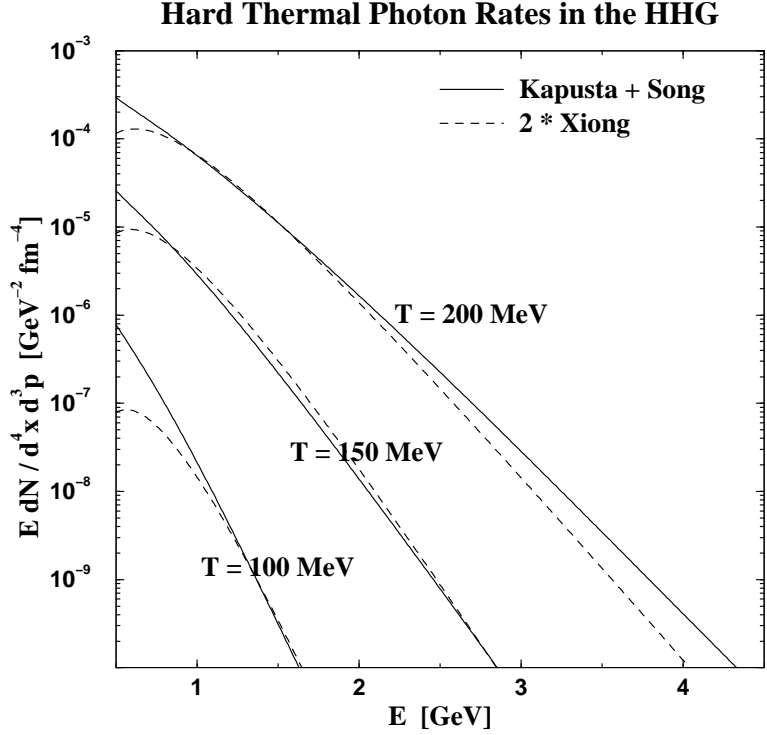}}}
\caption{Comparison of the pocket formula Eq. (\ref{hadrons}) for
HHG photon production rates (dashed line) with the rates derived
by Kapusta and Song (solid line) at $T=100$, 150, 200 MeV.}       
\protect\label{fig2.10}
\end{figure}

As a simple 
estimate the following expression for the HHG photon production rate
has been suggested \cite{Steffen01}\footnote{This expression is identical 
with the
one given by Xiong et al. \cite{Xiong92} for the $a_1$ contribution 
multiplied by a factor of 2 in order to take into account the contributions
from Ref.\cite{Kapusta91} and Ref.\cite{Song93}.}
\beq
E\frac{dN}{d^4xd^3p}[{\rm{fm}}^{-4}{\rm{GeV}}^{-2}]
\simeq 4.8\> T^{2.15}\> e^{-1/(1.35ET)^{0.77}}\> e^{-E/T}\>.
\label{hadrons}
\eeq
Alternatively the sum of the rates from Ref.\cite{Nadeau92,Song93} - the 
rates in Ref.\cite{Song93} are not given analytically - can be used.
Both approximations for the HHG rate agree at least within the uncertainties, 
discussed below, for relevant temperatures between 
100 and 200 MeV and photon energies of interest between 1 and 4 GeV 
(see Fig. \ref{fig2.10}).

\begin{figure}[hbt]
       \centerline{\resizebox{10cm}{!}{\includegraphics{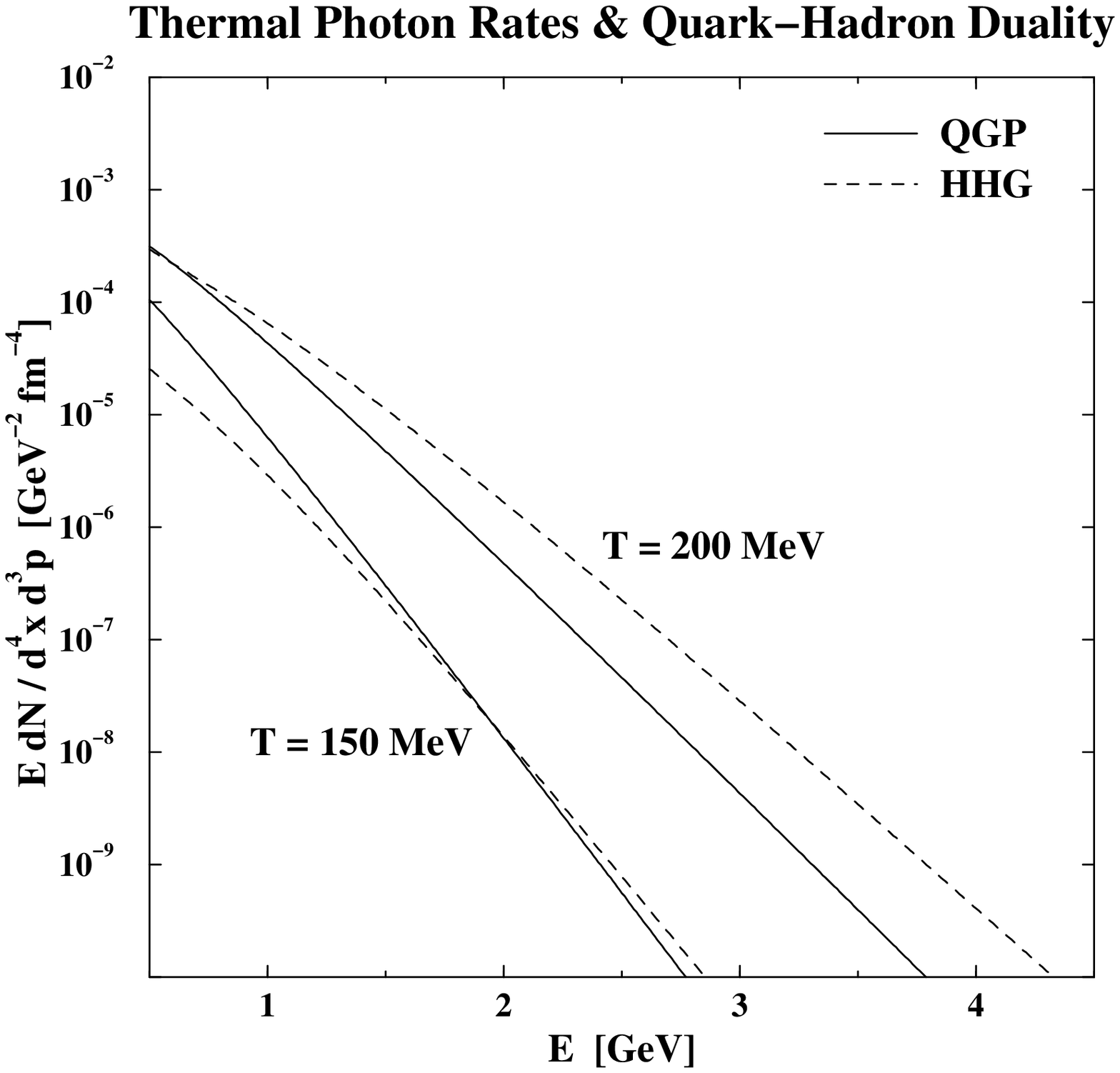}}}
\caption{Comparison of the photon production rate from the QGP and the 
hadron gas at $T=150$ and 200 MeV ($N_F=2$) \cite{Steffen01}.}       
\protect\label{fig2.11}
\end{figure}

In Fig. \ref{fig2.11} the thermal photon rate from the QGP Eqs. 
(\ref{1-loop}), (\ref{brems}), and (\ref{aws})
and the hadron gas Eq. (\ref{hadrons}) at the same temperature are compared.
Note that the rates from the two phases agree approximately at $T=150$ MeV, 
but not at 200 MeV. The 
approximate agreement of the QGP and the HHG rate at $T=150$ MeV appears to be 
accidental as the energy and temperature dependence of the HHG rate Eq. (\ref{hadrons}),
obtained from fitting numerical results, and of the
QGP rate Eqs. (\ref{1-loop}), (\ref{brems}), and (\ref{aws}), derived in the weak
coupling limit, are different. 

Recently, also the role of in-medium effects of vector mesons in the HHG
on the photon production has been investigated 
\cite{Sarkar98,Roy99,Halasz98,Song98}. For a review on this subject see 
Ref.\cite{Alam01}.
The results depend on the model used for implementing medium effects
on hadrons. Whereas the change of the width appears to be rather unimportant
for the photon production rate \cite{Sarkar98}, changes in the mass of
the vector and axial vector mesons could have significant consequences.
In particular, many models predict a dropping vector meson mass with
increasing temperature and baryon density, such as Brown-Rho scaling 
\cite{Brown91}. The reduction of the $\rho $ and $a_1$ masses
in the HHG is expected to cause an enhancement of the photon
production rate. Song and Fai \cite{Song98} predicted an enhancement
of the rate by an order of magnitude, whereas Sarkar et al. \cite{Sarkar98}
found only an enhancement by a factor of 3. Hal\'asz et al. \cite{Halasz98},
on the other hand, found a reduction of the $a_1$-contribution to the
photon production by a factor 2 - 3 
compared to scenarios without in-medium modifications of the masses
\cite{Xiong92,Song93,Kim96}. Their conclusion is based on using
the Hidden Local Symmetry model \cite{Bando88}, in which there
is a linear relation between the coupling and the vector meson masses.
Hence, a reduced mass leads to smaller coupling which suppresses
the photon rate. The photon production rate obtained in this way lies between
the one found by Kapusta et al. \cite{Kapusta91} and the one of
Song \cite{Song93}. However, it is not clear whether this reduction of
the $a_1$ contribution by medium effects is a physical effect or caused 
by the particular choice of the effective Lagrangian \cite{Halasz98}. 

The radiative decay of the axial vector mesons, $a_1\rightarrow \pi \gamma$,
$b_1\rightarrow \pi \pi^0 \gamma$, and $K_1\rightarrow K\gamma $ has been 
discussed by Haglin \cite{Haglin94}. These contributions appear to be 
important, i.e. comparable to the $\pi \rho \rightarrow \pi \gamma $ and
the $\pi \pi \rightarrow \rho \gamma$ contributions, for photon energies 
below 1.5 - 2 GeV and to be dominant for $E<1$ GeV. 

In the analysis of the photon emission rate from a HHG, based on 
constraints of data and general arguments (see above) by Steele,
Yamagishi, and Zahed \cite{Steele96}, also a finite pion chemical 
potential describing a dilute pion gas, i.e. a deviation from
chemical equilibrium, has been taken into account. 
Assuming $\mu_\pi =100$ MeV, an enhancement of the
photon production rate by about a factor of 2 compared to
an equilibrated pion gas at the same temperature has been observed
\cite{Steele96}. Furthermore, a finite 
baryon density corresponding to the presence of nucleons, as it is 
the case at SPS, has been shown to increase the rates further by about a 
factor of 1.5 below $E=1.5$ GeV \cite{Steele97}. 
The influence of strange mesons ($\eta$, $\Phi$, $K$) included in this 
investigation turned out to be negligible for the photon rate 
\cite{Lee98}. 

Finally, let us mention, that bremsstrahlung from the HHG seems to affect 
only the photon production rate at small energies below about 100 MeV
\cite{Jeon98,Eichmann00}.

Summarizing, there are still significant uncertainties in the photon 
production rate from the HHG in spite of intense effort during the last 
ten years. The photon production rate of the HHG is at best known up
to a factor of 3. Within this uncertainty it appears to be of the
same magnitude as the 2-loop HTL result for the photon rate from the QGP
in the relevant temperature regime. This statement is sometimes associated 
with the quark-hadron duality hypothesis
\cite{Rapp00,Gallmeister00}. However, even if the QGP and the HHG rates
are similar, the QGP 
might be distinguishable from the hadron gas in the photon spectrum due to
a different space-time evolution of the two phases as discussed below.

\subsubsection{Prompt Photons}\label{subsubsec:pro}

Besides thermal emission of photons from the QGP and the HHG there is 
another source for direct photons coming from hard parton collisions
in the initial non-ther\-mal stage of the heavy-ion collisions. These 
so-called prompt photons have to be subtracted as well as the thermal HHG 
photons for identifying the QGP radiation. On the other hand, prompt photons
in heavy-ion collisions may contain interesting information on nuclear
effects on the parton 
distributions. As a matter of fact, an enhancement in the pion and photon
production in $pA$ collisions compared to results from a 
simple scaling from $pp$ collisions has been observed experimentally. 
This nuclear effect, also called Cronin effect \cite{Antreasyan79}, 
is most relevant at transverse momenta between 3 and 6 GeV \cite{Brown96}.
There are also indications for a nuclear enhancement in the WA98 data
above $p_T\simeq 2.5$ GeV \cite{Dumitru01}.

The production rate of prompt photons from hard parton scatterings can be
computed similarly as the QGP rate. The amplitudes of the basic processes
(Compton scattering, quark-antiquark annihilation, and bremsstrahlung)
are folded with the parton distributions. The thermal distribution functions 
have now to be replaced by the parton distributions in the nuclei.
 
Let us first consider $pp$ collisions. Assuming the  
QCD factorization theorem, the photon production
cross section for the process $pp \rightarrow X\gamma $
is given by (see e.g. \cite{Apanasevich98})
\beq
\hspace*{-0.5cm}
E\frac{d\sigma}{d^3p}=\sum_{abc} \int dx_a dx_b\> f_a(x_a,Q)\> f_b(x_b,Q)\>
K\>  \frac{s}{\pi} \frac{d\sigma}{dt}(ab\rightarrow c\gamma)
\> \delta (s+t+u), 
\label{prompt}
\eeq
where $f_i$ are the parton distribution functions in the nucleons,
depending on the parton momentum fraction $x_i$ and the factorization 
scale $Q$, and $d\sigma /dt$ is the differential cross section for 
the elementary parton process ($ab\rightarrow c\gamma$), e.g. 
Compton scattering,
with the Mandelstam variables $s$, $t$, and $u$. 
The sum extends over all 
possible parton states and $K\simeq 2$ is a phenomenological
factor taking account of 
next-to-leading order effects. The integrals in  Eq. (\ref{prompt}) are
performed numerically using Monte-Carlo techniques. 

In order to explain the experimental data \cite{Vogelsang97}, two different 
approaches have been employed. The first approach is based on next-to-leading
order calculations of the cross sections, where the renormalization scale
$\Lambda_{\rm QCD}$ and the factorization scale $Q$ are determined in a way 
to optimize the agreement between theory and experiment \cite{Aurenche87}. 
%Using a single set of structure functions and a unique value for 
%$\Lambda_{\rm QCD}$ an excellent agreement with the data over the entire
%range of collision energies and photon momenta was obtained \cite{Gordon97}.
%Only the most recent data from E706 \cite{Apanasevich98} using nuclear 
%targets and from R806 (ISR) \cite{Annassontzis82} 
%could not be fitted in this way, which might indicate an inconsistency
%of the data \cite{Aurenche99}. 
This method has been improved further on by using a soft-gluon 
resummation and
considering next-to-next-to-leading order corrections \cite{Kidonakis00}.

The second approach uses non-optimized 
scales but introduces a phenomenological,
non-perturbative effect in the parton distribution, namely 
%% tp
% intrinsic
% $k_T$-broa\-den\-ing \cite{Fontannaz78,Hutson95}. 
a transverse momentum distribution of finite width, called intrinsic
$k_T$ \cite{Feynman78,Fontannaz78,Hutson95}. 
%% tp ende
For this purpose the parton distribution 
functions $dx_i\> f_i(x_i,Q)$ are replaced by $dx_i\> d^2{k_T}_i 
\> f_i(x_i,Q)  g({k_T}_i)$, where ${k_T}_i$ is the transverse
parton momentum of the parton in the nucleon. Then one has to integrate
additionally over ${k_T}_i$ in Eq. (\ref{prompt}). The transverse momentum
distribution is usually approximated by a Gaussian
\beq
g(k_T)=\frac{e^{-k_T^2/\langle k_T^2\rangle}}{\pi \, \langle
k_T^2\rangle},
\label{transverse}
\eeq
where the average square of the intrinsic transverse momentum 
of the parton in the initial state, $\langle k_T^2\rangle$, is a tunable
parameter. Using the uncertainty principle for the partons confined in
the nucleon with radius $r_N$ one finds 
$\sqrt{\langle k_t^2 \rangle } \approx \pi / 2r_N  \approx 0.37~\mbox{GeV}$
\cite{Dumitru01}. However, this value
is too small to explain the data, which requires $\langle k_T^2 \rangle
= 1$ - 1.5
GeV$^2$ \cite{Dumitru01,Hutson95}. 

%% tp
% Intrinsic $k_T$-broadening 
Intrinsic $k_T$ 
%% tp ende
can also be caused by multiple gluon radiation \cite{Lai98}, which makes
$\langle k_T^2\rangle $ energy dependent \cite{Papp00}. 
%% tp
% Intrinsic $k_T$-broadening 
Intrinsic $k_T$ 
%% tp ende
has also been applied successfully to explain muon, jet, and
hadron production in $pp$ collisions at Tevatron, such as $\pi^0$- and
$J/\psi $ production \cite{Papp00}. The cross section for photon production
is expected to be increased by a factor of 3 to 8 by 
%% tp
% intrinsic $k_T$-broadening 
intrinsic $k_T$ 
%% tp ende
\cite{Wong98}. Further improvement of fitting the 
data can be achieved by allowing for a $K$-factor dependence on the 
collision energy and photon momentum \cite{Wong98,Barnafoldi00}. 
 
Summarizing the status of prompt photons in $pp$ collisions, we quote
Ref.\cite{Aurenche99}: ''Despite many years of intense experimental and
theoretical efforts, the inclusive production of prompt photons in hadronic 
collisions does not appear to be fully understood.'' 

New effects and
further uncertainties
arise in the extrapolation of the prompt photon production rate from
$pp$ to $pA$ and heavy-ion collisions. The photon spectrum 
$E\, dN/d^3p$ for
prompt photons in $pA$ and $AA$ collisions follows from the cross section
for $pp$ collisions Eq. (\ref{prompt})
by introducing a nuclear thickness function and 
integrating over the impact parameter \cite{Dumitru01,Papp00}.
Nuclear effects on the parton distributions are expected to play an important 
role. For example, an additional $k_T$ broadening from soft nucleon 
collisions in the nucleus prior to the hard collision (Cronin effect) 
has been predicted \cite{Papp00}. Nuclear broadening has been observed, e.g. 
in the dimuon production in $pA$ collisions \cite{McGaughey99}. It also
allows an understanding of the $\pi^0$ production at SPS 
\cite{Gyulassy98,Wang98}.  Furthermore, it can lead to a strong enhancement of 
the prompt photon cross 
section in $AA$ collisions, because a part of the photon momentum can
be supplied by the incoming partons \cite{Dumitru01}. 
%Dumitru et al.
%\cite{Dumitru01} showed that the WA98 photon spectrum above $p_T=2.5$ GeV
%can be explained by prompt photons if a nuclear broadening of
%$\Delta k_T^2=\langle k_T^2\rangle_{AA} -\langle k_T^2\rangle_{pp}\simeq
%0.5$ - 1 GeV$^2$ is introduced. 

Other nuclear effects, which might play a role, are 
the parton energy loss and nuclear shadowing 
\cite{Jalilian00}. They are expected to lead to a
suppression of the prompt photon cross section 
of about 30\% at RHIC energy. At SPS energies, on the other hand, a small
enhancement of the photon production by antishadowing is expected
\cite{Dumitru01}.

Finally, a significant contribution (``strong flash of photons'')
to the photon production from
the early non-thermalized stage of the fireball in heavy-ion collisions
has been predicted using the parton cascade model \cite{Srivastava98a}.
These photons are produced from the fragmentation of time-like
quarks ($q\rightarrow q\gamma$), produced in semi-hard multiple
scatterings in the pre-equilibrium phase.
However, recently there have been some doubts raised on this result by one 
of the authors \cite{Srivastava01}.

Concluding, the production of prompt photons in heavy-ion collisions is
not well understood at the moment. As we will see below, this leads to 
controversial conclusions about the role of prompt photons in the photon
spectrum at SPS measured by WA98. In order to predict the prompt photon
spectrum at RHIC and LHC precise $pp$ and $pA$ data on photons at the
corresponding energies will be very helpful \cite{Papp00}.

Summarizing the status of the
theoretical investigations of the direct photon production 
rate in heavy-ion collisions, new methods for calculating the rate
from the QGP as well as improvements of the HHG and prompt photon
rates are necessary. Only then will it be possible to make reliable 
predictions which can be used for a comparison of theoretical and experimental
spectra at SPS as well as at RHIC and LHC.

\subsection{Hydrodynamics and Photon Spectra}\label{subsec:hps}

The static thermal photon production rates 
discussed above cannot be compared directly 
to the experiment, in which only spectra and yields of the photons
from the entire space-time evolution of the fireball can be observed. 
Therefore, one has to convolute the rates
with the space-time evolution to obtain the photon spectrum. 
In the present Section, we will 
consider the basic concepts and the theoretical description of the
space-time evolution of the fireball in relativistic heavy-ion collisions.
In particular, we will discuss hydrodynamical methods and their application
to the computation of photon spectra. The assumptions
and approximations of the hydrodynamical models 
are another source for uncertainties in
predicting the photon production, as we 
will see below.

\subsubsection{Space-Time Evolution of the Fireball}\label{subsubsec:hyd}

\begin{figure}[hbt]
\centerline{\resizebox{12cm}{!}{\includegraphics{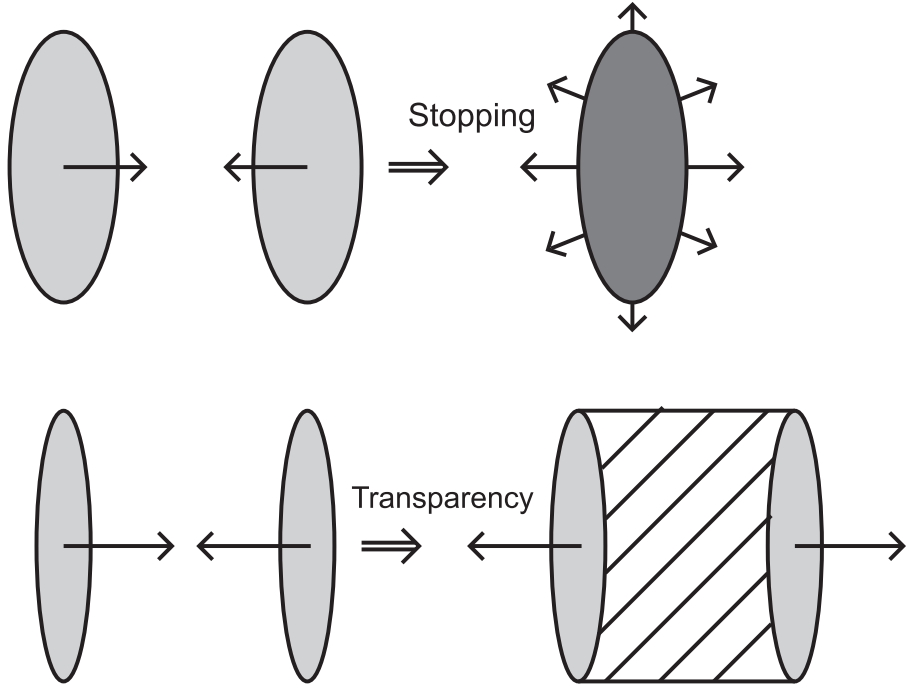}}}
\caption{Nucleus-nucleus collisions at $\sqrt{s}\ll 100$ GeV (stopping)
and at $\sqrt{s}\simeq 100$ GeV or larger (transparency).}       
\protect\label{fig2.12}
\end{figure}

\begin{figure}[hbt]
\centerline{\resizebox{10cm}{!}{\includegraphics{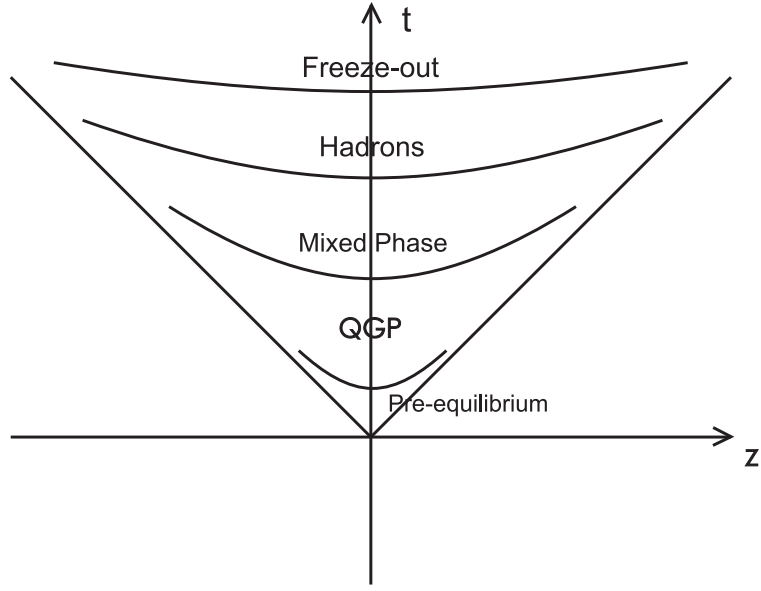}}}
\caption{Space-time diagram of a ultrarelativistic
heavy-ion collision in the Bjorken scenario.}
\protect\label{fig2.13}
\end{figure}

There are two basic scenarios for the space-time evolution of the fireball
in relativistic heavy-ion collisions \cite{Mueller85} as shown in 
Fig. \ref{fig2.12}.
For collision energies $\sqrt{s}\ll 100$ GeV (AGS, SPS), 
the nuclei are stopped 
in the collision to a large extent 
and a dense and hot expanding fireball with a finite 
baryon density (finite chemical potential)
is formed, which might be in the QGP phase initially if the
critical temperature $T_c$ of about 150 - 170 MeV is exceeded. 
The expansion leads to a temperature drop until $T_c$ is reached, at which
hadronization sets in. After a potential 
mixed phase and hadronic phase, the interactions
in the fireball will finally freeze out allowing the hadrons to
propagate freely.
In the second scenario, expected to be valid for $\sqrt{s} \simeq 100$ GeV
or larger (RHIC, LHC),
there is not enough time for the highly Lorentz contracted nuclei to be 
stopped in the nucleus-nucleus collision. Rather they propagate through 
each other approximately 
transparently. However, the vacuum between the receding nuclei
will be highly excited from the initial hard parton collisions and will
decay violently into a baryon-free (zero chemical potential)
parton gas by secondary parton collisions
or, in a non-perturbative picture, by string decay. The secondary collisions
will drive the parton gas to thermal equilibrium, corresponding to
the QGP stage. The system is mainly expanding in beam direction in a 
boost-invariant way (Bjorken scenario) \cite{Bjorken83,Gyulassy84}, 
accompanied by a cooling of the fireball. 
The various stages, mixed phase, hadronic phase, and freeze-out,
follow as in the first scenario. The space-time diagram of the second
scenario, showing the various stages, is sketched in Fig. \ref{fig2.13}. 
The $z$-axis agrees with the beam direction. At $t=0$, the maximum
overlap of the nuclei takes place. The produced particles in this diagram
lie above the light-cone due to causality. The hyperbolas denote
curves of constant proper time $\tau=\sqrt{t^2-z^2}$, on which the same 
physics, e.g. energy density and temperature, occurs, according to the 
boost-invariant Bjorken scenario \cite{Bjorken83}.

In order to speak of the QGP as a thermal system, we need a large volume 
and particle number, and a sufficiently long life-time of the equilibrated
system. Rough estimates give a sufficiently large volume of the order of 
1000 fm$^3$ at RHIC or
LHC, a large parton number up to about a several thousand, 
and a sufficiently long
life-time of the parton 
gas of  5 - 10 fm/c before hadronization sets in. For the formation time
of the QGP a typical value of the order of 0.5 - 1 fm/c has been accepted 
\cite{Satz92}. However, doubts have been raised, whether
the parton gas in a heavy-ion collision will reach a thermalized stage at 
all, at least by elastic scatterings as assumed usually \cite{Serreau01}.
Moreover, the realization of a chemical equilibrium between gluons and
light quarks appears to be questionable at RHIC and LHC \cite{Biro93,Xiong94}.

In order to describe the dynamical evolution of a many-particle system in
non-equi\-li\-bri\-um or equilibrium, transport models are adopted. Starting
from the Boltzmann equation \cite{Reif65}, describing the transport of 
different interacting hadron species semi-classically
\cite{Bertsch88}, particle production, e.g. photon production,
in heavy-ion collisions up to collision
energies of about 1 A$\cdot$GeV can be treated 
successfully \cite{Cassing90}. Transport models have also been used to 
describe the photon spectrum in relativistic heavy-ion collisions
\cite{Hui00}. However, in these approaches only a hadron gas but not a
QGP phase has been considered, which requires the transport theoretical 
description of a parton gas. Although such microscopic models for the 
parton-gas dynamics based on perturbative QCD exist \cite{Geiger95,Wang97}, 
they have not been applied to photon production from the QGP so far. 
Hence, no
transport theoretical predictions of photon spectra in relativistic
heavy-ion collisions taking into account a QGP phase are available.

To illustrate the hydrodynamical calculation of the photon spectrum, we will
consider in detail a simple hydrodynamical model in the following.
This model is certainly oversimplified as it neglects the transverse expansion
of the fireball and is based on an unrealistic EOS, which leads to a strong
first order phase transition and a long-lived
mixed phase in contradiction to lattice results \cite{Karsch01b}. More
realistic hydrodynamical descriptions including transverse flow and an 
improved EOS will be discussed subsequently.

Assuming a local thermal and chemical equilibrium hydrodynamical 
equations can be derived from the Boltzmann equation \cite{Reif65}. 
The relativistic hydrodynamical equations follow from the conservation 
of the baryon number, energy, and momentum. If we assume an ideal fluid, i.e.
neglect dissipative effects, the energy-momentum tensor is given by
\beq
T^{\mu \nu}=(\epsilon+P)\> u^\mu \, u^\nu -P\> g^{\mu \nu},
\label{ideal}
\eeq
where $\epsilon $ is the energy density, $P$ the pressure, 
$u^\mu=\gamma (1,{\bf v})=\partial x^\mu /\partial \tau$ 
($\gamma=1/\sqrt{1-{\bf v}^2}$) the (local) 4-velocity of the fluid, 
and $g^{\mu \nu}$ the Minkowski metric.
From the conservation of the energy-momentum tensor, 
$\partial _\mu T^{\mu \nu}=0$, multiplied by $u^\nu$, the relativistic Euler 
equation follows
\beq
u^\mu\> \partial_\mu \epsilon + (\epsilon +P) \partial_\mu u^\mu =0.
\label{Euler}
\eeq
Assuming for simplicity only a longitudinal boost-invariant
expansion, i.e. $u^\mu =x^\mu/\tau$, (Bjorken scenario \cite{Bjorken83}),
as it might be the case approximately at RHIC and LHC energies,
the Euler equation can be written as
\beq
\frac{d\epsilon}{d\tau}+\frac{\epsilon+P}{\tau}=0.
\label{Bjorken}
\eeq
For an ideal ultrarelativistic gas, such as the non-interacting
QGP, $\epsilon =3P$ holds and the evolution of the energy density,
depending only on time, can be determined easily: $\epsilon=\epsilon_0
\> (\tau_0/\tau)^{4/3}$. Furthermore, one obtains 
$T=T_0\> (\tau_0/\tau )^{1/3}$.
Here $\tau_0$, $\epsilon_0$, and $T_0$ are the initial time,
energy density, and temperature, respectively. 
They are determined by the time at which 
the local equilibrium has been achieved. 

The results of a hydrodynamical model depend strongly on the choice
of the initial conditions. Therefore a reliable determination of the initial 
conditions is crucial. The initial conditions can be taken in principle from 
transport calculations describing the approach to equilibrium,
such as the parton cascade model (PCM) \cite{Geiger95} or HIJING 
\cite{Wang97}, which treat the entire evolution of the parton gas from the
first contact of the cold nuclei to hadronization. However, there are no
unambiguous criteria for determining the completion of the equilibration process in
these transport models.  Another possibility for fixing the initial conditions
comes from relating observables
to the initial conditions. For example,
the initial temperature can be related to the particle multiplicity
$dN/dy$, assuming an ideal parton gas with an isentropic expansion
\cite{Hwa85}  
\beq
T_0^3=\frac{c}{4a}\> \frac{1}{V_0}\> \frac{dN}{dy}.
\label{initial}
\eeq
Here $V_0=\pi R_A^2\tau_0$ is the initial volume 
with the nucleus radius $R_A\simeq 1.3\> A^{1/3}$ fm,
For the initial time $\tau_0$, one usually assumes values of 
the order of 1 fm/c. Furthermore, $c=2\pi^4/(45\zeta(3))\simeq 3.6$
and $a=8\pi^2/45+7\pi^2N_F/60$ with $N_F$ light quark flavors.
%The initial energy density follows from using the Stefan-Boltzman law for 
%the ideal QGP, $\epsilon_0=3aT_0^4$.  

Another relation between the initial temperature and the initial time,
which is used frequently, is based on an argument using the uncertainty 
principle \cite{Kapusta92}. The formation time $\tau $ of a particle with an 
average energy 
$\langle E\rangle$ is given by $\tau \langle E\rangle \simeq 1$.
The average energy of a thermal parton is about $3T$. Hence, we find 
$\tau_0\simeq 1/(3T_0)$. However, the formation time of a particle, i.e. the 
time required to reach its mass shell, is not necessarily identical
to the thermalization time \cite{Kapusta92}. Consequently,
the determination of the initial conditions is far from being
trivial. However, if data for hadron production are available, such as at SPS, 
they can be used to determine
or at least constrain the initial conditions for a hydrodynamical
calculation of the photon spectra \cite{Huovinen99}. 

Another essential ingredient for a hydrodynamical model is the EOS.
Since we want to describe a fireball undergoing a phase transition, we need 
an EOS for the QGP as well as for the HHG. The QGP EOS has been determined in 
lattice QCD \cite{Karsch00,Karsch01}, which shows a clear deviation from 
an ideal QGP at temperatures accessible in heavy-ion collisions. In most 
hydrodynamical calculations, however, a simple bag model
EOS has been used for the QGP \cite{Csernai94}. 
For a vanishing chemical potential, the 
pressure and energy density in this model are given by
\bea
P_q &=& g_q\> \frac{\pi^2}{90}T^4-B,\nonumber \\
\epsilon _q &=& g_q\> \frac{\pi^2}{30}T^4+B, 
\label{QGP_EOS}
\eea
where the effective number of degrees of freedom is
\beq
g_q=2\> (N_C^2-1)+\frac{7}{8}\> 4\> N_C\> N_F
\label{dof}
\eeq
with the number of colors $N_C=3$. For two active quark flavors ($N_F=2$)
one gets $g_q=37$ and for three ($N_F=3$) $g_q=47.5$, 
respectively. The bag constant
$B$ is related to the critical temperature (see below) and typically of the 
order $B^{1/4}=200$ MeV. The EOS is given by 
$\epsilon_q=3P_q+4B$. 

For the HHG EOS usually an ideal hadron gas is adopted. However,
the number of hadron species included varies. Typically, all hadrons up
to masses of 2 or 2.5 GeV are taken. For illustration we will restrict 
ourselves to a massless pion gas \cite{Steffen99}. 
Then the pressure and energy density are
given by 
\bea
P_h &=& g_h\> \frac{\pi^2}{90}T^4,\nonumber \\
\epsilon _h &=& g_h\> \frac{\pi^2}{30}T^4 
\label{HHG_EOS}
\eea
with $g_h=3$ and $\epsilon_h=3P_h$. In fact, comparing the photon spectrum at
SPS energies, obtained by using this simple EOS, with results from 
using more realistic EOS, e.g. \cite{Srivastava00}, one finds that 
$g_h=3$ should be replaced by $g_h=8$ \cite{Steffen01}.

The two EOS are matched together by the Gibbs criteria: $T_c^q=T_c^h=T_c$
and $P_c^q=P_c^h=P_c$. Together with Eqs. (\ref{QGP_EOS}) and (\ref{HHG_EOS})
a relation between the bag constant and the critical temperature follows
\beq
T_c^4=\frac{90B}{(g_q-g_h)\pi^2}.
\label{bag}
\eeq
For instance, a bag constant of $B^{1/4}=200$ MeV implies $T_c=144$
MeV for $g_q=37$ and $g_h=3$. Now in addition to the initial conditions,
$\tau_0$ and $T_0$, there are two more parameters, namely the
critical temperature $T_c$ and the freeze-out temperature $T_f$, where the 
hydrodynamical evolution ceases. The critical temperature, predicted by
lattice QCD, is in the range $170 \pm 20$ GeV \cite{Karsch01}, and
the freeze-out temperature should be between 100 and 160 MeV \cite{Braun98}. 

\begin{figure}[hbt]
\centerline{\resizebox{10cm}{!}{\includegraphics{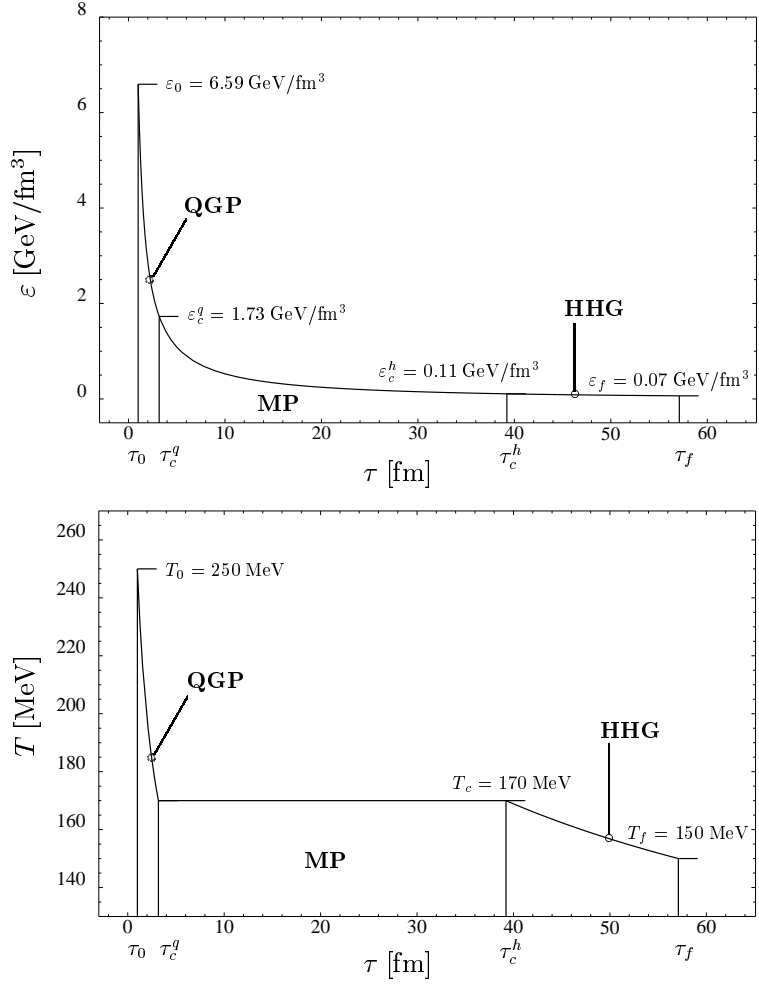}}}
\caption{Energy density and temperature evolution for $g_q=37$, $g_h=3$, 
$\tau_0=1$ fm, $T_0=250$ MeV, $T_c=170$ MeV, and $T_f=150$ MeV 
\cite{Steffen99}.}
\protect\label{fig2.14}
\end{figure}

\begin{figure}[hbt]
\centerline{\resizebox{10cm}{!}{\includegraphics{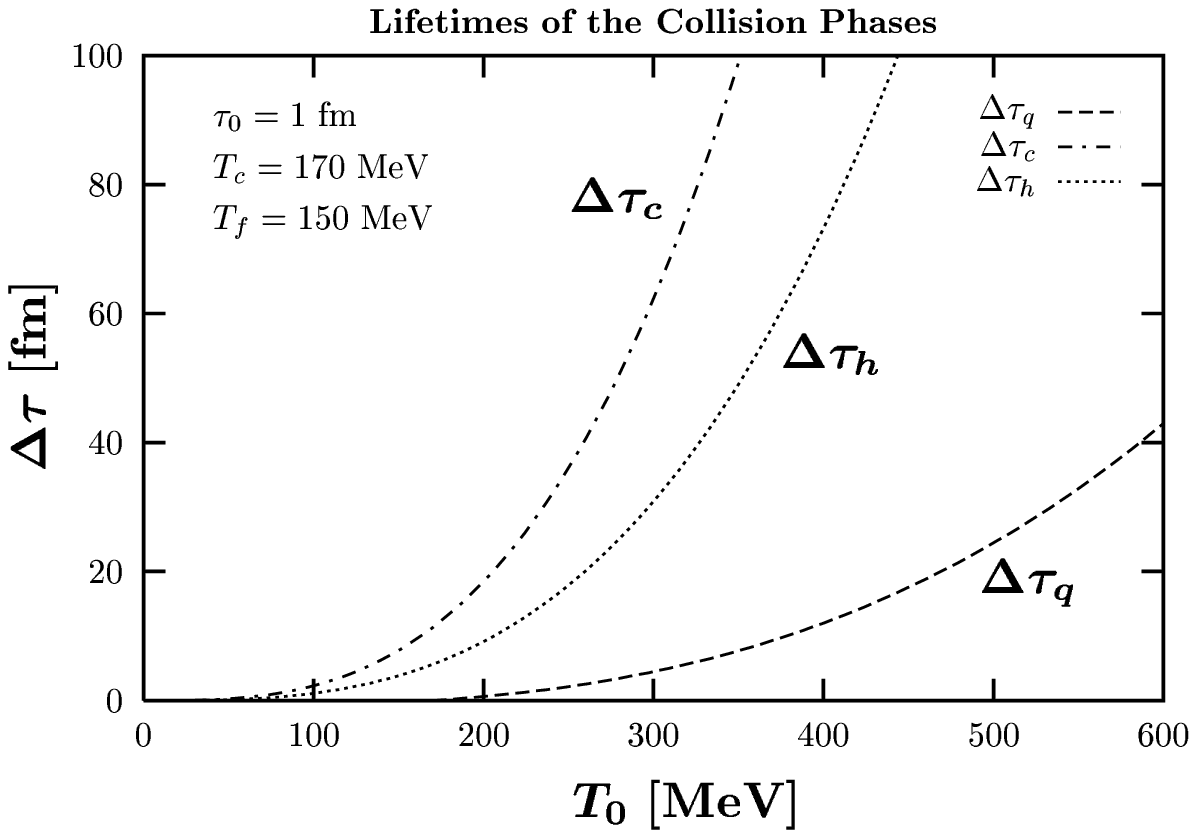}}}
\caption{Life-times of the QGP (dashed line), the mixed phase (dash-dotted 
line), and the HHG (dotted line) as a function of $T_0$ \cite{Steffen99}.}
\protect\label{fig2.15}
\end{figure}

\begin{figure}[hbt]
\centerline{\resizebox{10cm}{!}{\includegraphics{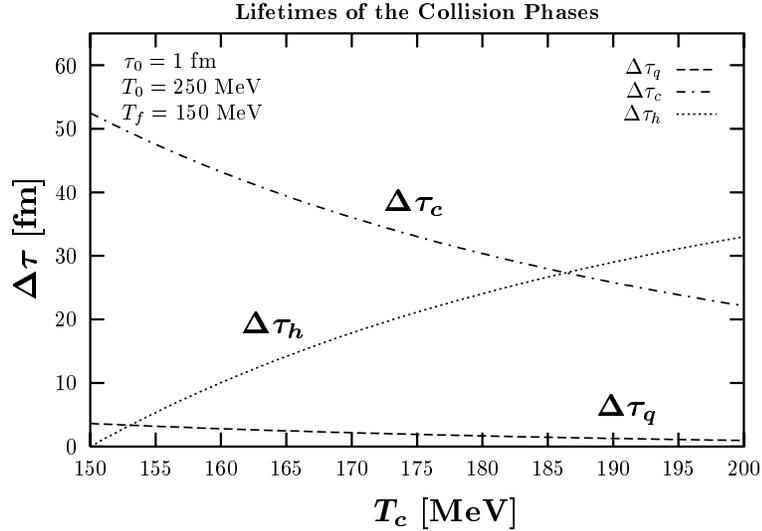}}}
\caption{Life-times of the QGP (dashed line), the mixed phase (dash-dotted 
line), and the HHG (dotted line) as a function of $T_c$ \cite{Steffen99}.}
\protect\label{fig2.16}
\end{figure}

The construction above implies the existence of a mixed phase corresponding
to a first-order phase transition (see Fig. \ref{fig2.14}). 
Although lattice QCD favors a continuous
phase transition instead of a first-order transition \cite{Karsch01b}, 
lattice calculations show also a rapid change in the energy density
similar as in the bag and pion gas model due to a large increase in the 
number of degrees of freedom going from the HHG to the QGP. 
The life-times of the different phases in our 
simple model 
are given by \cite{Steffen99}
\bea
\Delta \tau_q &=& \tau_0 \> \left [ \left (\frac{T_0}{T_c}\right )^3-1\right ]
,\nonumber \\
\Delta \tau_c &=& \tau_0 \> \left (\frac{T_0}{T_c}\right )^3\>
\left [\frac{g_q}{g_h}-1\right ], \nonumber \\
\Delta \tau_h &=& \tau_0 \> \left (\frac{T_0}{T_c}\right )^3\>
\left [ \left (\frac{T_c}{T_f}\right )^3-1\right ],
\label{life}
\eea
where $\Delta \tau_c$ denotes the life-time of the mixed phase,
during which the temperature $T=T_c$ stays constant. 
The life-times of the different phases as a function of the initial 
and the critical temperature are shown in Fig. \ref{fig2.15} 
and Fig. \ref{fig2.16}.
The simple EOS of a massless pion gas 
leads to a strong first-order transition and hence to a very long-living 
mixed phase. Hence, it is important to use a realistic EOS for the HHG.
In particular, in the no-phase-transition scenario, to which 
the phase-transition scenario has to be compared for predicting
signatures, a realistic EOS is essential. For example, the initial
temperature in the massless pion gas has to be chosen unrealistically
high ($T_0=578$ MeV), if the initial temperature of the QGP is
$T_0=250$ MeV and identical values for the initial time and the
entropy are assumed in both scenarios \cite{Srivastava94}. 

The hydrodynamical model presented above for illustration
is certainly oversimplified. 
A transverse expansion cannot be neglected, in particular in the later
stages of the fireball, changing the photon spectra significantly at
RHIC and LHC (see below). There are different hydrodynamical models for
relativistic heavy-ion collisions on the market, 
which describe the expansion of the fireball in 2 or 3 space dimensions
\cite{Clare86,Wong94}.
Of course, the hydrodynamical equations can only be solved by rather
elaborate numerical techniques in this case. Also it might not be justified 
to restrict to an ideal fluid, but dissipation might be important. For 
example, perturbative estimates of the viscosity of the QGP yielded a
large value \cite{Danielewicz85,Thoma94}. Hence, the Euler equation should be 
replaced by the Navier-Stokes or even higher order dissipative
equations. However, dissipative effects render the numerical 
treatment much more difficult and introduce new parameters such as 
viscosity \cite{Clare86}. After all, first attempts in this direction
have been undertaken already \cite{Muronaga01}. 

Under the simplifying assumption of an ideal fluid, the 
hydrodynamical equations can be solved numerically
using the respective EOS
for each of the two phases and the initial conditions, such as initial time and 
temperature, as input. The final results depend strongly on the input 
parameters as well as on other details of the model, as
in the simple 1-dimensional case. Also it is important to adopt a 
realistic EOS, in particular for the hadron gas, as discussed above.
%Furthermore, the high $p_T$ tail above 2 GeV of the spectra depends
%on assumptions about the density profile of the fireball \cite{Dumitru99}. 

Finally, the deviation from a chemically equilibrated QGP, which is 
expected to be important at RHIC and LHC \cite{Biro93,Xiong94}, should be taken
into account. It is expected that the parton gas at RHIC and LHC energies
will be thermalized rapidly on a time scale of
0.5 to 1 fm/c \cite{Geiger95,Wang97}. However, a chemical equilibration of 
the plasma might require much more time if it is realized at all
\cite{Biro93,Xiong94}. This means that the parton abundances are less than
their equilibrium value. In order to describe the deviation from
chemical equilibrium, a time-dependent
gluon or quark suppression factor $\lambda_{g,q}(\tau )$, sometimes called 
fugacity, is introduced. Then the non-equilibrium distribution functions
read $f_g(E,\tau)=\lambda_{g}(\tau )\, n_B(E)$ and $f_q(E,\tau)=
\lambda_{q}(\tau )\, n_F(E)$, respectively. For example, the energy density
at zero baryon density is now
given by $\epsilon=(\lambda_g a_g + \lambda_q a_q)\, T^4$,
where $a_g=8\pi^2/15$ and $a_q=7\pi^2N_F/20$. This expression can be used
in Bjorken's hydrodynamic equation Eq. (\ref{Bjorken}), which yields
$[\lambda_g+(a_q/a_g)\lambda_q]^{3/4}\> T^3\tau={\rm constant}$
\cite{Biro93}.

The parton phase space will be populated by inelastic parton
reactions producing quarks and gluons. To lowest order those
are: $gg\rightarrow q\bar q$ and $gg\rightarrow ggg$. The time
dependence of the fugacities can now be determined from
rate equations, which contain the cross sections for theses processes
\cite{Biro93}. As a subtle point, screening
masses (Debye screening, thermal quark mass) have to be used
to cut off IR singularities in these cross sections.
These screening masses also depend on the fugacities, as screening
is less efficient in a dilute system \cite{Biro93}.
Solving the rate equations together with Bjorken's hydrodynamic
equation, the time-dependence of the fugacities and the temperature
is obtained. The initial values for the temperature and the fugacities 
can be taken from PCM \cite{Geiger95} or HIJING \cite{Wang97} at the 
moment at which thermalization is completed, i.e., as soon as there
are approximately exponential and locally isotropic momentum distributions
\cite{Biro93}. Using HIJING initial conditions, the initial fugacities are far
from their equilibrium value, $\lambda_{g,q}^{\rm eq}=1$, 
in particular for the quark 
component, $\lambda_q^0\ll 1$. 
Larger initial fugacities follow from the PCM. Anyway, due to the larger
cross sections for gluon production, one expects
much more gluons than quarks in the early stages of the fireball,
which is called the ``hot glue'' scenario \cite{Shuryak92}. The fugacities
increase with time but might never reach their equilibrium value before
hadronization sets in, in particular at RHIC \cite{Biro93}. At the same time
the temperature of the fireball drops even more rapidly than in equilibrium
because the production of partons consumes energy. 

The above picture of 
chemical equilibration can also be incorporated in more realistic
hydrodynamical models containing also a transverse expansion 
\cite{Srivastava97}. It has been shown that the system evolves initially 
to chemical equilibrium but will be driven away from it at a later stage,
in which the transverse flow becomes important.

For predicting photon spectra from a chemically non-equilibrated 
QGP, it is not only necessary to modify the hydrodynamics, but also
the photon production rates change. Starting from Eq. (\ref{rate1}),
the equilibrium distribution functions have now to be replaced by
$f_{g,q}$, containing the fugacities. Also the fugacities have to 
be considered, for example, in the thermal quark mass in
Eq. (\ref{pre_htl}), $m_q^2=(\lambda_g+\lambda_q/2)\, g^2T^2/6$
\cite{Biro93}, serving as an IR cutoff. Modified rates for photon production
from Compton scattering, annihilation, and bremsstrahlung, obtained in this 
way, have been used to predict photon spectra for RHIC and LHC
\cite{Strickland94,Traxler96,Mustafa00,Dutta01,Srivastava97a}, 
which we will discuss in Section \ref{sec:cte}. 
A more sophisticated way is to calculate
the non-equilibrium rates by generalizing the HTL method to
chemical non-equilibrium \cite{Baier97,Carrington99}. Baier et al.
\cite{Baier97} have calculated the 1-loop HTL photon production rate 
in this way and found results similar to the ones of the simplified approach
of Ref.\cite{Traxler96}. 

Recently, Wang and Boyanovsky \cite{Wang00,Wang01} found a 
significant enhancement of the photon production for $p_T> 1.0$ - 1.5 GeV
due to the finite life-time of the QGP. Considering this non-equilibrium 
effect within the real time formalism, they obtained a power law spectrum for
the photons from off-shell quark bremsstrahlung, $q \rightarrow q\gamma $. 

\subsubsection{Photon Spectra}\label{subsubsec:spe}

As emphasized already a few times, photon spectra follow from convoluting
the photon production rates with the space-time evolution of the heavy-ion 
collision, for which usually hydrodynamical models are employed,
\beq
E\, \frac{dN}{d^3p} = \int d^4x\: E\, \frac{dN}{d^4x\, d^3p}.
\label{integ1}
\eeq
Here, the rate on the right-hand-side depends on the temperature, which 
depends in turn on the space-time coordinate in accordance with the 
assumption of a local equilibrium.

For illustration, but also because it is widely used, we will discuss
the calculation of the spectra using simple Bjorken hydrodynamics,
following Ref.\cite{Steffen99}. 
In this model the fireball is a longitudinally expanding
cylinder. Hence, we can write
\beq
\int d^4x = \pi\,R_a^2 \; \int dt\;dz,
\label{cylind}  
\eeq
where $R_A\simeq 1.3\> A^{1/3}$ fm and $z$ is the beam axis.
It is convenient to make a coordinate transformation to
proper time $\tau$ and rapidity $y'$ of the emitting fluid
cell, i.e. $t=\tau\> \sinh y'$ and $z=\tau \> \cosh y'$,
yielding
\beq
\int dt\;dz = \int_{\tau_0}^{\tau_f} d\tau\,\tau \; 
\int_{-y_{\rm nucl}}^{+y_{\rm nucl}} dy',
\label{integ2}
\eeq
where $\tau_0$ and $\tau_f$ are the initial and freeze-out times and
$y_{\rm nucl}=\mathrm{arcosh} [\sqrt{s}/(2A\cdot {\rm GeV})]$ \cite{Wong94} is the 
center-of-mass projectile rapidity. For SPS ($\sqrt{s} =17 A\cdot {\rm GeV}$)
one finds $y_{\rm nucl}=2.8$, for RHIC ($\sqrt{s} =200 A\cdot {\rm GeV}$)
$y_{\rm nucl}=5.3$, and for LHC ($\sqrt{s} =5500 A\cdot {\rm GeV}$)
$y_{\rm nucl}=8.6$. 
Using $E/d^3p=1/(d^2p_Tdy)$ with the transverse momentum $p_T$ and
the rapidity $y$ of the photon, one arrives at
\beq
\frac{dN}{d^2p_{\perp}\,dy}= 
\pi\,R_A^2 \; \int_{\tau_0}^{\tau_f} d\tau\,\tau \; 
\int_{-y_{\rm nucl}}^{+y_{\rm nucl}} dy'\, 
E\, \frac{dN}{d^4x\, d^3p}.
\label{spectrum}
\eeq
The thermal photon spectrum is defined in the center-of-mass system
and the photon rate on the right-hand-side in the local rest frame of the
emitting fluid cell, where the photon energy is given by
$E=p_T\> \cosh (y'-y)$. 

During the mixed phase the photon production rate is given as
\beq
E\, \frac{dN}{d^4x\, d^3p} 
= \lambda (\tau )\> \left(E \,\frac{dN}{d^4x\,d^3p} \,\right)_{QGP}
+ [1-\lambda (\tau )] \> \left(E ,\frac{dN}{d^4x\,d^3p} \,\right)_{HHG},
\label{mixed}
\eeq
where $\lambda (\tau)=V_{\rm QGP}(\tau)/V_{\rm tot}(\tau)$ 
is the QGP volume fraction.

\begin{figure}[hbt]
\vspace*{-2cm}
\centerline{\resizebox{17cm}{!}{\includegraphics{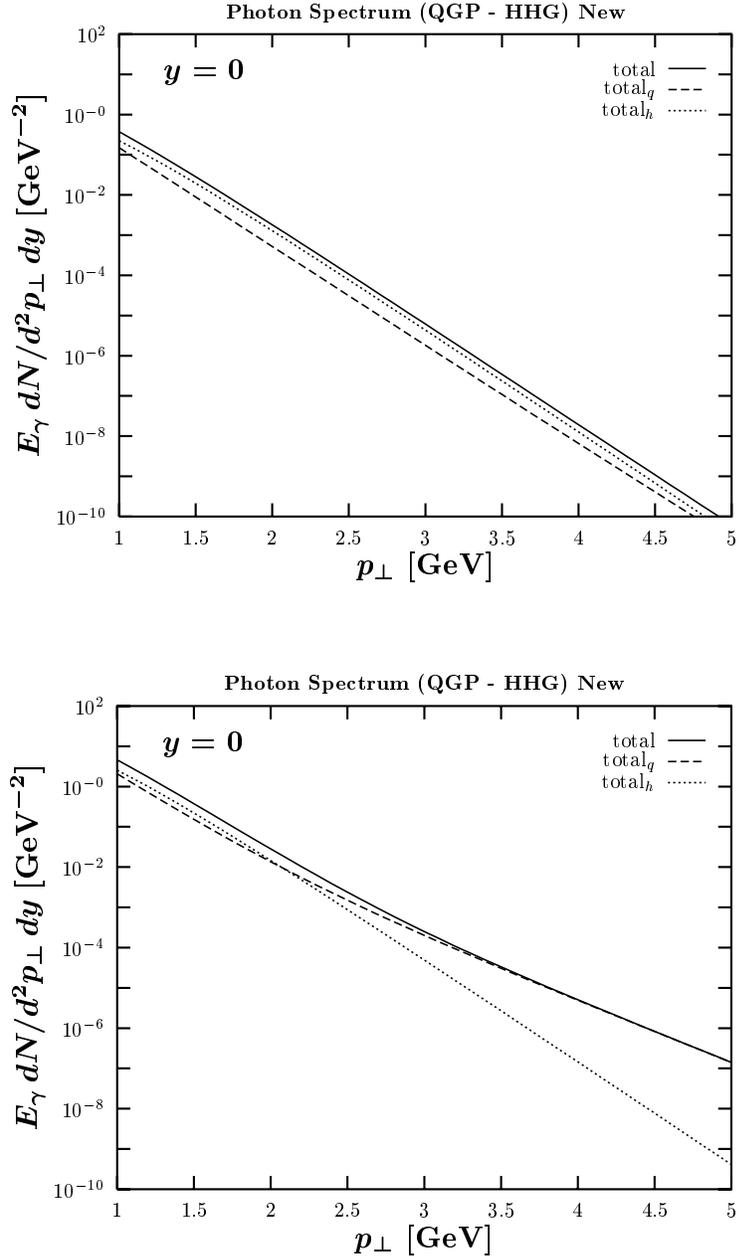}}}
\vspace*{-4cm}
\caption{Photon spectra at an initial temperature $T_0=200$ MeV (above)
and 300 MeV (below). The dashed line corresponds to the spectrum from
the QGP, the dotted one to the spectrum from the HHG, and the solid one
to the sum of both \cite{Steffen99}.}
\protect\label{fig2.17}
\end{figure}

\begin{figure}[hbt]
\centerline{\resizebox{10cm}{!}{\includegraphics{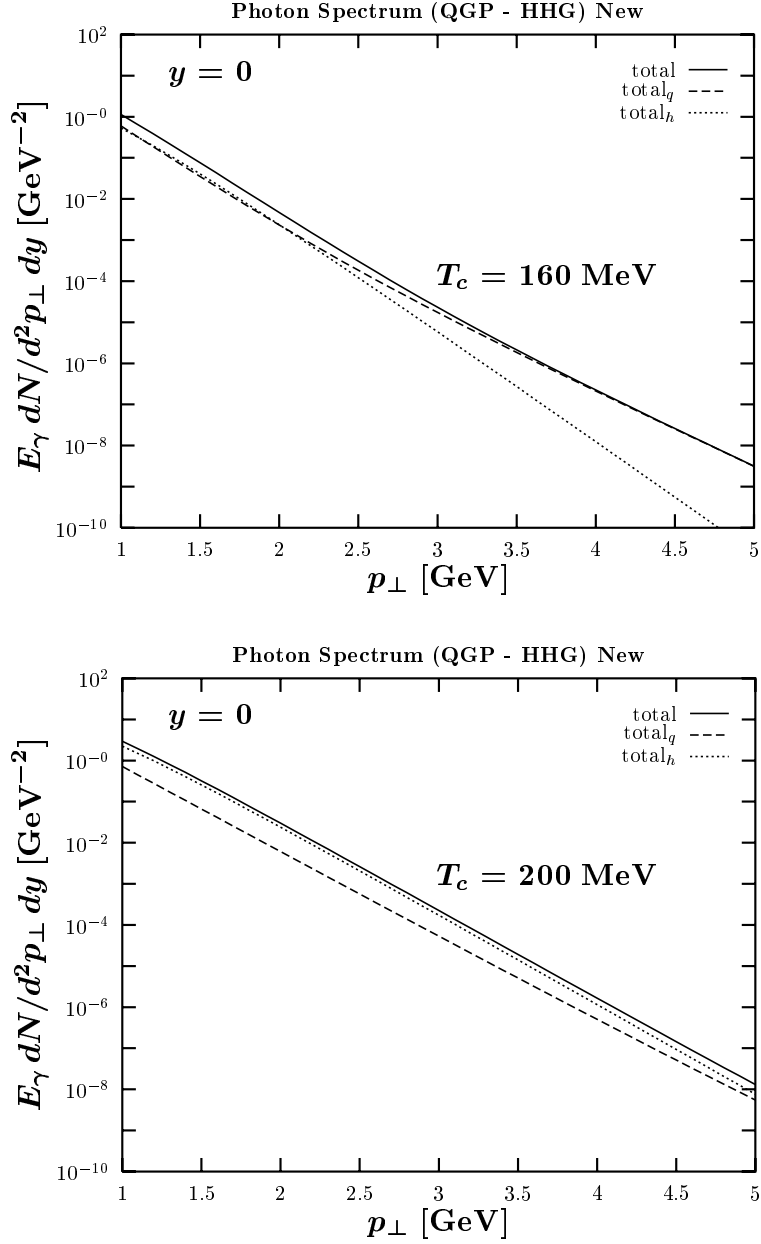}}}
\caption{Photon spectra at a critical temperature $T_c=160$ MeV (above)
and 200 MeV (below), where the same notation as in Fig. \ref{fig2.17} is
used \cite{Steffen99}.}
\protect\label{fig2.18}
\end{figure}

Eq. (\ref{spectrum}) together with the estimates for the rates from the 
QGP and the HHG allows a systematic investigation of the photon spectrum,
depending on the mass number $A$, the projectile rapidity $y_{\rm nucl}$,
the thermalization time $\tau_0$, the initial temperature $T_0$, the
critical temperature $T_c$, and the freeze-out temperature $T_f$. In the 
following all spectra are calculated for photons at mid-rapidity $y=0$.
The following results have been obtained \cite{Steffen99}:
The photon spectrum is proportional to $R_A^2$ (see Eq. (\ref{spectrum})),
i.e. $dN/(d^2p_Tdy)\sim A^{2/3}$. The dependence of the spectra on the
limits of the rapidity integration $y_{\rm nucl}$ is very weak since
photons from a fluid cell with $y'$ far away from zero do not contribute
to mid-rapidity photons because they must have a large energy in the
local rest frame of the fluid cell and are therefore exponentially 
suppressed in the rate. However, the collision energy, from which 
the projectile rapidity follows, determines the initial time and
temperature, which have an important influence on the spectrum. In fact,
one can show that $dN/(d^2p_Tdy)\sim \tau_0^2$ \cite{Steffen99}. 
At higher collision energies,
smaller thermalization times are expected \cite{Biro93}. At the 
same time the initial temperature is increased, which is the dominating
factor. For example, an increase of the initial temperature from
$T_0=200$ MeV to 300 MeV increases the thermal photon yield by more than
a magnitude for a fixed initial time \cite{Steffen99}. In particular,
high $p_T$ photons are enhanced, i.e., the spectrum gets flatter
corresponding to a higher temperature. The reason for this enhancement of the 
yield is the longer life-time of the fireball and the larger rates at higher
$T_0$ (see Fig. \ref{fig2.15}). 
The dependence of the spectrum on $T_0$ is exemplified 
in Fig. \ref{fig2.17}. As can be seen from Fig. \ref{fig2.16},
an increase of $T_c$ will result
in a decrease of the life-times of the QGP and the mixed phase 
and an increase of the
HHG life-time. Depending on the rates from the different phases, this 
affects the spectrum. Using the rates discussed above one observes an
increase of the spectrum by about a factor of 5 going from $T_c=160$ MeV
to $T_c=200$ MeV (see Fig. \ref{fig2.18}), 
due to a higher mean temperature in the latter case.
A lower value of $T_f$ implies a longer life-time of the HHG. However, since
the rates are small at low temperatures, a change of the freeze-out
temperature is negligible in this simple model. However, taking into
account a transverse expansion, this statement will be changed.

The role of a finite chemical potential on the photon spectra will be 
discussed in connection with the comparison of the theoretical spectra
with SPS data from WA80 and WA98 in Section \ref{sec:cte}. 
The modifications of the spectra due to a 
chemical non-equilibrium will be considered in the predictions of spectra
for RHIC and LHC (Section \ref{sec:cte}). 

So far, different aspects of the hydrodynamical calculation of photon spectra
in relativistic nucleus-nucleus reactions have been 
investigated. The various results for SPS, RHIC and LHC, using different 
hydrodynamical models and rates, will be reviewed in
Section \ref{sec:cte}, 
where also the prompt photon spectrum will be considered.
However, a systematic and comprehensive hydrodynamical calculation of
photon spectra together with dilepton and hadronic spectra
from SPS to LHC energies, considering the most recent rates,
a realistic EOS, a reasonable fixing procedure for the initial conditions,
transverse expansion, and chemical non-equilibrium, is missing.
Hence, besides the problems with the rates, 
the ambiguities in the description of the fireball evolution
are another main source
for uncertainties in the theoretical prediction of photon spectra
in relativistic heavy-ion collisions \cite{Peressounko00}. 
After all, simple hydrodynamical models, e.g. with 
a simplified EOS and without transverse expansion, can be useful to
study systematically certain aspects such as the dependence on different 
parameters (initial time and temperature, critical  
temperature, etc.) or the relative importance of different contributions
to the spectrum \cite{Steffen99,Steffen01}.

\section{Experiments}\label{sec:exp}

The detection of electromagnetic radiation may involve very different 
technologies, just because for the possible energies or wavelengths 
very different physics is relevant, ranging from atomic and molecular 
processes at low energy to particle physics concepts at high 
energy, the latter being important for the subject of this report, 
where we deal with photon energies in the GeV range. In 
this regime one 
concentrates on measuring individual quanta, i.e. photons, and their 
four-momenta. Huge detectors weighing up to several hundred tons are 
employed to perform this measurement of individual photons. For 
essentially all of these detectors, the photons have to be converted 
into charged particles which eventually will be measured.

\subsection{Experimental Methods}\label{subsec:em}

In the following Section the experimental foundations of direct 
photon measurements are discussed. For this purpose we present  
different methods for inclusive photon detection and discuss in the 
following the further requirements for the identification of the 
direct fraction of the measured photons.

\subsubsection{Photon Detection}\label{subsubsec:pd}

High energy photon detectors can be divided into two main 
categories depending on the detection principle:
\begin{enumerate}
    \item  \emph{electromagnetic calorimeters}, which attempt to measure the 
    total energy which has been deposited in a given amount of 
    material by an electromagnetic cascade following the first 
    conversion and

    \item  \emph{conversion detectors}, which combine the photon 
    conversion into e$^+$e$^-$ with the subsequent momentum 
    measurement of the charged leptons by tracking.
\end{enumerate}

In its ideal forms these two different types of detectors have 
complementary merits. On the one hand, 
the energy measurement in a calorimeter is 
affected by statistical number fluctuations in the electromagnetic 
shower which become less and less important for photons of higher 
energy, while the momentum determination of the conversion products in 
a tracking system has to deal with the measurement of smaller 
deflections of tracks 
with increasing e$^+$,e$^-$-momenta. Calorimeters are therefore 
generically better suited for very high energy photons.
In addition, converters are required to be relatively thin to allow 
for precise momentum measurement from just the two conversion products 
so that the detection probability (which includes the conversion 
probability) is relatively low, while calorimeters usually have 
detection probabilities of essentially $100\%$. Calorimeters are also 
intrinsically fast which allows to use them for triggering purposes.

On the other hand, the momentum measurement from tracking devices 
yield pointing capabilities far superior to calorimeters which in 
their basic configurations have almost none --- this can help 
considerably to reduce potential backgrounds of particles not 
originating from the reaction studied, as e.g. cosmic rays or beam 
halo. The conversion 
measurement should also suffer less from misidentification of other 
particles, which might still yield signals in a calorimeter. This is 
most important for relatively small photon energies. Furthermore, 
calorimeters have inherent limits with regard to the separation of 
two particles with small opening angles because of the finite lateral 
dimensions of showers in the detector material, while even for very 
small angular separations of the photons the e$^+$e$^-$-pairs may have 
very well separated tracks allowing them to be individually detected in 
a conversion/tracking device. This may be of importance for more 
sophisticated measurements like two-photon-correlations, but will also 
have an impact on detection capabilities in a high multiplicity 
environment or for extremely high momenta, where hadron decay photon 
pairs start to merge in a calorimeter.

Calorimeters have been most widely used for the detection of high 
energy photons because of their advantages sketched above. One should, 
however, keep in mind that the conversion method may be advantageous 
in special situations, and that for very precise measurements the 
combination of both methods may be considered, as they should be 
affected by very different sources of systematic error.
Electromagnetic calorimeters are constructed as either \emph{homogeneous} 
or as \emph{sampling} calorimeters. Sampling calorimeters are built 
using a high-$Z$ material (e.g. $Pb$) as absorber and an active material 
(very frequently light generating like plastic scintillator) in consecutive 
layers. Homogeneous calorimeters use one material which serves as 
both absorbing and active material. They can in principle achieve much 
better detector performance than sampling calorimeters because of the 
additional \emph{sampling fluctuations} in the latter. 

Examples of high energy photon detectors used include
\begin{itemize}
    \item  lead-scintillator-sandwich calorimeters 
    \cite{prd:dem:87,prd:dem:90,zpc:bon:88:1,zpc:bon:88:2,zpc:bon:89,plb:ali:92,zpc:ans:88,plb:alb:88,prl:abe:94,prl:abe:93},            

    \item  lead glass calorimeters which measure the Cherenkov light 
    emitted by shower particles via photomultiplier tubes 
    \cite{plb:bal:79,plb:ang:80,npb:ang:89,plb:ada:95,Albrecht:1996,Aggarwal:2000},

    \item  calorimeters out of scintillating crystals like 
    $NaI$ \cite{plb:ake:85,sjnp:ake:90,zpc:ake:86} or 
    $BGO$ \cite{Aggarwal:1997} readout via either 
    phototubes or photodiodes,

    \item  liquid-argon calorimeters which are sampling detectors 
    measuring the specific ionization 
    of charged shower particles in wire chambers filled with liquid 
    argon \cite{zpc:ann:82,zpc:kou:82,prl:mcl:83,pr:alv:92,prl:aba:96,prl:abb:00},

    \item  lead-proportional tube sampling calorimeters
    \cite{plb:bal:93}            
    and

    \item  converters combined with magnetic electron-positron-pair 
    spectrometers \cite{zpc:bad:86,plb:bad:85,zpc:ake:90,zpc:bau:96}. 
\end{itemize}

Homogeneous calorimeters made of scintillating crystals are generally 
superior in energy resolution due to their high light output. Sampling 
calorimeters suffer from additional sampling fluctuations but are much 
less expensive, especially if detectors of large thickness are 
required to allow containment of very high energy photon showers.
Of course, the higher the photon energy, the less important is the 
intrinsic energy resolution of a calorimeter. 

Another important figure of merit is the suppression of hadrons. Most 
calorimeter materials have much longer hadronic interaction length 
compared to their radiation length. Hadrons are very unlikely to 
deposit a considerable fraction of their energy and are effectively 
suppressed at high energies. A similar effect helps in  
suppressing a large fraction of muons. Further hadron 
suppression can be achieved by 
exploiting the differences in shape of showers induced by hadrons or 
photons, with electromagnetic showers being usually much better 
contained. A fine lateral segmentation allows to discriminate showers 
from their lateral width, which is larger for hadronic showers. 
Longitudinal segmentation can sample the depth profile -  
hadronic showers penetrate deeper into the detector.

A combination of sufficient longitudinal and lateral segmentation can 
even provide pointing capabilities which help to suppress background particles 
which do not originate from the interaction vertex.

\begin{sloppypar}
For further suppression of charged particles additional tracking 
detectors can be used. These will also help to reduce the 
contamination by the usually small fraction of electrons and positrons 
produced, i.e. charged particles generating electromagnetic showers in 
calorimeters. The detectors may either be specialized \emph{charged 
particle veto detectors}, like multiwire proportional chambers 
\cite{prd:dem:87,prd:dem:90,zpc:bon:88:1,zpc:bon:88:2,zpc:bon:89,plb:ada:95} or 
streamer tube detectors \cite{Albrecht:1996,Aggarwal:2000}, 
providing the point of impact of charged 
particles just in front of the calorimeter or full \emph{magnetic 
spectrometers} 
\cite{plb:ali:92,zpc:ans:88,plb:alb:88,prl:abe:94,prl:abe:93,prl:aba:96,prl:abb:00}.
\end{sloppypar}

For direct photon 
measurements other requirements, like e.g. knowledge of the detector 
response or particle discrimination capabilities, become more and more 
important. Thus the detector choice may well depend on the particular 
measurement ``strategy'' that is used.

% Optimal photon detectors in current high energy physics experiments 
% have to be optimized according to different requirements. 
% Good resolution at medium and high energies 
% which requires 

\subsubsection{Direct Photon Identification}\label{subsubsec:dpi}

In particle physics experiments one usually tries to achieve an 
\emph{individual identification} of direct photons. The environment 
is such that photons from meson decays (mostly from pions) will be 
\emph{both} detected with high probability in a certain energy range. 
Photons detected as \emph{single}, i.e. no appropriate other photon 
is measured which would combine with the candidate to a known 
hadron, may be treated as \emph{direct} 
photons and only small corrections for detection efficiency or finite 
geometrical acceptance have to be applied. Random coincidences of two 
particles which fake a photon combination originating from a hadron 
decay are rare and can be neglected. 
In collider experiments (see 
e.g. \cite{prl:abe:94,prl:abe:93,prl:aba:96,prl:abb:00}) one usually requires 
photons to be \emph{isolated}, i.e. not accompanied by any other 
cluster in the calorimeter. In addition to the suppression of 
multi-photon background from hadron decays, this also suppresses 
direct photons produced by Bremsstrahlung processes of hard quarks, 
which would be observed close to the jet fragmentation 
products\footnote{This suppression may in general cause a bias of the 
measured 
distribution and has to be accounted for when comparing to theoretical 
calculations.}.
For very high photon energies 
the discrimination of single photons 
against neutral mesons, esp. pions, decaying into two photons with a 
small opening angle becomes extremely important. For these cases the 
two photon showers can not be separated. Two different methods have 
been used to statistically estimate the contamination
\cite{prl:abe:94,prl:abe:93,prl:aba:96,prl:abb:00}:
\begin{enumerate}
    \item[a)]  The shower shape is measured in a dedicated detector and 
    compared with simulated shower shapes from direct photons and from 
    the expected background.

    \item[b)]  The higher conversion probability in the first layers of 
    the detector for two collinear photons compared to a single photon 
    is used to estimate the background from the longitudinal 
    distribution of the shower.
\end{enumerate}

Uncertainties of the direct photon measurement in particle physics experiments 
include:
\begin{enumerate}
    \item  Background from beam halo (muons), beam-gas events or 
    other no-target contributions, 

    \item  multi-event pile-up,

    \item  trigger efficiency,

    \item  misidentified charged and neutral hadrons (important in 
    calorimeters at lower energies),

    \item  contamination from merged neutral pions (important in 
    calorimeters at higher energies),

    \item  single photons from hadron decays (important for small 
    detector acceptance and/or asymmetric decays),

    \item  energy scale uncertainty,

    \item  conversion probability in conversion measurements,
    
    \item  loss of photons from conversion and
    
    \item  uncertainty of luminosity, target thickness etc. (overall 
    scale uncertainty) \label{item10}.
\end{enumerate}

\medskip

Heavy-ion experiments have the additional difficulty of dealing with 
high multiplicities already at relatively low energies. It is 
therefore not possible to identify individual direct photons. One has 
to resort to a \emph{statistical identification} of direct 
photons.\footnote{This is also true for some particle physics experiments 
using detectors with limited resolution.} For any 
given photon there is a high probability that another random photon 
is detected which would combine well with the first to a potential 
hadron decay photon pair. The only possible strategy is therefore to 
measure all inclusive photons regardless of their origin and to 
subtract the photons originating from hadron decays (see e.g. 
\cite{Aggarwal:2000}). For most situations, where the fraction of 
direct photons compared to all photons is relatively small, this implies 
that the direct photon result 
is obtained by subtracting two large numbers. Therefore such analyses 
are even more sensitive to systematic errors. To the sources of 
systematic uncertainty cited above have to be added:
\begin{enumerate}
    \setcounter{enumi}{10}
    \item  effects of a multiplicity dependent detector response. 
    \label{item11}
\end{enumerate}
This includes e.g. the influence of randomly overlapping showers in a 
calorimeter.

Other crucial sources of systematic error in this situation are 
related to the determination of the decay photons from hadrons, most 
importantly neutral pions and $\eta$ mesons. The preferred option is 
of course to measure the production of neutral mesons simultaneously in the same 
data set as the inclusive photons. Such a measurement is usually 
performed by a two-photon invariant mass analysis 
\cite{Aggarwal:2000}. Even if the two-photon decay mode is not the 
only decay producing background photons (there is e.g. the Dalitz 
decay $\pi^{0} \rightarrow \mathrm{e^{+}e^{-}}\gamma$), the known 
branching ratios allow to calculate all relevant decay contributions.
In this case the following systematic uncertainties 
on the relative amount of meson decay photons
may contribute: 
\begin{enumerate}
    \setcounter{enumi}{11}
    \item  Uncertainty of the subtraction of combinatorial background,

    \item  effects of multiplicity dependent detector response on 
    the meson reconstruction (different from item \ref{item11}),

    \item  geometrical acceptance,

    \item  loss of mesons from photon conversion,

    \item  loss of mesons from merging photon showers and

    \item  those effects from the energy scale uncertainty, 
    from beam-gas events or 
    other no-target contributions which differ from the effects on 
    inclusive photons.
\end{enumerate}
Any uncertainty on the overall normalization of the cross sections 
(like those mentioned under item \ref{item10})
does not enter here, if the meson spectra are determined from the same 
data sample.
Those hadrons decaying into photons, which are not measured, have to 
be estimated from other sources. This is usually the case for heavier 
mesons (e.g. $\omega$, $\eta^{\prime}$), which contribute only a very 
small fraction to the total number of decay photons. In conversion 
measurements, however, already the detection of neutral pions is very 
difficult and can very often not be done, because the single photon 
detection probability is low. This may introduce a very 
large systematic uncertainty of the decay background to be subtracted 
and has to be treated with great care.

\subsection{Experimental Results}\label{subsec:er}

\subsubsection{$pp$ and $pA$ Experiments}\label{subsubsec:ppe}

\begin{figure}[tb]
    \centering
    \includegraphics{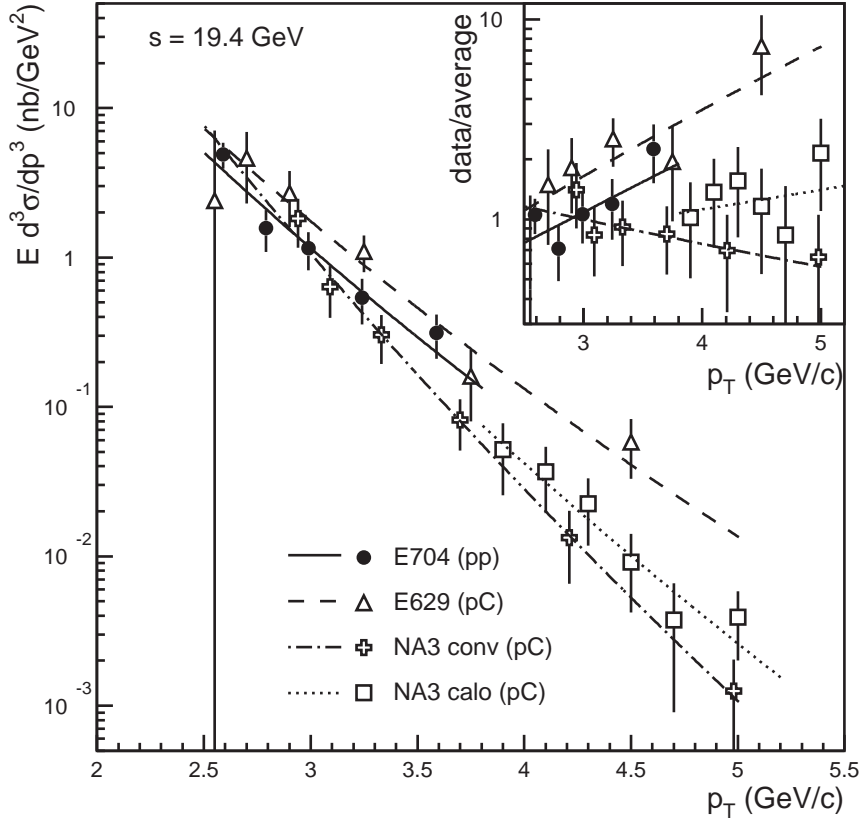}
    \caption{Direct photon cross section per nucleon for $p$-induced 
    reactions at $\sqrt{s} = 19.4 \, \mathrm{GeV}$. Data are from 
    experiments E704 \cite{plb:ada:95}, E629 \cite{prl:mcl:83} and 
    NA3 \cite{zpc:bad:86}. The inset shows the ratios of 
    experimental data to a simultaneous fit to all data sets.}
    \label{fig:ppexp1}
\end{figure}

\begin{figure}[tb]
    \centering
    \includegraphics{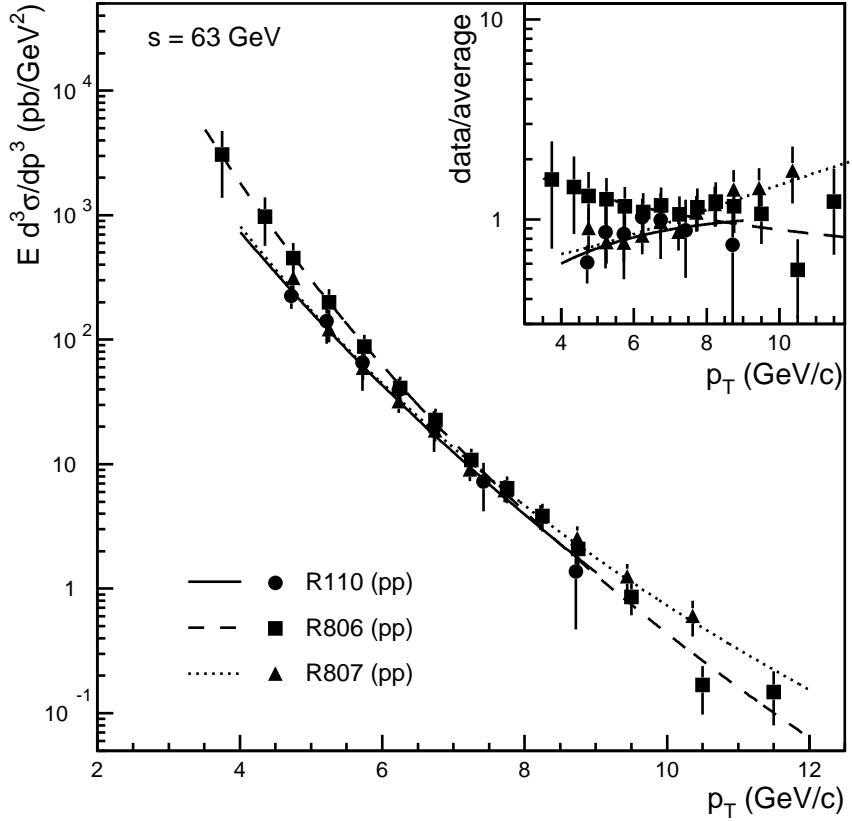}
    \caption{Direct photon cross section for $pp$ reactions at $\sqrt{s} = 
    63 \, \mathrm{GeV}$. Data are from 
    experiments R110 \cite{npb:ang:89}, R806 \cite{zpc:ann:82} and 
    R807 (AFS) \cite{sjnp:ake:90}. The inset shows the ratios of 
    experimental data to a simultaneous fit to all data sets.}
    \label{fig:ppexp2}
\end{figure}

\begin{figure}[tb]
    \centering
    \includegraphics{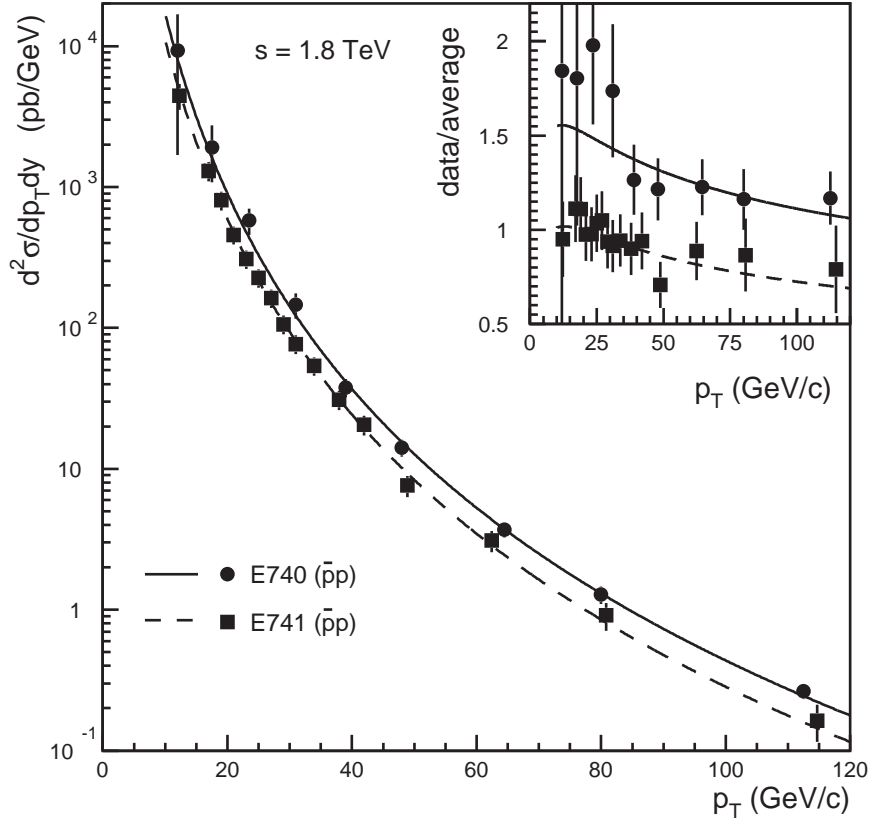}
    \caption{Direct photon cross section for ${\bar{p}p}$ reactions 
    at $\sqrt{s} = 1.8 \, \mathrm{TeV}$. Data are from 
    experiments E740 (D0) \cite{prl:abb:00} and 
    E741 (CDF) \cite{prl:abe:94}. The inset shows the ratios of 
    experimental data to a simultaneous fit to all data sets.}
    \label{fig:ppexp3}
\end{figure}

\begin{figure}[tb]
    \centering
    \includegraphics{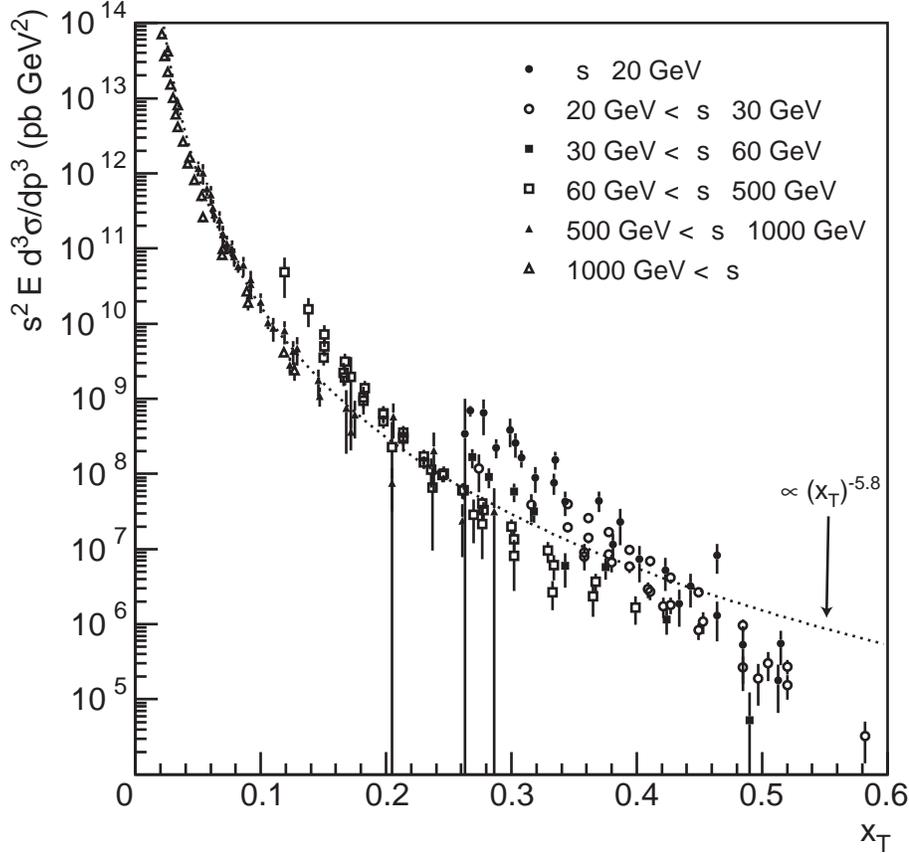}
    \caption{Direct photon cross section scaled with $s^{2}$ 
    for $pp$ and $\bar{p}p$ reactions 
    at different energies. The dashed line shows a fit of a power law 
    according to Eq. (\protect\ref{eq:scaling}) to all data sets.}
    \label{fig:ppexp4}
\end{figure}

\begin{figure}[tb]
    \centering
    \includegraphics{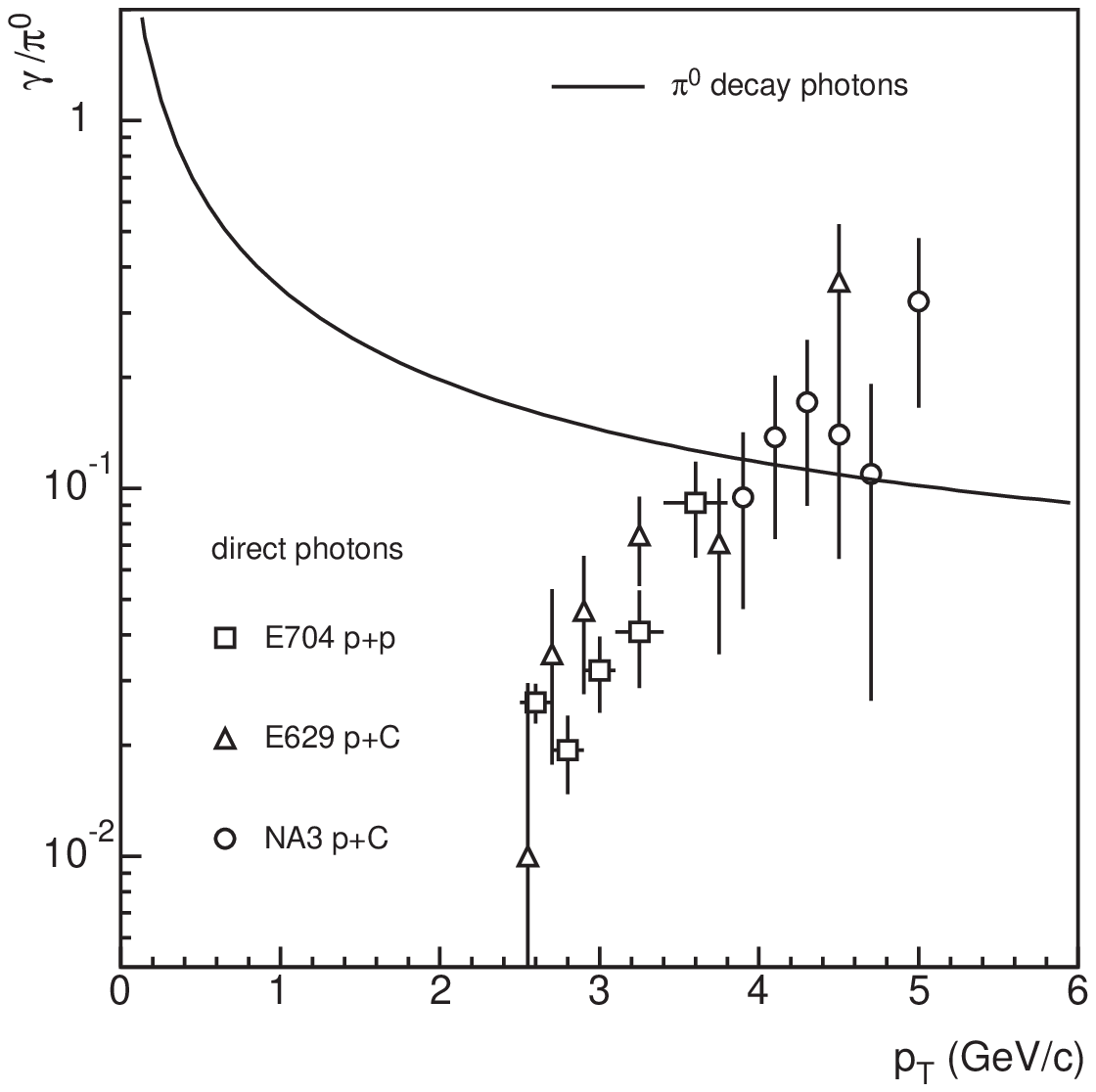}
    \caption{$\gamma / \pi^{0}$ ratio as a function of $p_{T}$ for 
    $\sqrt{s} = 19.4 \, \mathrm{GeV}$. The 
    open symbols show the values for direct photons 
    extracted in $pp$ and $pC$ experiments. 
    The solid line shows an estimate of the ratio for decay photons 
    from $\pi^{0}$.}
    \label{fig:ppexp5}
\end{figure}

A great wealth of experimental results from $pp$ and $pA$ experiments is 
available. For a compilation of these results see e.g. 
\cite{Vogelsang97}. The beam energies range from 
$\sqrt{s} = 19.4 \, \mathrm{GeV}$ to $\sqrt{s} = 1.8 \, \mathrm{TeV}$. 
The lowest energy data (up to $\sqrt{s} = 30.6 \, \mathrm{GeV}$) have 
been measured in fixed-target experiments at CERN 
\cite{prd:dem:87,zpc:bon:88:1,zpc:bon:88:2,zpc:bon:89,prd:dem:90,plb:bal:79,plb:bal:93,zpc:bad:86,plb:bad:85},            
and Fermilab \cite{plb:ada:95,prl:mcl:83,pr:alv:92} while the data at 
higher energies are from collider experiments at ISR 
\cite{plb:ang:80,npb:ang:89,plb:ake:85,sjnp:ake:90,zpc:ake:86,zpc:ann:82,zpc:kou:82},
SPS \cite{plb:ali:92,zpc:ans:88,plb:alb:88} 
and the Tevatron \cite{prl:abe:94,prl:abe:93,prl:aba:96,prl:abb:00}.

We will only discuss some of the experimental results in different energy 
regimes as examples for the experimental achievements in this field. 
We have chosen energies for which different data sets exist which may 
be compared. 
The lowest beam energy at which direct photon data are available is 
$\sqrt{s} = 19.4 \, \mathrm{GeV}$. At this energy cross sections for 
direct photons are measured in $pp$ reactions by the Fermilab experiment 
E704 \cite{plb:ada:95}, and by two 
experiments in $p$+$C$ reactions, E629 \cite{prl:mcl:83} at Fermilab and NA3 
\cite{zpc:bad:86} at CERN. These data are shown in 
Fig. \ref{fig:ppexp1} together with fits
% \footnote{These fits are only used to facilitate the comparison of 
% different data sets. No particular importance is given to the 
% numerical values of fit parameters, and thus they are not stated.} 
of the phenomenological 
function \cite{rnc:hagedorn:83}:
\begin{equation}
        E \frac{d^{3}\sigma}{dp^{3}} = A \cdot \left( \frac{p_{0}}{p_{T} + 
        p_{0}} \right)^{n}.
        \label{eq:hage}
\end{equation}
The NA3 experiment has performed measurements using two independent 
tech\-ni\-ques, i.e. measurements using a calorimeter (label ``calo'') and conversion 
measurements (label ``conv''). There are considerable discrepancies 
between the different data sets --- these can be judged more easily 
from the inset, where the different data sets are normalized to a 
common fit to all data points. The results from E629 are 
considerably above the other data sets, especially the measurement at 
the highest $p_{T}$, which is about a factor of 10 higher. All data 
sets show different slopes, and the overall discrepancies, even 
ignoring the high $p_{T}$ point of E629, are of the order of a factor 
of 3 which is mostly covered by the experimental error bars.

As the discrepancies appear to be smallest at low $p_{T}$ and increase 
with higher $p_{T}$ it is likely that energy scale uncertainties play 
a role here. Still the variations even between the two data sets 
within the same experiment indicate that other systematic 
uncertainties like background contamination are not negligible. One 
should however note that some of these results were among the first direct photon 
measurements available, and some progress has been made concerning 
e.g. background estimates and detector simulations. Nevertheless this 
comparison gives a first hint of the difficulties of direct photon 
measurements. 

Fig. \ref{fig:ppexp2} shows direct photon cross sections for $pp$ 
reactions at $\sqrt{s} = 63 \, \mathrm{GeV}$ from CERN ISR 
experiments R110 \cite{npb:ang:89}, R806 \cite{zpc:ann:82} 
and R807 (AFS) \cite{sjnp:ake:90}. Again fits with 
Eq. (\ref{eq:hage}) are provided, and the inset shows the ratio of 
data to an overall fit. The relative variation of the different data 
sets is smaller than at lower energies which may in part be due to 
the higher photon energies, which can be more reliably measured with 
a calorimeter. Still there is a significant difference especially in 
the slope of the different measurements which might e.g. be related to 
uncorrected non-linearities. In addition, these experiments have required 
isolation cuts which make the interpretation more difficult.

As some of the more recent examples, Fig. \ref{fig:ppexp3} shows direct 
photon cross section for $\mathrm{\bar{p}p}$ reactions 
at $\sqrt{s} = 1.8 \, \mathrm{TeV}$ from 
experiments E740 (D0) \cite{prl:abb:00} and 
E741 (CDF) \cite{prl:abe:94}. Measured photon energies reach values 
beyond 100~GeV. While a calorimetric measurement is well suited for 
such a task in general, one of the major uncertainties in this case 
lies in the (in)ability to discriminate single photons from merged 
photons originating from $\pi^{0}$ decay. The spectra of the two 
experiments are very similar in shape, while there appears to be a 
difference of about $50 \% $ in absolute normalization. Still the 
relative agreement between the two sets is good relative to earlier 
direct photon measurements.

The parameters of the fits of Eq. (\ref{eq:hage}) to the Tevatron data 
are in the range $p_{0} = 1.7 - 2.1 \, \mathrm{GeV}/c$ and $n = 5.8 - 
6.1$. Fits to jet cross sections \cite{d0:jets} in a similar $p_{T}$ 
range yield $p_{0} \approx 16 \, \mathrm{GeV}/c$ and $n \approx 
8.5$ -- decreasing the parameter $p_{0}$ yields slightly smaller 
values of the power $n$, they are, however, still significantly larger 
than the values obtained for the direct photon cross section\footnote{Pure 
inverse power law fits yield also lower powers $n$, 
however, they provide only much worse descriptions of the data.}. 
These values are larger than the simple parton model scaling 
prediction \cite{rmp:owens:87}, which would result in $n = 4$. This 
discrepancy is not surprising, as the strong $x_{T}$ dependence of the 
structure functions is expected to modify this behavior.
The fits to the data at lower $\sqrt{s}$ yield consistently larger 
values of $n$, indicating that the deviation from the simple parton 
scaling becomes more important at lower energies.

Dimensionality arguments in the context of the parton model suggest 
that the cross section may be parameterized as \cite{rmp:owens:87}:
\begin{equation}
    E d^3\sigma_{\gamma}/dp^3 = f(x_T,\theta)/s^2, 
    \label{eq:scaling}
\end{equation}
where $x_T=2p_T/\sqrt{s}$ and $\theta$ is the emission angle of the 
photon. We have therefore 
attempted to combine photon measurements at midrapidity 
at all available energies by 
plotting $s^{2} E d^3\sigma_{\gamma}/dp^3$ as a function of $x_{T}$ 
in Fig. \ref{fig:ppexp4}. The experimental range of $x_{T}$ spans 
from 0 to roughly 0.6, where high energy data contribute at low 
$x_{T}$ and low energy data at high $x_{T}$. 
This recipe provides an astonishingly good 
universal representation of all the photon data. 
Fig. \ref{fig:ppexp4} also shows a fit of a power law $f(x_{T}) = a 
\cdot x_{T}^{b}$ which yields an exponent of $b = -5.79$. Looking into 
more of the details in Fig. \ref{fig:ppexp4} one can see, however, 
that the individual data sets are not perfectly described. Especially 
at lower beam energies and low photon transverse momenta the data 
deviate from the universal curve. Possible interpretations of these 
discrepancies will be discussed in Section \ref{subsec:cpp}.

Since the early measurements of direct photons it has been 
customary to investigate first the ratio of photons to neutral 
pions $\gamma/ \pi^{0}$. For the experimentalist, this is a convenient 
quantity, as some of the systematic errors drop out in such 
ratios. Neutral pions are the major source of decay photons which 
provide a background to the direct photon measurement. 
In particular, the inclusive photon yield 
at a given transverse momentum is dominated by photons from asymmetric 
decays of $\pi^{0}$ which carry approximately the same transverse 
momentum. The $\gamma/ \pi^{0}$ ratio is therefore a good indicator of 
the difficulty to extract a direct photon signal.

More quantitatively one can 
obtain an approximation to the $\gamma_{\pi^{0}\mathrm{-decay}}/ 
\pi^{0}$ ratio, 
which is the dominant contribution to the decay photon background as:
\begin{equation}
    \frac{\gamma_{\pi^{0}\mathrm{-decay}}}{\pi^{0}} \approx 
    \frac{2/p_{T} \, \int_{p_{T}}^{\infty}dp_{T}^{\prime} \, 1/p_{T}^{\prime} \, 
    dN_{\pi^{0}}/dp_{T}^{\prime}}{1/p_{T} \, 
    dN_{\pi^{0}}/dp_{T}}.
    \label{eq:gampidecay}
\end{equation}
This formula follows directly, if one assumes that for a given $p_{T}^{\prime}$ 
of the neutral pions the decay photons are uniformly distributed in 
$p_{T}$ between 0 and $p_{T}^{\prime} $.

In Fig. \ref{fig:ppexp5} the $\gamma/ \pi^{0}$ ratio is displayed for 
reactions at $\sqrt{s} = 19.4 \, \mathrm{GeV}$. Included are the 
experimental results\footnote{The NA3 data obtained with the conversion trigger have not 
been included, as the neutral pion spectra from this data sample do 
not agree with the general trend of the other data sets.} 
for $pp$ and $pC$ as in Fig. \ref{fig:ppexp1} 
 and the estimate of the $\pi^{0}$ 
decay photons according to Eq. (\ref{eq:gampidecay}). The relative 
variation between the different data sets appears to be smaller in the 
ratio compared to the direct photon cross section, which may indicate 
that some of the systematic errors do actually cancel. Even at the 
highest $p_{T}$ the decay photons are still not negligible compared to 
the direct photons. The direct photon data essentially stop around 
3~GeV/c, where they tend to go below the $10 \% $ level relative to the decay 
photons. At low $p_{T}$ the extraction of direct photons appears to be 
hopeless as the decay photon background increases and at the same 
time the direct photon signal is expected to be very small.

\subsubsection{Heavy-Ion Experiments at the SPS}\label{subsubsec:hie}

Experimental data on direct photons in heavy-ion collisions is 
scarce, as the extraction is much more difficult due to the much 
higher particle multiplicity. The highest available energy in heavy-ion 
collisions so far at the CERN SPS has been approximately at the lowest energy 
where direct photons could be measured in $pp$.

Using the relatively light ion beams of $^{16}O$ and $^{32}S$ at a 
beam energy of 200~$A$~GeV, corresponding to a nucleon-nucleon center of 
mass energy of $\sqrt{s_{NN}} = 19.4 \, \mathrm{GeV}$, the experiments 
WA80 \cite{Albrecht:1996,Albrecht:1991}, HELIOS (NA34) \cite{zpc:ake:90} 
and CERES (NA45) \cite{zpc:bau:96} have attempted to measure direct 
photons. All these measurements have been able to deliver upper 
limits of direct photon production. 

HELIOS has studied $p$-, $^{16}O$- and $^{32}S$-induced reactions 
\cite{zpc:ake:90} with a conversion method. Photons convert in an 
iron plate with a thickness of $5.7 \% $ of a radiation length. The 
electron-positron pairs are tracked in one drift chamber each before 
and after a magnet with a momentum kick of $\approx 80 \, 
\mathrm{MeV}/c$. Two planes of multiwire proportional chambers 
bracket the converter and help in localizing the conversion point.
The authors estimate the 
ratio of the integrated yields of inclusive photons and neutral pions:
\begin{equation}
    r_{\gamma} = \frac{N_{\gamma}}{N_{\pi^{0}}}
    \label{eq:helios}
\end{equation}
for $p_{T} > 100 \, \mathrm{MeV}/c$. They calculate the neutral 
pion yield from the number of negative tracks in their magnetic 
spectrometer. Their results (with $4-11 \% $ 
statistical and $9 \% $ systematic uncertainty) and their estimate of decay 
photons (with $9 \% $ systematic uncertainty) agree within these errors. 
An analysis of the $^{32}S$-induced data with a higher cutoff of 
$p_{T} = 600 \, \mathrm{MeV}/c$ yields a comparable result. However, 
the results are of limited value in the context of both prompt and thermal 
direct photons, as they are dominated by the lowest $p_{T}$, where the 
expected direct photon emission would be negligible.

A similar measurement has been performed by the CERES experiment, 
which has studied $^{32}S$~+~$Au$ reactions \cite{zpc:bau:96}. 
Photons are measured when 
they convert in the target, the e$^{+}$-e$^{-}$-pairs are 
reconstructed by tracking in the two RICH detectors. The RICH 
detectors operate with a high threshold ($\gamma \approx 32$) to 
effectively suppress background from charged hadrons. Momentum and 
charge information are obtained from the deflection in a 
superconducting double solenoid between the RICH detectors. Photon 
conversions are identified by requiring a vanishing opening angle in 
the first RICH (unresolved double ring) and a larger opening angle 
from the magnetic field in the second RICH (two distinct rings). The 
measured photon spectra have to be corrected for reconstruction 
efficiency --- the correction factor ranges from $\approx 2$ at the 
largest $p_{T}$ to $\approx 6$ at $p_{T} = 0.5 \, \mathrm{GeV}/c$ and 
increases dramatically below $p_{T} = 0.5 \, \mathrm{GeV}/c$.
They obtain 
inclusive photon spectra in central $^{32}S$~+~$Au$ reactions in
$0.2 \, \mathrm{GeV}/c \le p_{T} \le 2.0 \, \mathrm{GeV}/c$. The 
results agree within errors with their hadron decay generator, which 
is tuned to reproduce charged and neutral pion spectra from different 
heavy-ion experiments. They estimate a similar ratio of integrated 
yields:
\begin{equation}
    r_{\gamma}^{\prime} = \left( \frac{dN_{ch}}{d\eta} \right)^{-1}
    \int_{0.4 \, \mathrm{GeV}/c}^{2.0 \, \mathrm{GeV}/c}
    \frac{dN_{\gamma}}{dp_{T}}dp_{T},
    \label{eq:ceres}
\end{equation}
which they use --- again by comparing to the generator 
--- to establish an upper limit ($90 \% $ CL) of $14 \% $ for 
the contribution of direct photons to the integrated inclusive photon 
yield. One of the uncertainties which is difficult to control in 
this analysis relates to the fact that they use simulated hadron 
yields in their generator which are tuned to other measurements with 
different trigger biases and systematic errors, and that especially the 
neutral pions have not been measured within the same data set.

In addition, the CERES experiment has utilized another method to extract 
information on a possible direct photon contribution. As in 
naive pictures of particle production in these reactions the 
direct photon multiplicity is proportional to the square of the 
initial multiplicity while the hadron multiplicity should be 
proportional to the initial multiplicity, they have studied the 
multiplicity dependence of the inclusive photon production. Their 
upper limit on a possible quadratic contribution is slightly lower 
than the above limit on direct photons from $r_{\gamma}^{\prime}$, 
its relation to the direct photon contribution 
is however dependent on the model of particle production.
Similar to the HELIOS measurements both these results are dominated 
by the low $p_{T}$ part of the spectra, so the result is consistent 
with the expectation of a very low direct photon yield at low $p_{T}$.

\begin{figure}[tb]
    \centering
    \includegraphics{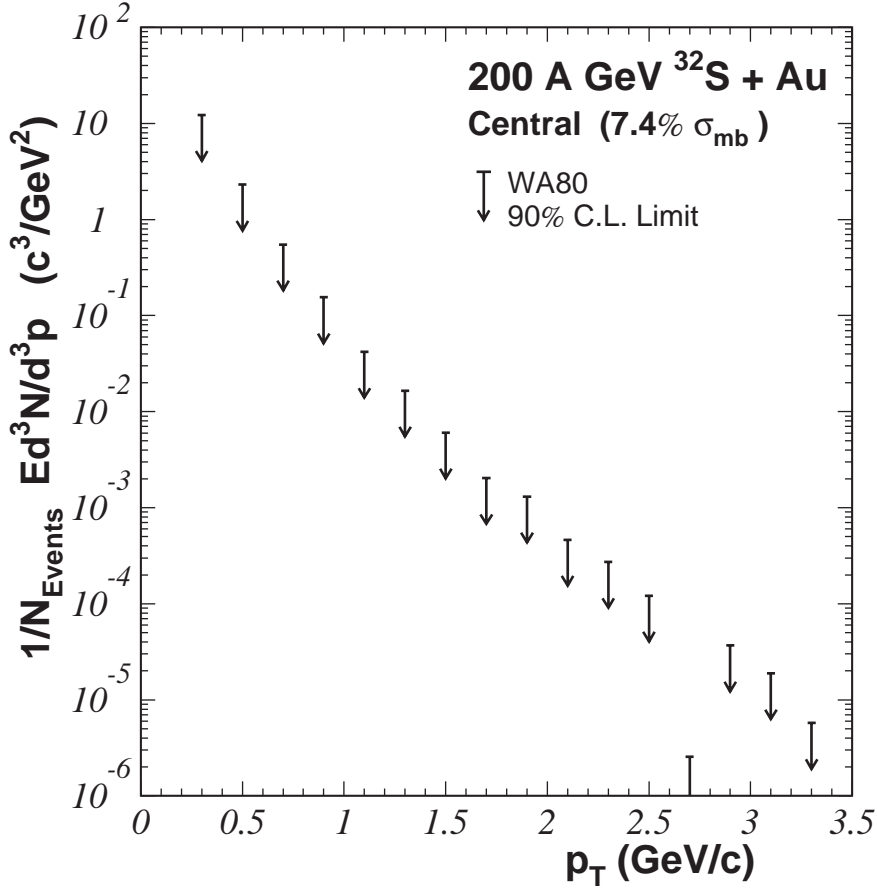}
    \caption{Upper limits ($90 \% $ CL) of the direct photon 
    multiplicity as a function of  $p_{T}$ in central 
    reactions of $^{32}S$~+~$Au$ for 
    $\sqrt{s} = 19.4 \, \mathrm{GeV}$.}
    \label{fig:wa80}
\end{figure}

\begin{figure}[tb]
    \centering
    \includegraphics{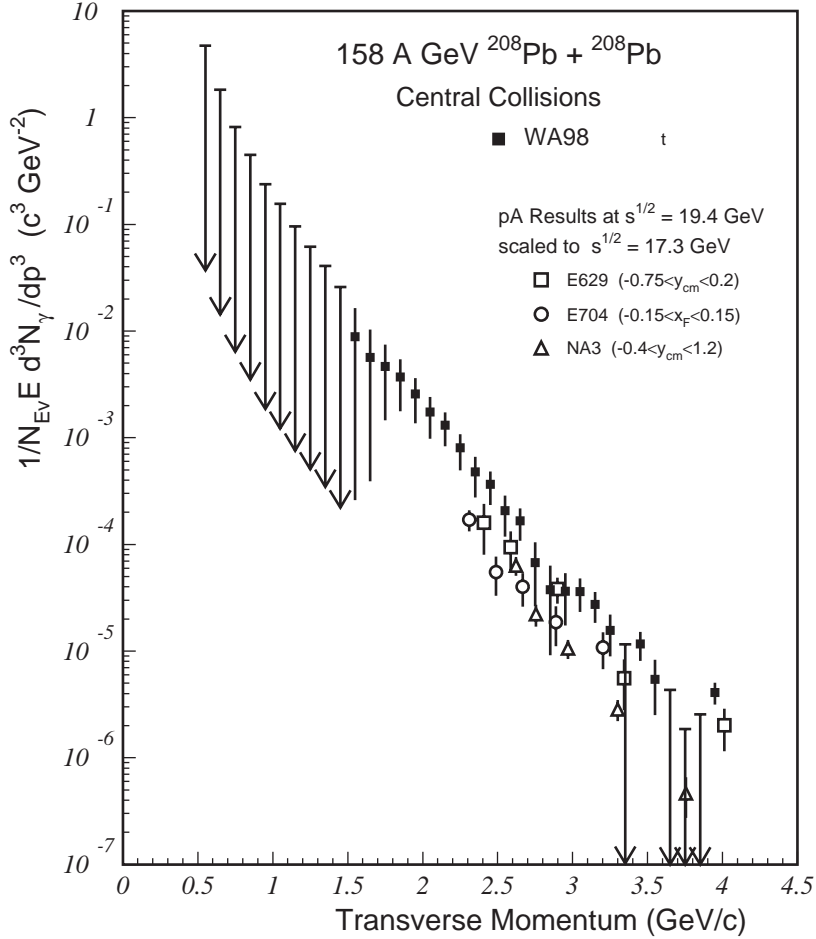}
    \caption{Invariant direct photon 
    multiplicity as a function of  $p_{T}$ in central 
    reactions of $Pb$~+~$Pb$ for 
    $\sqrt{s} = 17.3 \, \mathrm{GeV}$. The error bars correspond to 
    combined statistical and systematic errors, the data points with downward 
    arrows indicate $90 \% $ CL upper limits. For comparison scaled direct 
    photon results from $p$-induced reaction are included (see text).}
    \label{fig:wa98dat}
\end{figure}

The WA80 experiment has performed measurements with 
$^{16}O$ \cite{Albrecht:1991} and 
$^{32}S$ \cite{Albrecht:1996} beams using a lead glass calorimeter 
for photon detection. Most hadrons deposit very little energy 
and are thus effectively suppressed. A shower shape discrimination 
further reduces the hadron background, and charged hadrons are in 
addition
rejected by the help of a charged particle veto detector consisting of 
streamer tubes with pad readout. The contamination by neutrons and 
antineutrons is estimated by Monte Carlo simulation --- it is 
expected to be small. The raw spectra are corrected for a multiplicity dependent 
reconstruction efficiency. The systematic errors are check\-ed by 
performing the analysis with a number of different choices of 
experimental cuts.
Inclusive photons and $\pi^{0}$ and $\eta$ 
mesons have been measured in the same data samples, which helps to 
control the systematic errors. WA80 reports no significant direct 
photon excess over decay sources in peripheral and central collisions 
of $^{16}O$~+~$Au$ and $^{32}S$~+~$Au$. The average excess in central 
$^{32}S$~+~$Au$ collisions in the range 
$0.5 \, \mathrm{GeV}/c \le p_{T} \le 2.5 \, \mathrm{GeV}/c$ is given 
as $5.0 \% \pm 0.8 \% $ (statistical) $\pm 5.8 \% $ (systematic). A 
$p_{T}$ dependent upper limit ($90 \% $ CL) of direct photon 
production as shown in Fig. \ref{fig:wa80} has been obtained, which 
gives more information than the integrated limits, as it can constrain 
predictions at higher $p_{T}$, where a considerable direct photon 
multiplicity may be expected.

For $Pb$~+~$Pb$ collisions at 158~$A$~GeV ($\sqrt{s_{NN}} = 17.3 \, 
\mathrm{GeV}$) the WA98 experiment has performed photon measurements
\cite{Aggarwal:2000}
using similar detectors and analysis techniques as WA80. In 
peripheral collisions no significant direct photon excess was found. 
In central collisions the observed photons cannot entirely be 
explained by decay photons, implying the first observation of direct 
photons in high energy heavy-ion collisions. The extracted direct 
photon spectrum is shown in Fig. \ref{fig:wa98dat}. The only other 
direct photon measurements at a similar energy are from $p$-induced 
reactions as discussed in Section \ref{subsubsec:ppe}. Data from 
$pp$ reactions by E704 \cite{plb:ada:95} 
and from $p$+$C$ reactions by E629 \cite{prl:mcl:83} 
and NA3 \cite{zpc:bad:86} at $\sqrt{s} = 19.4 \, \mathrm{GeV}$ have 
been converted to the lower energy $\sqrt{s} = 17.3 \, \mathrm{GeV}$ 
assuming a scaling according to Eq. (\ref{eq:scaling}) and have 
been multiplied with the average number of binary nucleon-nucleon 
collisions in the central $Pb$~+~$Pb$ reactions (660). These scaled 
$p$-induced results are included in Fig. \ref{fig:wa98dat} for comparison. 
They are considerably below the heavy-ion results which indicates that 
a simple scaling of prompt photons as observed in $pp$ is not sufficient 
to explain the direct photons in central $Pb$~+~$Pb$ reactions. 
Interpretations of this discrepancy will be discussed in the following 
chapter.

\begin{figure}[tb]
    \centering
    \includegraphics{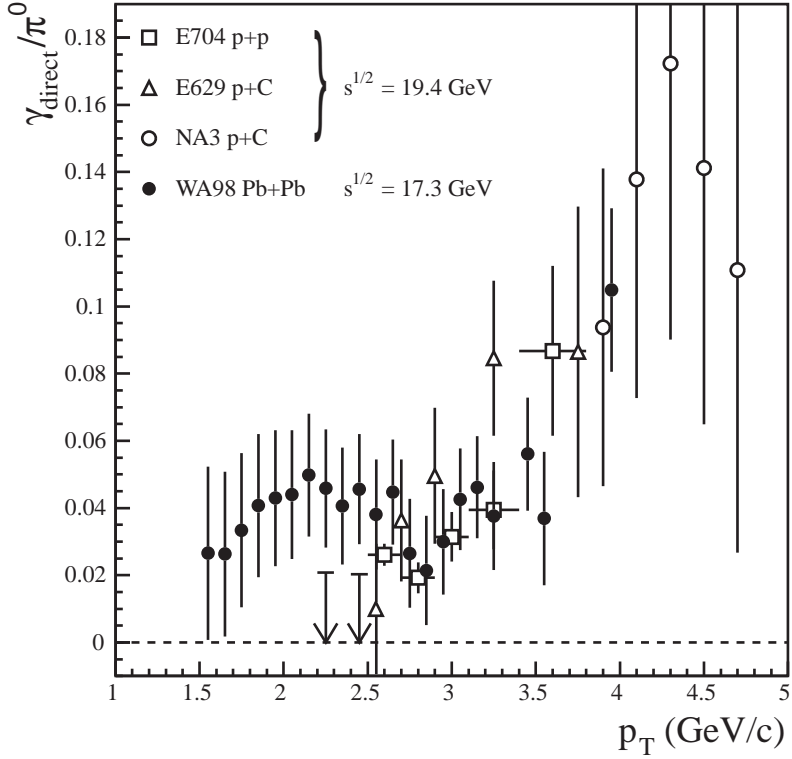}
    \caption{$\gamma_{direct} / \pi^{0}$ ratio as a function of $p_{T}$ 
    for $pp$ and $pC$ reactions
    $\sqrt{s} = 19.4 \, \mathrm{GeV}$ as in 
    Fig. \protect\ref{fig:ppexp5} (open symbols) 
    and for central $Pb$~+~$Pb$ reactions at 
    $\sqrt{s} = 17.3 \, \mathrm{GeV}$ (filled symbols).}
    \label{fig:wa98gampi}
\end{figure}

We would like to close our discussion of existing direct photon data 
by comparing the $\gamma / \pi^{0}$ ratio extracted from heavy-ion 
data to those from $pp$ and $pC$ in Fig. \ref{fig:wa98gampi}. The value 
in heavy-ion data is $\approx 3-5 \% $ in most of the $p_{T}$ range, 
which is similar to the lowest values extracted in the proton data. 
This may be taken as a hint that such levels of direct photons 
approach the feasibility limit of such measurements. Still lower 
levels will be very hard or impossible to detect.

\subsection{Outlook for RHIC and LHC}\label{subsec:orl}

In summer 2000 experiments at the Relativistic Heavy Ion Collider 
(RHIC) at BNL started to take data in collisions of Au nuclei at 
$\sqrt{s_{NN}} = 130 \, \mathrm{GeV}$, continuing with a beam energy 
of $\sqrt{s_{NN}} = 200 \, \mathrm{GeV}$ from 2001 on. First 
results of the RHIC experiments have already been presented 
\cite{QM2001}, however results on direct photons are not available at 
this early stage.

One of the major goals of the PHENIX experiment \cite{daveQM97,billQM01} 
at RHIC is the measurement of direct photons in the central detector 
arms at midrapidity. Photon measurements and 
neutral meson reconstruction are performed 
with electromagnetic cal\-orimeters using 
two different technologies, a lead glass detector, which consists of 
the transformed and updated calorimeter used in WA98 and a 
lead-scintillator sampling calorimeter. In addition, the sophisticated 
electron detection capabilities should also allow to measure inclusive 
photons via the e$^{+}$-e$^{-}$-pairs from conversions.
The central detectors cover $90^{\circ}$ in azimuth and the 
pseudorapidity range $ |\eta| < 0.35$. A central magnet provides an axial 
field, and tracking and momentum measurement is performed in 
three different sub-systems: pad chambers (PC), drift chambers (DC)
and time-expansion chambers (TEC). Electron identification is 
achieved by simultaneously using a ring imaging Cherenkov counter 
(RICH) for $p < 4.7 \, \mathrm{GeV}/c$, 
electromagnetic energy measurement in the calorimeters for
$p > 0.5 \, \mathrm{GeV}/c$ and 
$dE/dx$ measurement in the TEC for $p < 2 \, \mathrm{GeV}/c$. 
A planned upgrade of the TEC to a transition 
radiation detector (TRD) will further strengthen the electron 
identification. Photons converting in the outer shell of the 
multiplicity and vertex detector (MVD) can be identified as electron 
pairs with a small, but finite apparent mass\footnote{This finite 
mass is an artefact of the assumption of particle emission from the 
collision vertex.}. It is planned to add a converter plate to the 
experiment for part of the data taking to minimize uncertainties of 
the conversion probability and the location of the conversion point.
Photons with $p > 1 \, \mathrm{GeV}/c$ will be identified in the 
calorimeters with hadron suppression from the smaller deposited energy 
and additional rejection by time-of-flight (for slow hadrons) and shower 
shape analysis. Furthermore, charged hadrons will be identified by the 
tracking detectors in front of the calorimeters. The calorimeters will 
also measure $\pi^{0}$ and $\eta$ production necessary for the 
estimate of the decay photon background.

The different technologies should provide an excellent measurement of 
direct photons with independent checks of systematic errors. In 
addition, as RHIC is a dedicated heavy-ion accelerator, a much higher 
integrated luminosity is expected, which, together with the expected 
higher photon production rates, will make the RHIC measurements 
superior to the existing lower energy heavy-ion data.

Pb beams at even higher energies ($\sqrt{s_{NN}} = 5.5 \, 
\mathrm{TeV}$) will be available at the future Large Hadron Collider 
at CERN, which is supposed to deliver heavy-ion beams to physics experiments in 
2007. There will be one dedicated heavy-ion experiment, ALICE 
\cite{alice}, which is planning to measure direct photons. In 
addition, one of the $pp$ experiments, CMS \cite{cms,cmsecal} may also attempt 
to measure direct photons at very high transverse momenta in heavy 
ion collisions. Of course, as LHC is primarily a proton machine and 
the heavy-ion beam time will be limited, the measurement conditions 
are not as favorable as at RHIC, the expected photon rates, however, 
should still be higher and compensate partially for this.

The photon measurement in ALICE will be performed with the photon 
spectrometer PHOS \cite{phos}, which consists of a calorimeter out of 
$PbWO_{4}$ crystals read out by avalanche photodiodes. $PbWO_{4}$ has 
recently received a lot of attention as a relatively radiation hard, 
dense crystal suited for photon detection at colliders --- it will 
also be used for the electromagnetic calorimeter of the CMS 
experiment. The dense material allows to use small cross sections of 
individual modules and thus yields excellent position resolution and 
low double hit probability. The energy resolution of the detector is 
expected to be below $2 \% $ for $E_{\gamma} > 4 \, \mathrm{GeV}/c$. 
The detector will be operated at $T = -25^{\circ} \mathrm{C}$, which 
results in a considerably higher light output compared to room 
temperature. It will cover $ |\eta| < 0.12$ and $100^{\circ}$ in 
azimuthal angle.
Different 
options of using either a charged particle veto detector or a 
pre-shower detector in front of PHOS are still being investigated. 
A pre-shower detector would provide much better hadron 
rejection capabilities at higher costs compared to a pure charged particle 
veto.

The dynamic range of the photon measurements at RHIC should extend 
over the range $1.0 \, \mathrm{GeV}/c \le p_{T} \le 30 \, 
\mathrm{GeV}/c$, discrimination of high $p_{T}$ photons from merging 
$\pi^{0}$ should be possible up to $p_{T} = 25 \, \mathrm{GeV}/c$. The 
ALICE PHOS has been optimized for photons in the range
$0.5 \, \mathrm{GeV}/c \le p_{T} \le 10 \, \mathrm{GeV}/c$, while 
measurements should be possible up to $p_{T} = 40 \, \mathrm{GeV}/c$ 
with $\pi^{0}$ rejection at least up to 
$p_{T} = 30 \, \mathrm{GeV}/c$.

\section{Comparison of Theory and Experiment}\label{sec:cte}

\subsection{Comparisons of prompt photons in $pp$ and
$pA$ collisions}\label{subsec:cpp}

\begin{figure}[hbt]
\centerline{\resizebox{10cm}{!}{\includegraphics{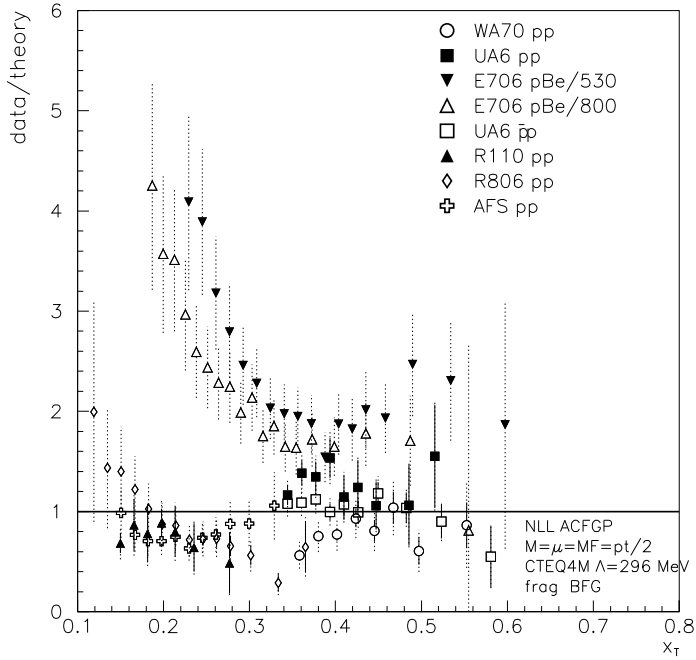}}}
\caption{Normalized ratio of data to theory for various experiments as
a function of $x_T=2p_T/\sqrt{s}$ \cite{Aurenche99}.}
\protect\label{fig4.1}
\end{figure}

The inclusive production of prompt photons in $pp$ collisions
is not fully understood \cite{Aurenche99},
as we mentioned already in Section \ref{subsubsec:pro}.
Using perturbative QCD together with an ``optimization'' prescription
\cite{Aurenche87}, in which the renormalization scale $\Lambda_{\rm QCD}$
and parameters of the
parton structure functions are optimized, an excellent agreement between
theory \cite{Aurenche88,Aurenche89,Vogelsang95,Gordon97}
and experiments until 1997 \cite{Vogelsang97}
over the entire range of collision energy $\sqrt{s}$ and transverse
photon momentum $p_T$
was obtained,
choosing a single set of structure functions
and a unique value for $\Lambda_{\rm QCD}$. However, new data
from E706 ($p-Be$ and $\pi -Be$) at $\sqrt{s}=31.6$ GeV and
$\sqrt{s}=38.8$ GeV \cite{Apanasevich98} 
%and some older R806 data from ISR \cite{Annassontzis82} 
cannot be explained in this way,
as shown in Fig. \ref{fig4.1}.

\begin{figure}[hbt]
\centerline{\resizebox{10cm}{!}{\includegraphics{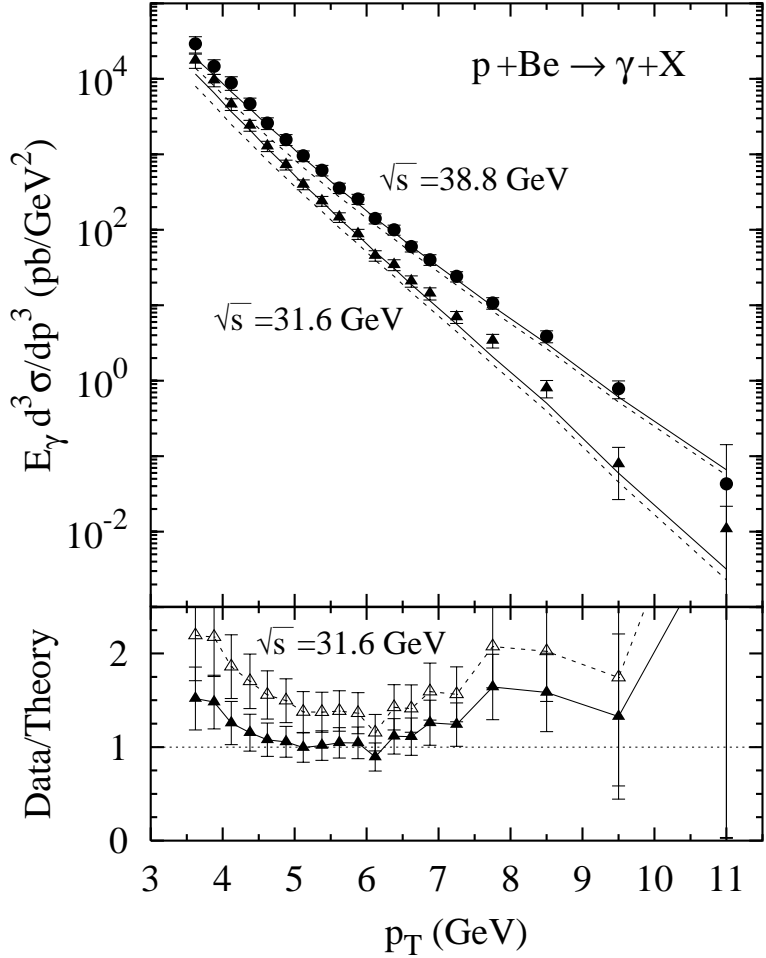}}}
\caption{Comparison of data from E706 with theoretical calculations
with (solid line) and without (dashed line) Cronin effect \cite{Papp00}.}
\protect\label{fig4.2}
\end{figure}

The disagreement between theory and experiment persists even if
the theoretical description is improved by a soft-gluon resummation and
by including next-to-next-to-leading-order corrections \cite{Kidonakis00}.
An agreement between theory and data can be achieved only by introducing
an intrinsic transverse parton momentum $k_T$ as
a new phenomenological energy-dependent parameter \cite{Fontannaz78,Hutson95}.
However, choosing $k_T=0.7$ GeV for fitting the UA6 data and
$k_T=1.2$-1.3 GeV for E706 implies $k_T>1.5$ GeV for the ISR data,
which destroys the agreement with data sets of WA70 and ISR
\cite{Aurenche99}. On the other hand, for the $pA$ data from E706
the introduction of an additional $k_T$ broadening
by nuclear effects (Cronin effect \cite{Antreasyan79}) provides
a possible explanation of the data \cite{Papp00} as shown in Fig. \ref{fig4.2}.

\begin{figure}[hbt]
\centerline{\resizebox{10cm}{!}{\includegraphics{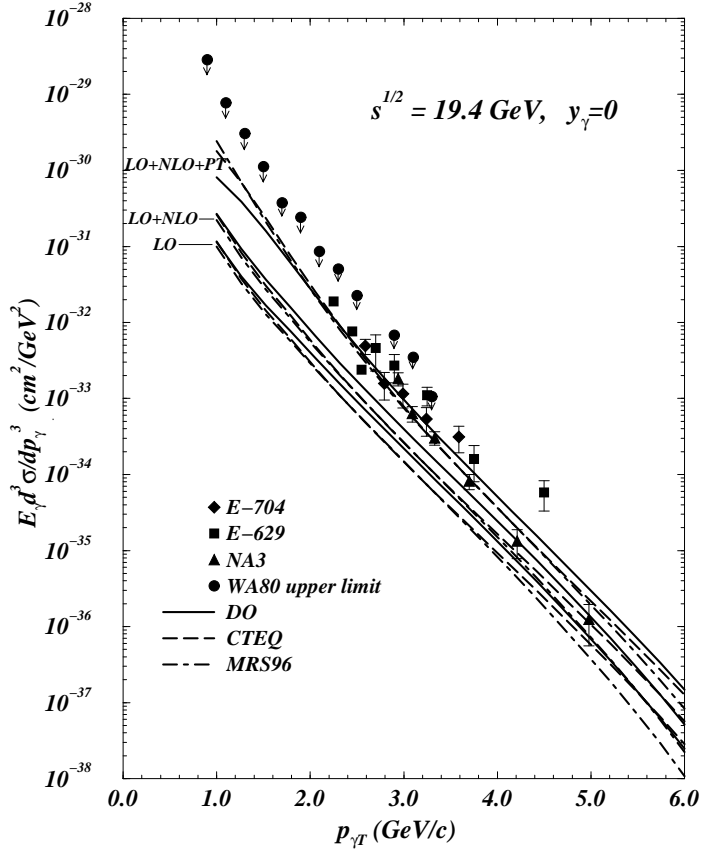}}}
\vspace*{-1cm}
\caption{Comparison of data at $\sqrt{s}=19.4$ GeV with theoretical 
calculations using next-to-leading order corrections (NLO), an intrinsic
transverse momentum (PT), and different parton distributions (DO, CTEQ, MRS96)
\cite{Wong98}.}
\protect\label{fig4.2a}
\end{figure}

Similarly, Wong and Wang \cite{Wong98} concluded that most experimental data 
can be explained if an intrinsic transverse momentum of the partons is taken 
into account. For example, at $\sqrt{s}=19.4$ GeV the photon invariant cross
section is enhanced by a factor of 2 if next-to-leading order corrections
are included and by a factor of 4 to 8 due to the transverse momentum effect
(see Fig. \ref{fig4.2a}). Also shown is the small
influence of using different parametrizations (DO, CTEQ, MRS96) of the parton 
distributions on the cross section.

Furthermore, Apanasevich et al. \cite{Apanasevich01} 
compared theoretical models including
next-to-leading order and intrinsic momentum effects with ratios
of $\gamma /\pi^0$ yields, in which various experimental and theoretical
uncertainties cancel. They conclude that the theory agrees reasonably with
data at $\sqrt{s}>30$ GeV, whereas at lower energies deviations 
between theory and experiments as well as between different data sets 
appear.

\subsection{Comparison with SPS Heavy-Ion Experiments}\label{subsec:cse}

% start of changes (tp)
\subsubsection{Comparison to Limits of WA80}\label{subsubsec:cl}

The first confrontations of theoretical calculations
\cite{Srivastava94,Shuryak94,Neumann95,Arbex95,Dumitru95}
to experimental
heavy-ion data were performed with the preliminary WA80 data
\cite{Santo94} which indicated a significant excess of direct photons
in central $S$+$Au$ reactions.
We will not discuss this in detail, as the final publication of the
WA80 data \cite{Albrecht:1996} did only provide an upper limit for direct
photon production. All of these publications have been able to
describe the preliminary WA80 data with scenarios including a phase
transition using 1-loop HTL rates for the QGP. To achieve this,
Shuryak and Xiong \cite{Shuryak94} had to assume
a surprisingly long-living mixed phase. Furthermore, Srivastava and
Sinha \cite{Srivastava94} and Dumitru et al. \cite{Dumitru95}
concluded that a pure hadron gas scenario could be excluded because
it would overpredict the observed photon spectra. While the
interpretation found in Ref. \cite{Shuryak94} could no longer be sustained
from the final WA80 data, the conclusions of
Refs. \cite{Srivastava94} and \cite{Dumitru95} are related to some
of the simplifying assumptions, mainly the unrealistic equation of
state used for the HHG.

\begin{figure}[hbt]
\centerline{\resizebox{10cm}{!}{\includegraphics{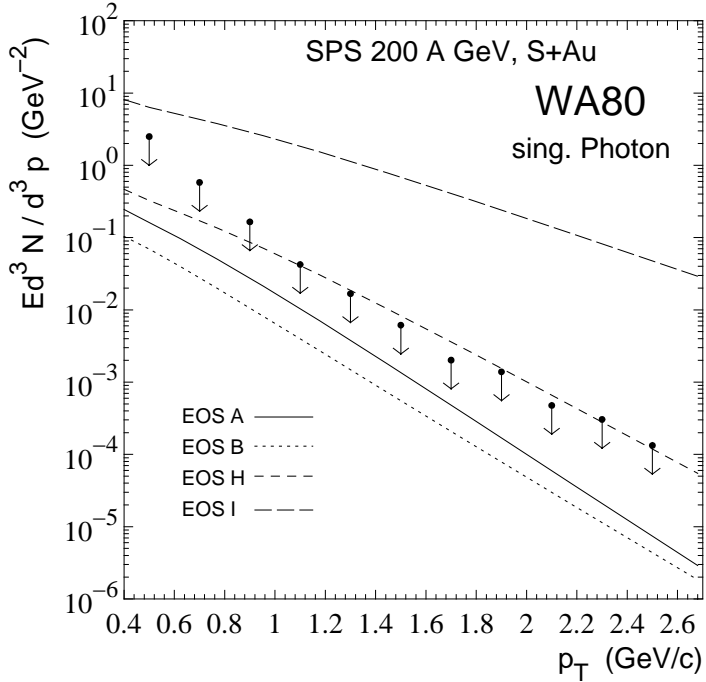}}}
\vspace*{-7cm}
\caption{Comparison of the WA80 upper limits \cite{Albrecht:1996}
with hydrodynamical calculations
using different EOS \cite{Sollfrank97}. EOS A and B contain a phase
transition, EOS H corresponds to a HHG of massive mesons and baryons and
EOS I to an ideal pion gas.}
\protect\label{fig4.4}
\end{figure}

\begin{figure}[hbt]
\centerline{\resizebox{10cm}{!}{\includegraphics{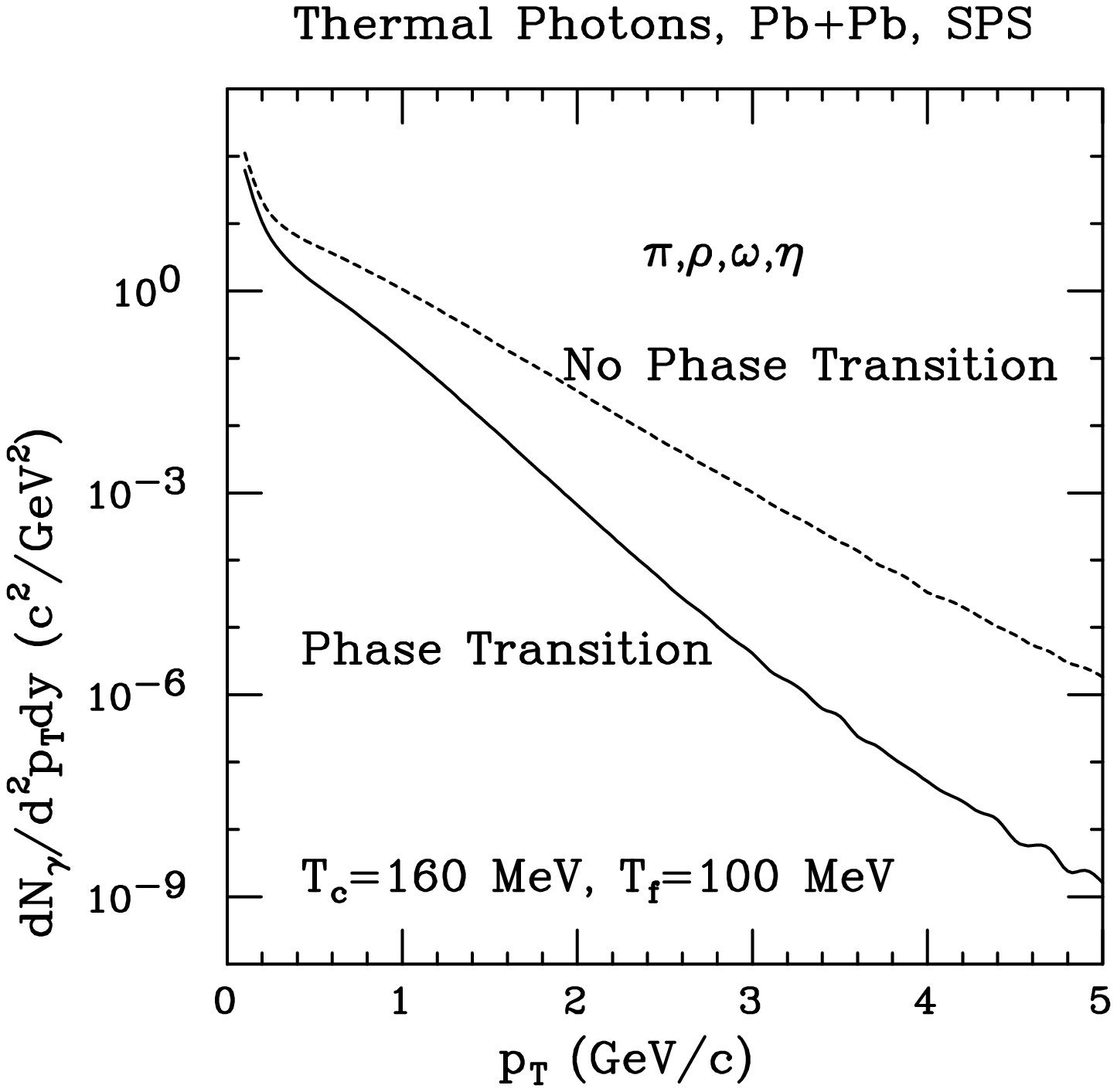}}
\hspace*{-3cm}\resizebox{10cm}{!}{\includegraphics{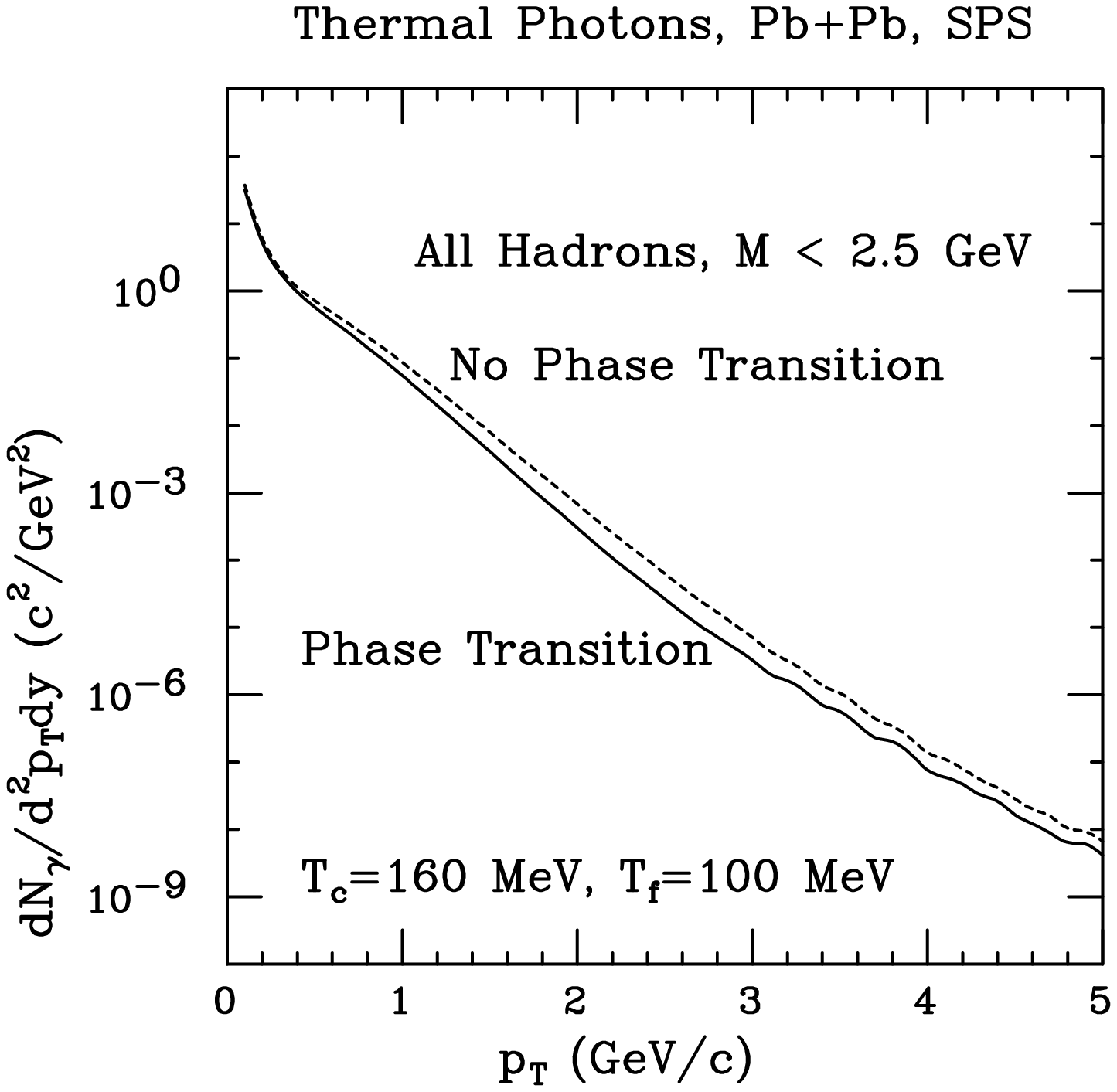}}}
\caption{Comparison of photon spectra with a simplified HHG EOS
(left) and a realistic one (right) \cite{Cleymans97}.}
\protect\label{fig4.3}
\end{figure}

Sollfrank et al. \cite{Sollfrank97} employed a 2+1-dimensional hydrodynamical
model to calculate hadronic and electromagnetic spectra at the same time.
They investigated various EOS with and without phase transition. They
concluded that the WA80 limits only exclude an ideal pion gas for the EOS
of the HHG. A no-phase-transition scenario with a more complicated HHG EOS,
on the other hand, is not ruled out (see Fig. \ref{fig4.4}) for initial
temperatures below $T_0=250$ MeV. Also in Ref.~\cite{Sarkar99}, it was found
that the WA80 upper limits for direct photons cannot distinguish between
a phase transition and no phase transition. Furthermore, it was argued that
medium effects on hadron masses will reduce the high $p_T$ spectrum
by about a factor of 2.

\begin{figure}[b]
\centerline{\resizebox{7cm}{!}{\includegraphics{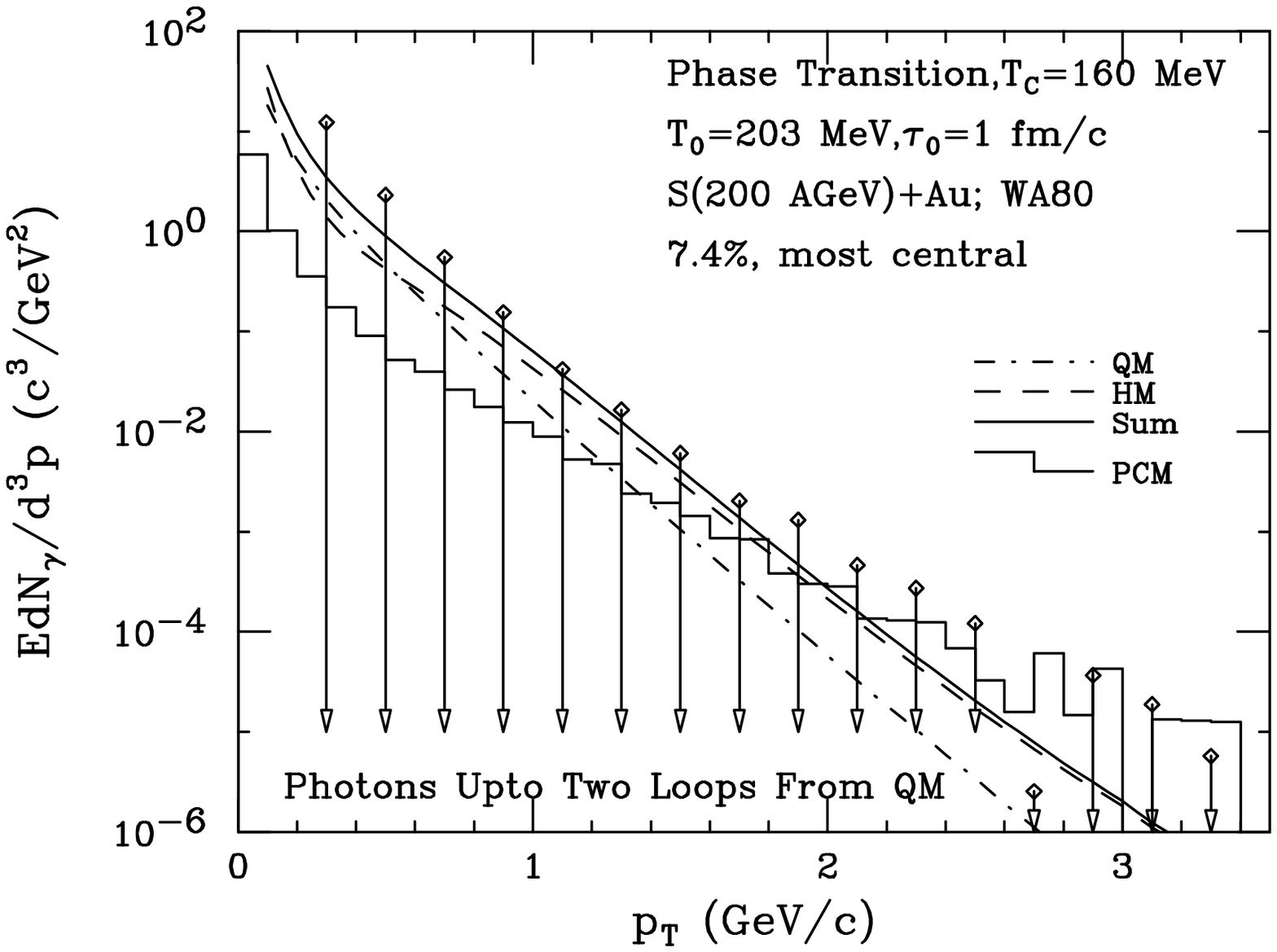}}
\resizebox{7cm}{!}{\includegraphics{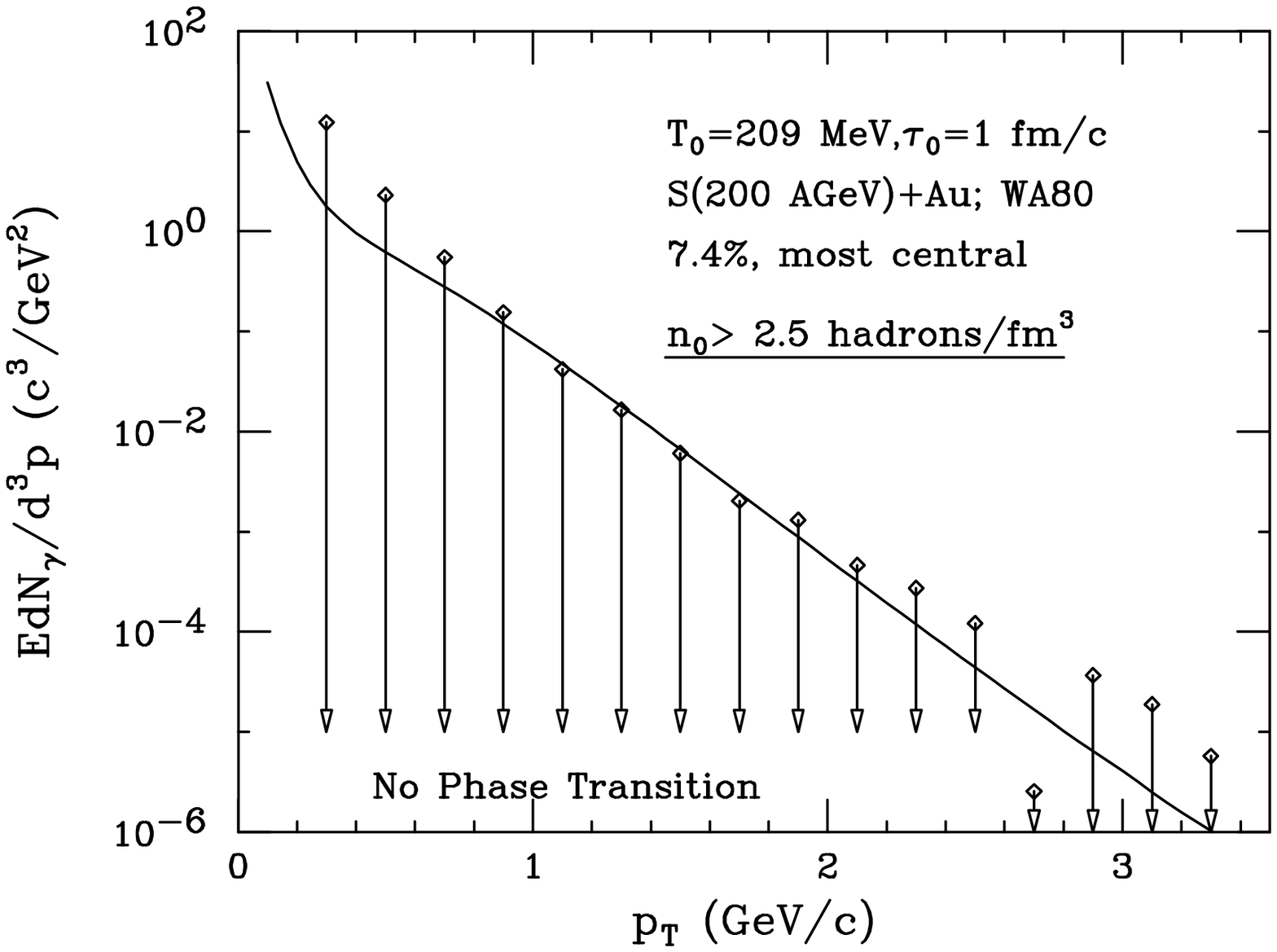}}}
\caption{Comparison of the WA80 upper limit \cite{Albrecht:1996}
for the direct
photon spectrum to a calculation with (left) and without (right)
phase transition \cite{Srivastava00a}.}
\protect\label{fig4.3.5}
\end{figure}

Recently, Srivastava and Sinha repeated their
calculations using the 2-loop HTL results \cite{Aurenche98,Steffen01}
and a HHG EOS with all hadrons with masses up to 2.5 GeV
\cite{Cleymans97,Cleymans98}. The importance of a realistic EOS for the HHG,
discussed already in Section \ref{sec:ts},
is demonstrated in Fig. \ref{fig4.3}, where one observes that a
HHG EOS with only $\pi$-, $\rho$-, $\omega$-, and $\eta$-mesons leads to
a large overprediction of the photon spectrum at SPS energies,
whereas a realistic EOS including hundreds of resonances
is similar to the spectrum with a phase transition. The new
conclusion by Srivastava and Sinha
was that the phase transition as well as the no-phase transition
scenario agree with the upper limits from WA80 \cite{Srivastava00a}
(see Fig. \ref{fig4.3.5}). However, they argue that
a hadron density of several hadrons per fm$^3$ is needed at the
initial time, which seems to contradict the assumption of a HHG.

Summarizing, the upper limits for direct photons from WA80 can be explained
with and without phase transition and, therefore, do not allow a
conclusion about the existence of a QGP phase. However, they have
triggered investigations of some of the simplifications used in
earlier calculations, as e.g. unrealistic EOS for the HHG.

\subsubsection{Comparison with WA98}\label{subsubsec:cwa98}

Recently the WA98 Collaboration presented the first data on direct photons
in relativistic heavy-ion collisions \cite{Aggarwal:2000}. Different groups
comparing their calculations with these data arrived at different conclusions,
which we will review in the following.

\begin{figure}[hbt]
\centerline{\resizebox{12cm}{!}{\includegraphics{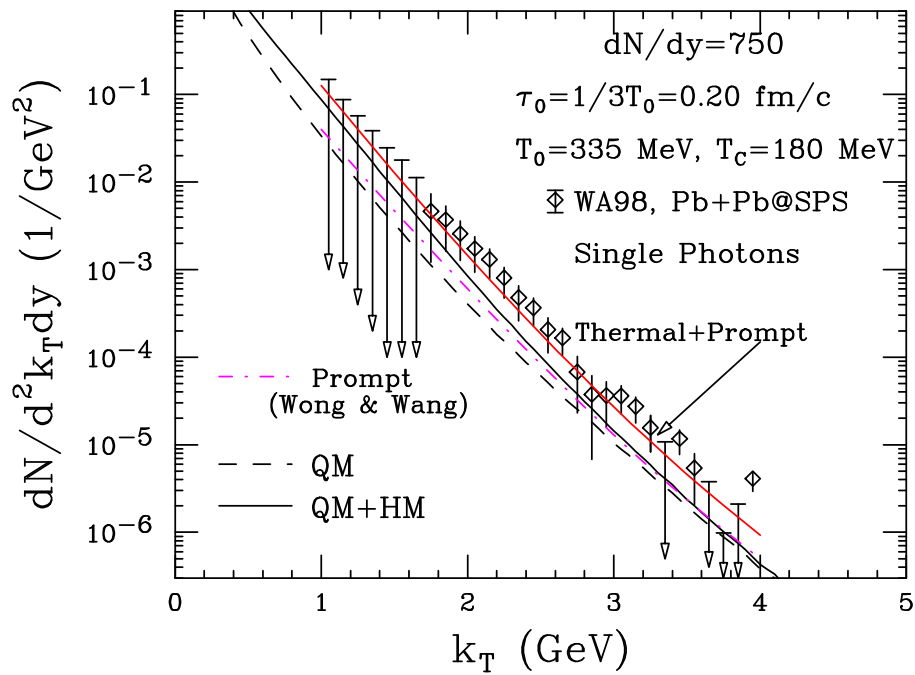}}}
%\vspace*{-7cm}
\caption{Comparison of the WA98 data with a hydrodynamical calculation
by Srivastava and Sinha \cite{Srivastava00}.}
\protect\label{fig4.5}
\end{figure}

\begin{figure}[hbt]
\centerline{\resizebox{14cm}{!}{\includegraphics{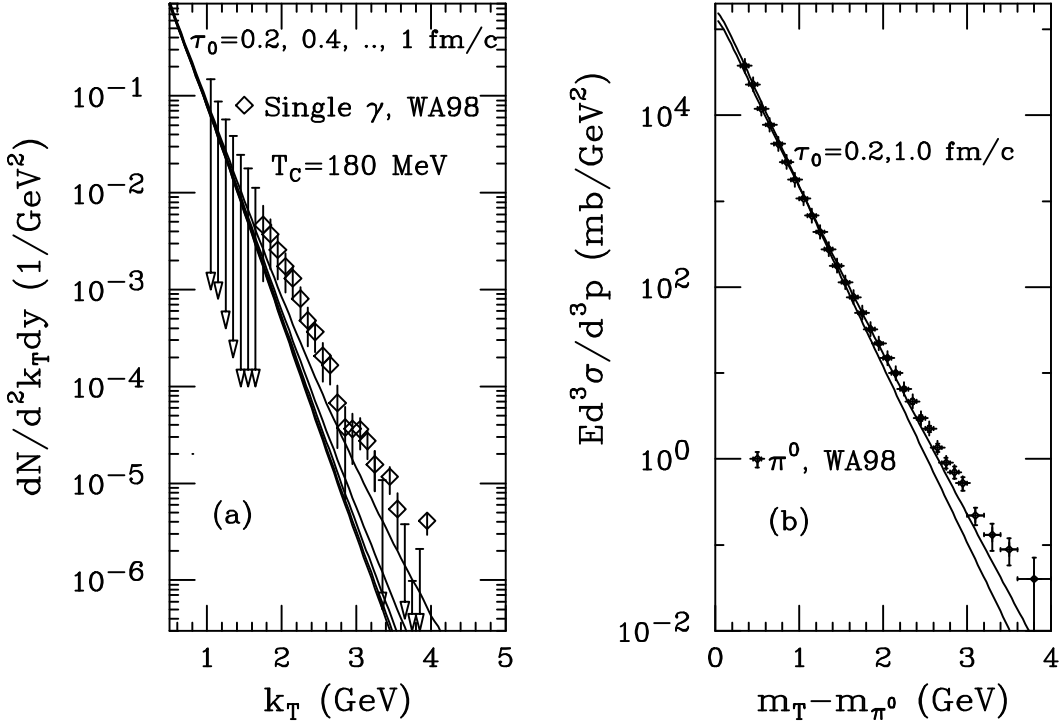}}}
%\vspace*{-7cm}
\caption{Comparison of the SPS photon and pion spectra with
hydrodynamical calculations
using different initial times \cite{Srivastava00}. $\tau_0=0.2$ fm/c
corresponds to the upper lines.}
\protect\label{fig4.6}
\end{figure}

Srivastava and Sinha \cite{Srivastava00} argued, using the 2-loop HTL rate
for the QGP contribution and a realistic EOS for the HHG,
that the QGP is needed to explain the data.
Their conclusion
is based on the use of a very high initial temperature ($T_0=335$ MeV)
and very small initial time ($\tau_0=0.2$ fm/c), which could explain
the observed flat photon spectrum for transverse photon momenta $p_T>2$ GeV
(see Fig. \ref{fig4.5}).
Their estimate of the initial conditions follows from the isentropy condition
Eq. (\ref{initial}) \cite{Hwa85} together with the use of the uncertainty
principle \cite{Kapusta92}, $\tau_0\simeq 1/3T_0$. However, the uncertainty
relation, giving the parton formation time, might underestimate the
thermalization time \cite{Kapusta92}. However, the authors argue that
such a small thermalization time also provides a very good description
of the intermediate mass dilepton excess measured by NA50 \cite{Srivastava01a}.
Also, if later thermalization times are assumed, one should add a
pre-equilibrium contribution to the photon spectrum.
Using more conservative
initial conditions ($T_0=196$ MeV, $\tau_0=1$ fm/c), the data are clearly
underestimated in particular at $p_T>2$ GeV (see Fig. \ref{fig4.6}).
Srivastava and Sinha
have also included prompt photons from the work by Wong and Wang
\cite{Wong98}, which follow from a next-to-leading order perturbative QCD
calculation, where an intrinsic parton momentum of $\langle k_T^2 \rangle
=0.9$ GeV$^2$ has been used. Srivastava and Sinha found that the thermal
photons contribute half of the total photon spectrum and that in particular
at large $p_T$ most of the thermal photons are due to the QGP contribution.
However, one should keep in mind that the question of the QGP photon rate
is not yet settled and that the 2-loop HTL calculation might be an
overestimation, since the LPM-effect is neglected there \cite{Aurenche00a}.
Furthermore, Srivastava and Sinha neglected a finite baryo-chemical
potential, which reduces the QGP contribution to the photon spectrum. For
example, even a small baryo-chemical potential of $\mu_B=100$ MeV,
corresponding to a quark-chemical potential $\mu_q=300$ MeV, reduces
the 1-loop HTL photon production rate by more than a factor of 3 at
$p_T=3$ GeV \cite{Traxler95}. On the other
hand, the photon production from nucleons and other baryons might enhance
the rate \cite{Steele97}.

Also Alam et al. \cite{Alam01a}
favor a QGP contribution for explaining the WA98 data. However, they claim
that due to the uncertainties in the rates and the hydrodynamical parameters
a definite conclusion is not possible at SPS energies. Alam et al. used a
EOS of the HHG with much less degrees of freedom than Srivastava and Sinha.
However, they considered in-medium modifications of hadron masses
which lead to an enhancement of the photon spectrum at SPS energy for photon
momenta above $p_T=2.5$ GeV. At $p_T=4$ GeV this enhancement amounts to an
order of magnitude.
In addition, they introduced an initial radial velocity $v_0$, which renders
the photon spectrum flatter at $p_T>2$ GeV even for moderate
values of $v_0=0.2 c$. Owing to these effects, Alam et al. obtained
a much flatter spectrum at high $p_T$, which allows an explanation
of the WA98 data without prompt photons within the phase-transition
as well as the no-phase-transition scenario, even for conservative
initial conditions, $T_0\simeq 200$ MeV and $\tau_0=1$ fm/c.
However, in a number of papers
hydrodynamical calculations were able to reproduce hadronic as well as
electromagnetic spectra (see e.g. \cite{Huovinen99})
and flow patterns \cite{Kolb99,Kolb00}
without assuming an initial radial velocity.
In particular, the high $p_T$ component above 2 GeV in the $\pi^0$-spectrum,
which has been used to argue about an initial radial velocity
\cite{Peressounko00a}, might also come from hard processes
and should not be treated in a hydrodynamical model \cite{Wang99}.

\begin{figure}[hbt]
\centerline{\resizebox{10cm}{!}{\rotatebox{-90}{\includegraphics{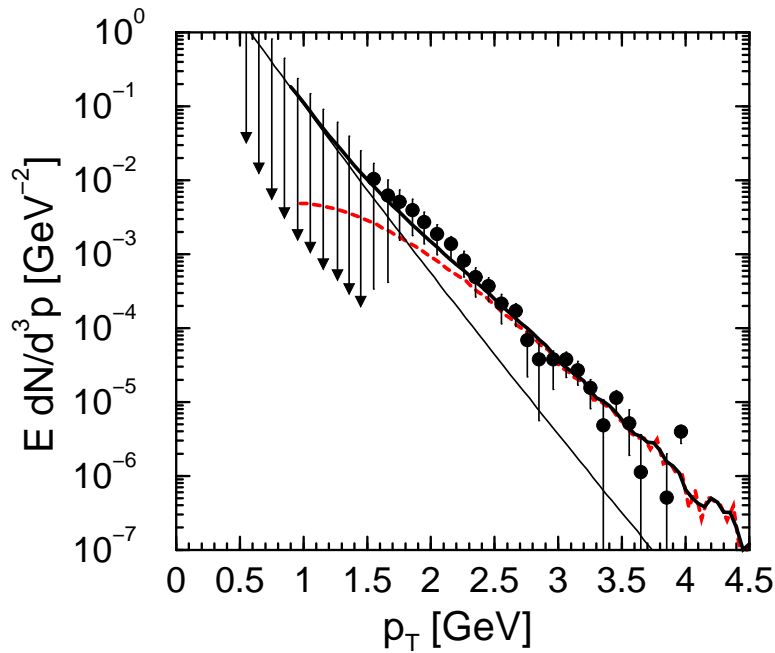}}}}
\caption{Comparison of the WA98 data with a theoretical spectrum
(upper solid line)  including
thermal (lower solid line) and prompt photons (dashed line)
\cite{Gallmeister00}.}
\protect\label{fig4.7}
\end{figure}

\begin{figure}[hbt]
\centerline{\resizebox{10cm}{!}{\includegraphics{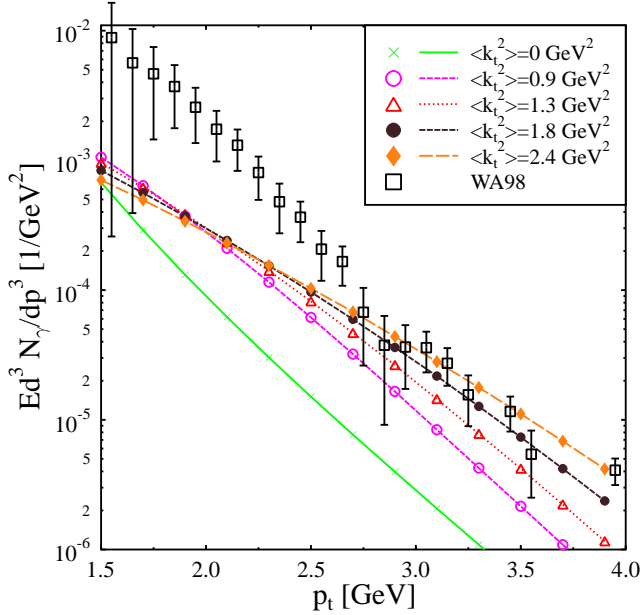}}}
\caption{The photon spectrum calculated for different intrinsic transverse 
momenta in comparison to the WA98 data \cite{Dumitru01}.}
\protect\label{fig4.7a}
\end{figure}

Gallmeister et al.
\cite{Gallmeister00}, on the other hand, argued that 
the low momentum part of the WA98 spectrum ($p_T<2$ GeV) is consistent 
with a thermal source, either QGP or HHG, which also describes dilepton data.
The hard part ($p_T>2$ GeV), on the other hand, agrees
with the prompt photon spectrum if its absolute value is normalized to the
data, corresponding to a large effective $K$-Factor of 5
(see Fig. \ref{fig4.7}). It should be noted that the prompt photon production
at $\sqrt{s}=17.3$ GeV is uncertain, in particular for low $p_T$.
Gallmeister et al. used a simple hydrodynamical model with
only a radial expansion and a fixed average temperature,
which reproduces dilepton data from SPS
\cite{Gallmeister00a}. In addition, they adopted only the 1-loop HTL
rate for the entire evolution of the fireball according to the
quark-hadron duality hypothesis. If in addition a transverse flow
of $v=0.3$ is included, the theoretical and experimental spectra agree.
In other words, according to
the investigation by Gallmeister et al., the WA98 spectrum
can be explained by a thermal source (QGP or HHG) plus prompt photons
and there is no necessity of a QGP phase.
Similarly, Dumitru et al.
\cite{Dumitru01} showed that the WA98 photon spectrum above $p_T=2.5$ GeV
can be explained by prompt photons if a nuclear broadening of
$\Delta k_T^2=\langle k_T^2\rangle_{AA} -\langle k_T^2\rangle_{pp}\simeq
0.5$ - 1 GeV$^2$ is introduced. For low $p_T<2.5$ GeV, however, 
prompt photons fail to reproduce the WA98 data regardless of the amount of
nuclear broadening employed (see Fig. \ref{fig4.7a}).

\begin{figure}[hbt]
\centerline{\resizebox{10cm}{!}{\includegraphics{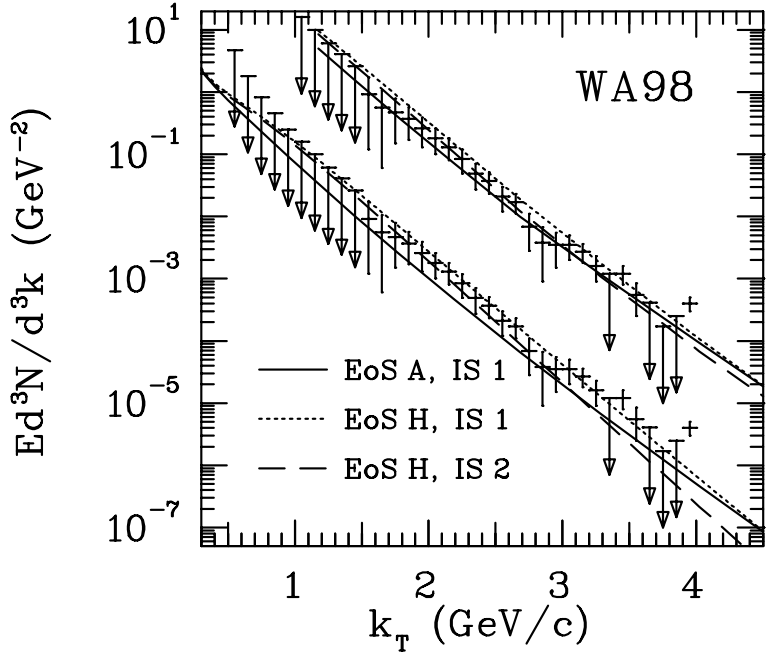}}}
\caption{The photon spectrum calculated for different EOS and initial 
conditions 
with prompt photons (upper set, scaled by a factor 100)
and without (lower set) in comparison to the WA98 data
\cite{Huovinen00}. EoS A, IS 1 contains a phase transition at $T_c=165$
MeV, an average initial temperature $T_0=255$ MeV, and a local maximum 
temperature $T_{\rm max}=325$ MeV. EoS H describes only a HHG with
$T_0=234$ MeV, $T_{\rm max}=275$ MeV (IS 1) and $T_0=213$ MeV, 
$T_{\rm max}=245$ MeV (IS 2).}
\protect\label{fig4.8}
\end{figure}

Huovinen et al. \cite{Huovinen99}, fixing
the initial conditions ($T_0=210$ - 250 MeV) in their hydrodynamical model
partly by a comparison with hadron spectra and using the most recent
results for the QGP photon rate by Arnold et al. \cite{Arnold01a},
were able to describe the data equally well with or without a phase
transition \cite{Huovinen00} (see Fig. \ref{fig4.8}).
They were able to fit the WA98 data without a high initial
temperature\footnote{It appears, however, that 
locally an initial temperature of $T_{\rm max} > 240$ MeV is required to
fit the flat slope of the data \cite{Huovinen01}.}, an initial radial 
velocity, prompt photons, or
in-medium hadron masses. This might be caused partly by a strong flow
at later stages since they do not assume a boost-invariant
longitudinal expansion. However, the different results and conclusions
between the work by Srivastava and Sinha \cite{Srivastava00} and
by Huovinen and Ruuskanen \cite{Huovinen00} cannot be explained
in this way \cite{Huovinen01}.
The conclusion, that the WA98 data can be explained with or
without a phase transition, was also obtained in
Refs.\cite{Peressounko00a,Chaudhuri00}.
However, there an initial
radial velocity (see above) had to be introduced in order
to obtain a quantitative description of the
WA98 data.

\begin{figure}[hbt]
\centerline{\resizebox{10cm}{!}{\includegraphics{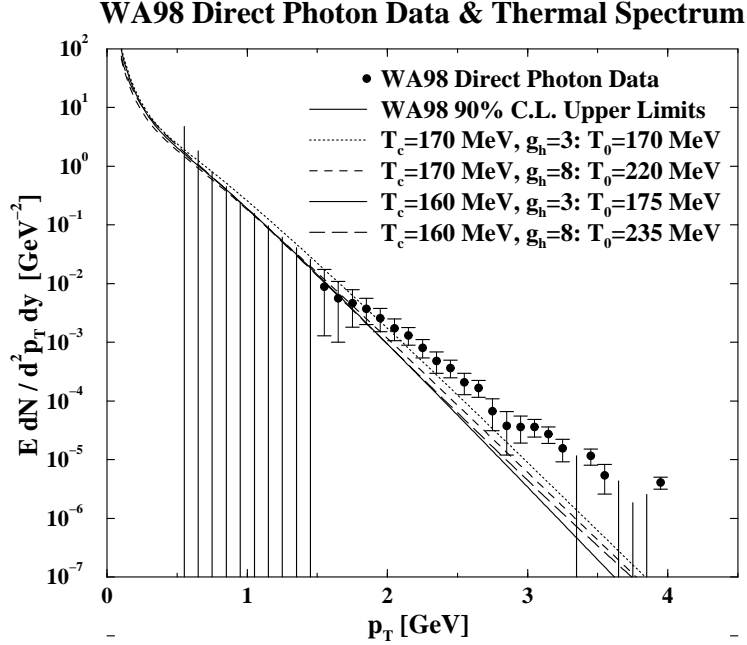}}}
\caption{The photon spectrum calculated for different critical
and initial temperatures and degrees of freedom in the HHG
in comparison to the WA98 data
\cite{Steffen01}.}
\protect\label{fig4.9}
\end{figure}

Steffen and Thoma \cite{Steffen01}, using the corrected
2-loop HTL rate, the parametrization of the HHG rate
Eq. (\ref{hadrons}), and
the simple 1-dimensional Bjorken hydrodynamics, found that the
thermal photons underestimate the WA98 data for $p_T>2$ GeV
for reasonable initial temperatures between 170 and 235 MeV
(see Fig. \ref{fig4.9}).
These high $p_T$ photons, however,
might be explained by prompt photons. Using a massless
pion gas ($g_h=3$) for the HHG EOS, the computed spectrum exceeds
the WA98 data at $p_T<2$ GeV already for initial temperatures
as low as $T_0=170$ MeV. However, if one increases the number of
massless pions to $g_h=8$, which provides an agreement with photon spectra from
the HHG using more realistic EOS \cite{Srivastava00}, the life-time
of the mixed phase is shortened, which allows higher initial temperatures,
up to $T_0=235$ MeV.

Summarizing, WA98 found a rather flat photon spectrum above
$p_T=2$ GeV, which cannot be easily explained by conservative
models. It requires either a high initial
temperature, a large prompt photon contribution, an initial
radial velocity, in-medium modifications of the hadron masses
and/or a strong flow at later stages.
Furthermore, the effect of a finite baryo-chemical potential, which
should be important at SPS energies, has not been investigated so far.
Also the effect of a finite life-time of the QGP, which flattens the
high $p_T$ spectrum \cite{Wang00,Wang01}, has only be considered for
RHIC energies (see below).
At the moment, we think it is fair to say that the uncertainties
and ambiguities in the
hydrodynamical models and in the rates do not allow to
decide from the WA98 photon spectra about the presence of a QCD phase
transition in SPS heavy-ion collisions.

\subsection{Predictions for RHIC and LHC}\label{subsec:prl}

\begin{figure}[hbt]
\centerline{\resizebox{12cm}{!}{\includegraphics{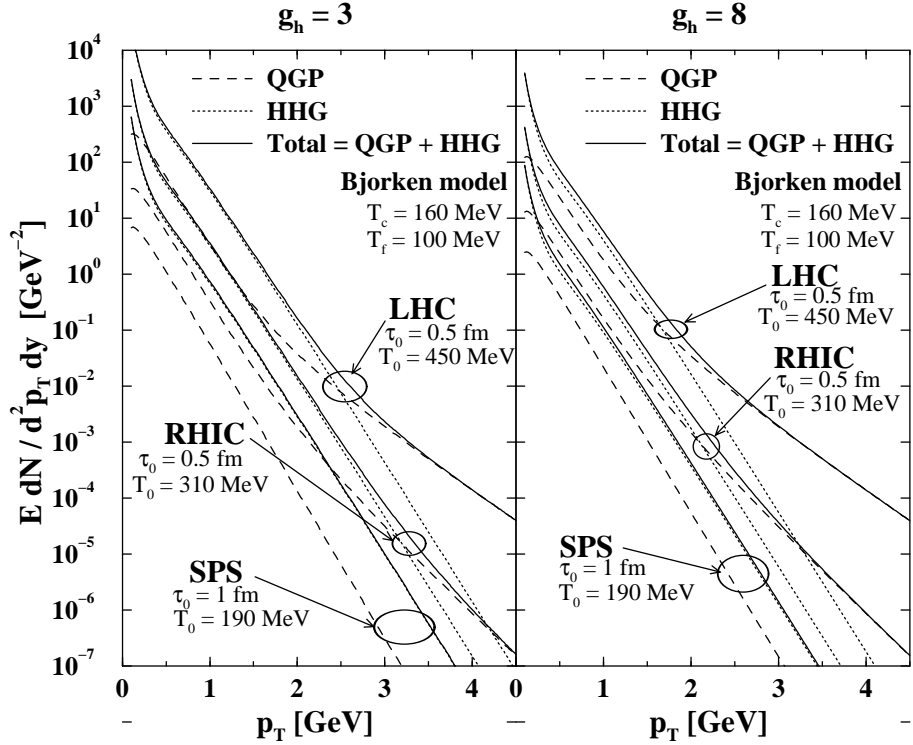}}}
\caption{QGP and HHG photon spectra at SPS, RHIC and LHC $Pb+Pb$
collisions for an ideal pion gas with $g_h=3$ (left) and
$g_h=8$ (right) degrees of freedom \cite{Steffen01}.}
\protect\label{fig4.9.5}
\end{figure}

There are a few predictions for the photon spectra for RHIC and LHC.
Although they suffer from large uncertainties coming from the unknown
initial conditions, at least for LHC in most cases
a window in the photon momentum is predicted,
where a thermal QGP contribution should be visible if the decay photons
are subtracted. Simple 1+1-dimensional models \cite{Steffen01,Steffen99}
show a dominance of
the QGP over the hadron gas contribution for $p_T>3$ GeV (RHIC) and
$p_T>2$ GeV (LHC), respectively, due to the flat spectrum
of the early QGP phase having a high initial temperature ($T_0>300$
MeV), as shown in Fig. \ref{fig4.9.5}.

Taking into account the transverse expansion, the photon spectrum
from the hadron gas becomes also flat due to a strong flow in the late stages
of the fireball. But even in this case, the QGP might
outshine the hadronic phase. However, at which collision energies and
photon momenta this happens,
is controversial. Hammon et al. \cite{Hammon98} predict that the
QGP should not be visible at RHIC, where the prompt photons dominate
the spectrum, but at LHC for $p_T\simeq 2$-5 GeV. They used the following
initial conditions for RHIC, $T_0=533$ MeV and 300 MeV and $\tau_0=0.1236$
fm/c and 0.5 fm/c, and for LHC, $T_0=880$ and 650 MeV and $\tau_0=0.1$
fm/c and 0.25 fm/c, together with a simple HHG EOS and the 1-loop HTL
result for the QGP and HHG rates.

\begin{figure}[hbt]
\centerline{\resizebox{7cm}{!}{\includegraphics{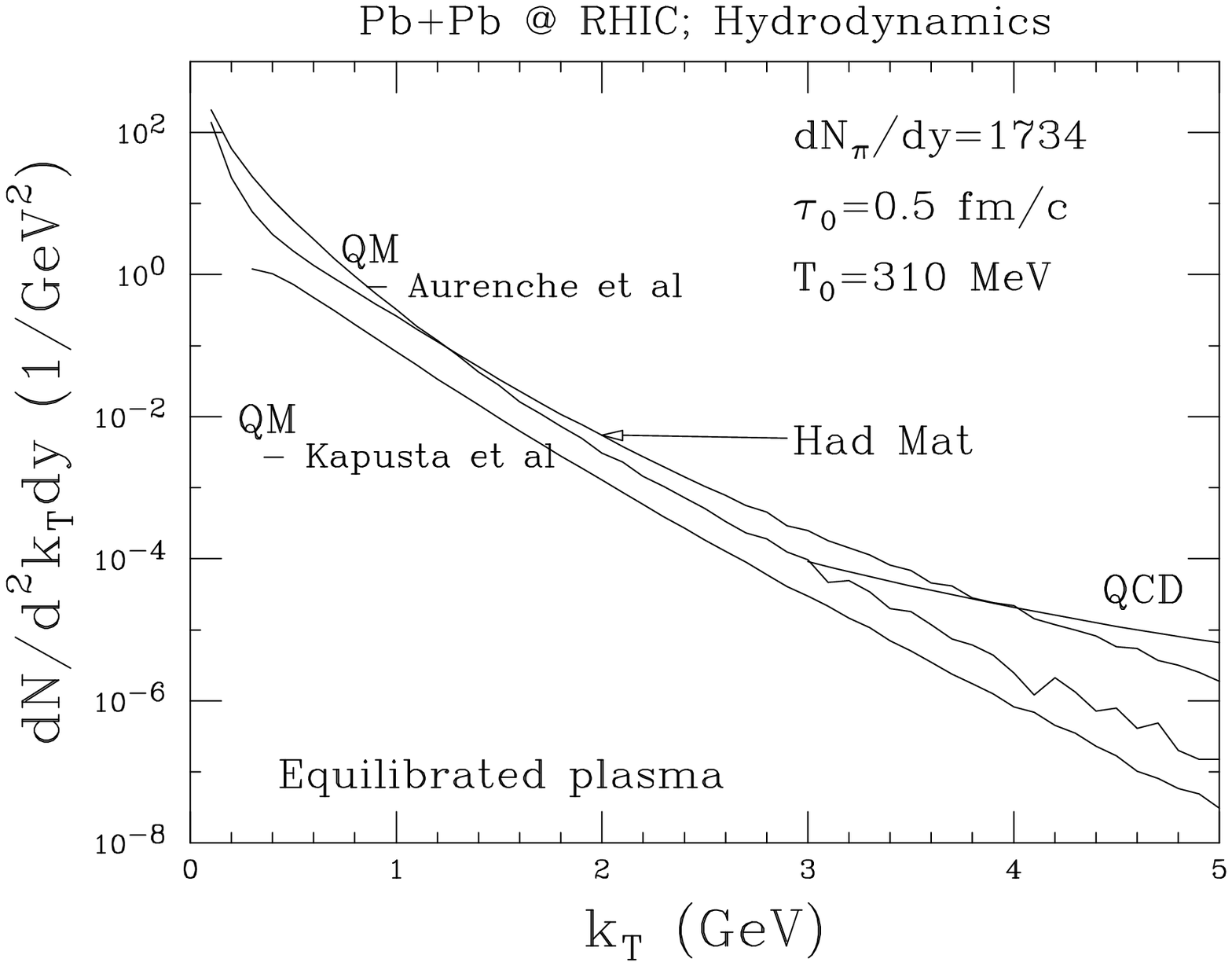}}
\resizebox{7cm}{!}{\includegraphics{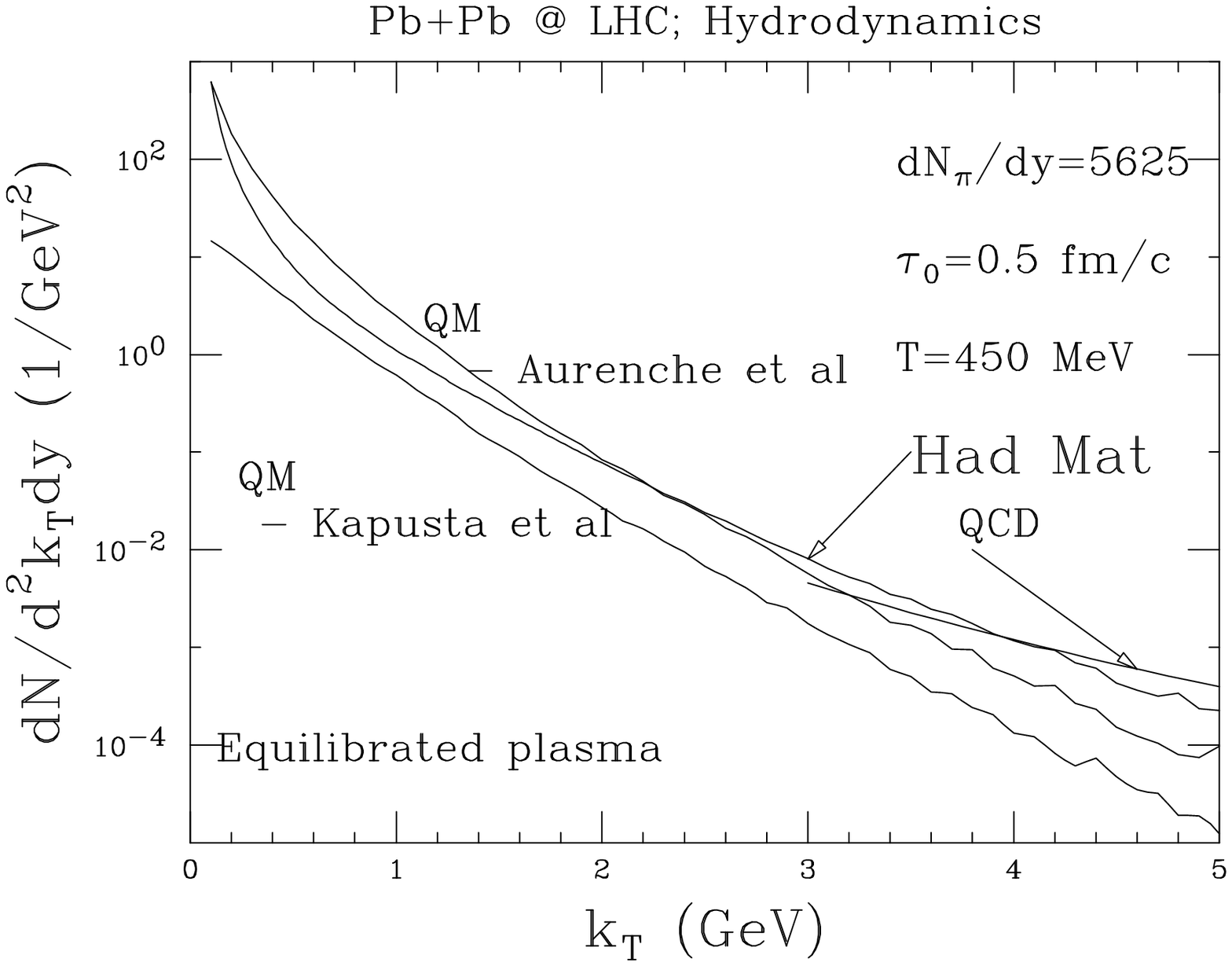}}}
\caption{Predictions for the photon spectrum at RHIC (left) and
LHC (right) \cite{Srivastava99}.}
\protect\label{fig4.10}
\end{figure}

Srivastava
\cite{Srivastava99}, taking into account the corrected
2-loop HTL rate,
predicts that the QGP should be observable for $p_T<1$ GeV at RHIC
and $p_T<2$ GeV at LHC (see Fig. \ref{fig4.10}).
Above these momenta the
sum of the photons from the
thermal hadron gas,
enhanced by flow, and of the prompt photons dominates.
Using the old 2-loop HTL rate, which is too large by a factor of 4,
Srivastava found that the QGP photons dominate already for
$p_T<3$ GeV at RHIC and $p_T<4$ GeV at LHC, which shows the sensitivity
of the predictions to the rates.

Peressounko and Pokrovsky \cite{Peressounko00} predict a ratio
of direct to decay photons of 0.2-0.3 at LHC, which is much larger than
the 5\% limit for direct photons to be detectable at ALICE.
Whether
the photons from the QGP or from the HHG dominate, depends on rates
that are adopted in their calculations.

Alam et al. \cite{Alam01a}, using $T_0=300$ MeV, $\tau_0=0.5$ fm/c
together with the incorrect 2-loop HTL rate for the QGP, predict
that for $p_T<2$ GeV most of the photons are thermal
photons. The photons from the QGP are an important source,
in particular around 2 GeV, but less important than HHG photons. The effect
of the transverse expansion is small in the QGP phase but large in the
late stages of the HHG, leading to a strong population of high $p_T$
photons together with prompt photons.

Dumitru et al. \cite{Dumitru01} have investigated the role of prompt photons
at RHIC. They showed that the effect of an intrinsic $k_T$ is
less important at RHIC than at SPS, but still leads to an
enhancement of the prompt photons by a factor of 3 at $p_T=3$-4 GeV. Below
$p_T<2$-3 GeV thermal photons should be visible.

\begin{figure}[hbt]
\centerline{\resizebox{7cm}{!}{\includegraphics{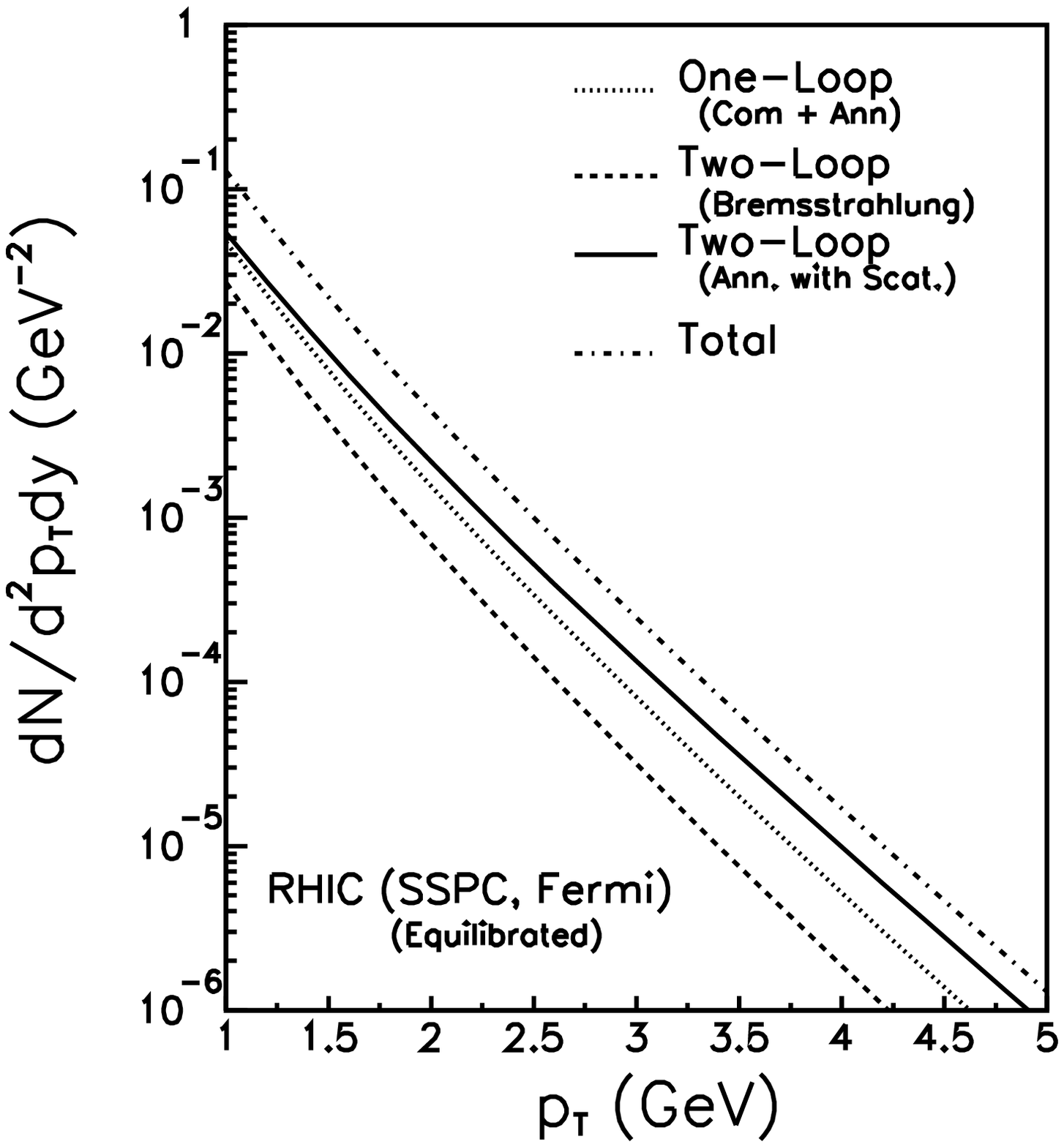}}
\resizebox{7cm}{!}{\includegraphics{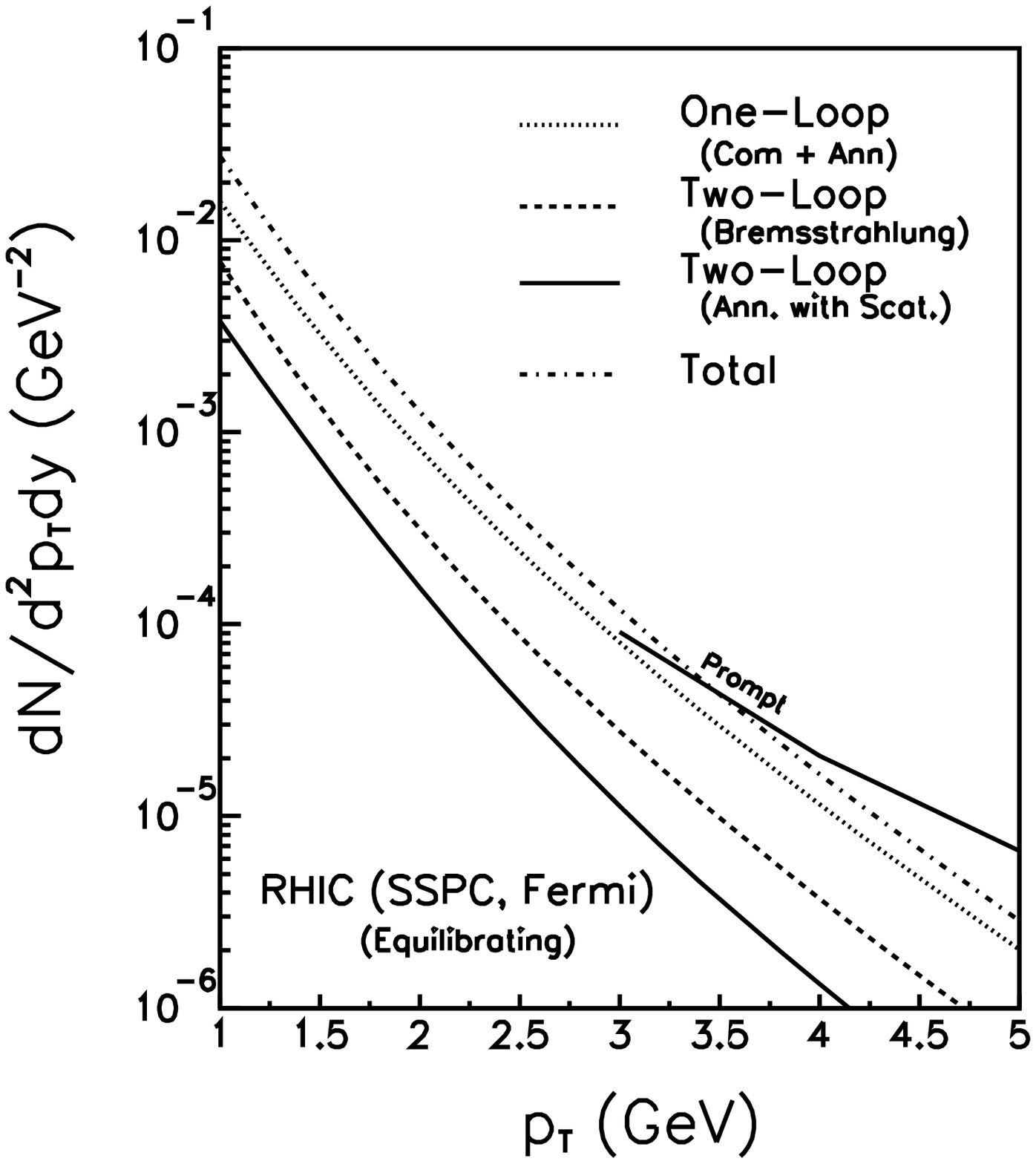}}}
\vspace*{-2cm}
\caption{Comparison of the the photon spectrum at RHIC
from an equilibrated QGP
(left) and a chemically non-equilibrated QGP (right)
at the same initial energy density $\epsilon_0=61.4$ GeV/fm$^3$
\cite{Mustafa00}.}
\protect\label{fig4.11}
\end{figure}

\begin{figure}[hbt]
\centerline{\resizebox{7cm}{!}{\includegraphics{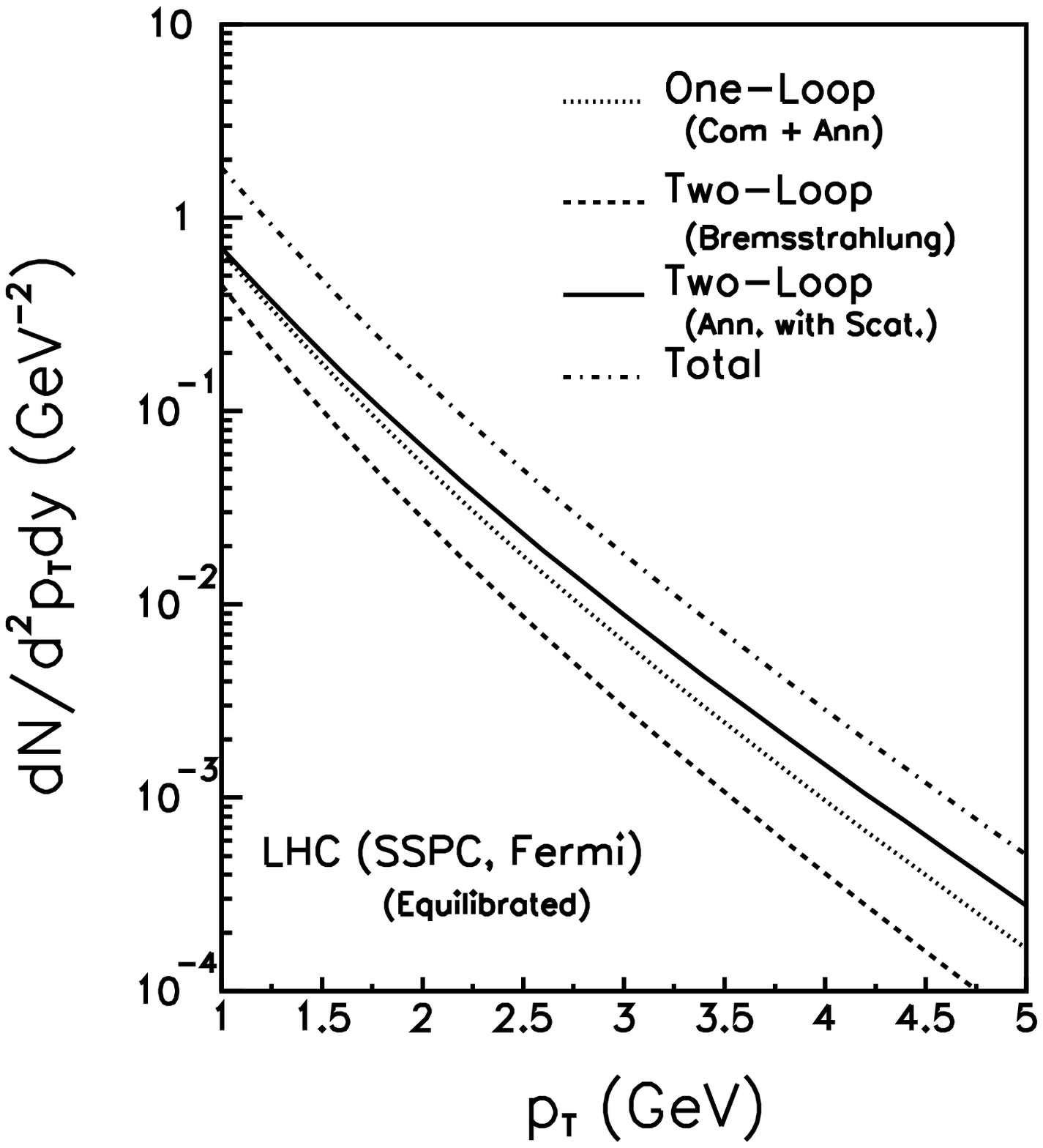}}
\resizebox{7cm}{!}{\includegraphics{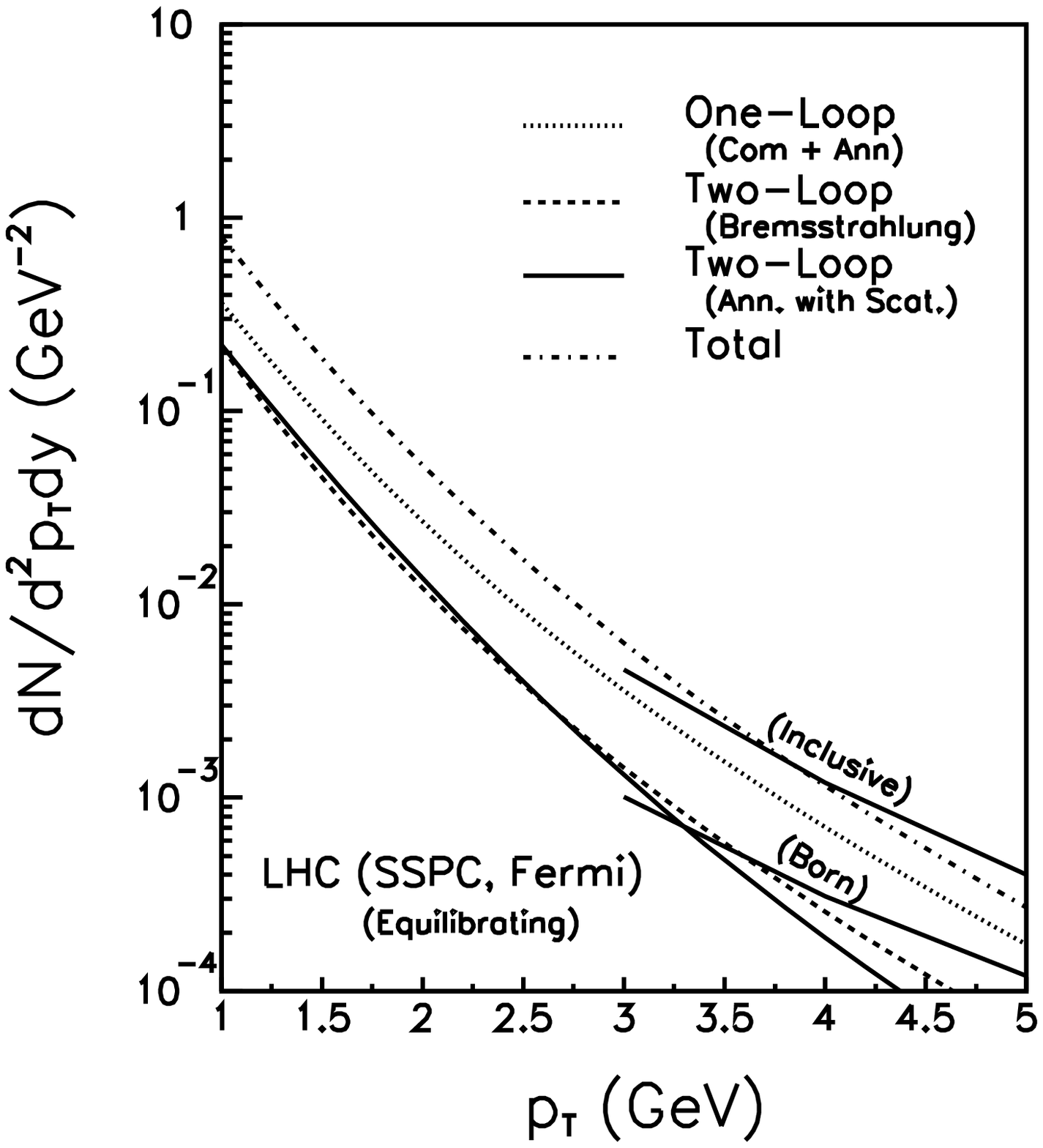}}}
\vspace*{-2cm}
\caption{Comparison of the the photon spectrum at LHC
from an equilibrated QGP
(left) and a chemically non-equilibrated QGP (right)
at the same initial energy density $\epsilon_0=425$ GeV/fm$^3$
\cite{Mustafa00}.}
\protect\label{fig4.12}
\end{figure}

All these investigations did not take into account deviations from a
chemical equilibrium, which could be important at RHIC
and LHC \cite{Biro93}. In particular, a gluon dominated, hot early
phase is expected, in which quarks are strongly suppressed compared to
their equilibrium abundance \cite{Shuryak92}. This would lead to a
suppression of about two orders of magnitude
of high $p_T$ photons from the QGP. However,
extending the rates as well as the hydrodynamical model to chemical
non-equilibrium, one observes that
this strong suppression is
compensated to a large extent by the much higher initial temperature of the
dilute parton gas compared to an equilibrated QGP at the same initial
energy density, as shown by Mustafa and Thoma \cite{Mustafa00}
(see Figs. \ref{fig4.11} and \ref{fig4.12}). Here a transverse expansion was
included, and initial conditions for the temperature and the fugacities
from the self-screened parton cascade model
(SSPC) \cite{Eskola96} ($T_0=668$ MeV, $\lambda_g^0=0.34$,
$\lambda_q^0=0.064$ for RHIC and $T_0=1020$ MeV, $\lambda_g^0=0.43$,
$\lambda_q^0=0.082$ for LHC) together with a Fermi-like nuclear profile
have been used. At RHIC, the non-equilibrium compared to
the equilibrium photon spectrum is suppressed by a factor of
5 at $p_T=1$ GeV but enhanced by a factor of 2 at 5 GeV. At LHC,
the non-equilibrium yield is smaller by about a factor of 3 for all
momenta. Note that in the non-equilibrium case the annihilation-with-scattering
contribution to the 2-loop HTL rate is largely reduced because it
has at least two quarks in the initial channel, which are suppressed
by the small quark fugacities. Hence, the relative importance of the various
contributions (Compton, annihilation, bremsstrahlung, and
annihilation-with-scattering) is different in the non-equilibrium compared
to the equilibrium scenario.
Also prompt photons have been included (see Figs. \ref{fig4.11} and
\ref{fig4.12}), and a dominance of the thermal
contribution over the prompt photons for $p_T<3.5$ GeV at RHIC and LHC
has been found.

It should be noted, however,
that predictions of the photon yield at RHIC and LHC suffer also from
a large uncertainty in the initial conditions (temperature, quark and gluon
fugacities) for the chemical equilibration, which are
predicted very differently in different transport models.
For example, the HIJING model \cite{Wang97} predicts much smaller
initial temperatures and fugacities, which lead to a much stronger
suppression of the photon yield in the non-equilibrium case \cite{Mustafa00}.

Moreover, effects of the finite life-time of the QGP, leading to a strongly
enhanced non-exponential photon spectrum at high $p_T$, could help
to identify the QGP contribution at RHIC and LHC \cite{Wang00,Wang01}.

It is interesting to note that at SPS even simple hydrodynamical models
without a phase transition work due to the small
life-time of the QGP phase \cite{Gallmeister00}.
At RHIC and LHC, on the other hand, we expect
that the QGP contribution becomes significant, which might allow to
distinguish between hydrodynamical calculations using different
EOS, initial conditions and photon rates as input.

In conclusion, the transverse expansion of the later stages in the HHG
phase is important at RHIC and LHC and renders the photon spectrum
rather flat at large $p_T$. Hence the hot QGP contribution to the spectrum,
presuming the formation of a QGP phase,
will be covered partly by these HHG photons and prompt photons. Whether
there is a $p_T$ window, where the QGP photons dominate depends on details
of the rates and the hydrodynamical models, which are not yet under control.
Hadronic and dilepton spectra will help to reduce these uncertainties by
constraining the initial conditions. Anyway, the QGP contribution to the
photon spectrum at RHIC and LHC will be significant and might be extracted
from the observed spectra by comparing with calculations.
Moreover, the experimental photon spectra at RHIC
and LHC will provide interesting information on the initial conditions
and the chemical equilibration of the fireball.

\section{Summary and Conclusions}\label{sec:sc}

Energetic photons from high-energy hadronic and nucleus-nucleus collisions
provide important information about fundamental aspects of the particles
involved and their interactions. In particular, they probe the parton
distributions in hadrons and nuclei. In relativistic heavy-ion collisions,
they serve as a direct probe for all stages of the fireball since they
leave the system without further interactions due to their large mean free
path. Most important, the thermal radiation from the fireball might allow
to extract information on the EOS of the matter produced in the collision.
Hence, the direct photon production provides one of the most promising
signatures for the QGP, a new state of matter likely to be created in
ultrarelativistic heavy-ion collisions.

Direct photons have been observed in $pp$ and $pA$ reactions at collision
energies from $\sqrt{s}=19.3$ GeV to 1.8 TeV. In these experiments, they can
be distinguished from the decay photons, coming from hadronic decays,
within the experiment directly. At the present stage, the results
for direct photons are controversial, especially at low collision energies.

Direct photons from high-energy nucleus-nucleus collisions cannot be
identified experimentally due to the high hadron multiplicity.
Rather, one has to subtract the dominating background of the decay photons
by reconstructing them from the measured hadrons, mainly $\pi^0$ and $\eta$.
In this way, WA98 were able to observe direct photons at the SPS in
central $Pb+Pb$
collisions at a beam energy of 158 GeV per nucleon, corresponding to
$\sqrt{s_{NN}}=17.3$ GeV, whereas WA80 gave only an upper limit
for direct photons in $S+Au$ collisions at a beam energy of 200 A GeV.
% The extracted direct photon spectrum from WA98 shows a clear excess
% over the background for transverse photon momenta between 1.5 and
% 3.5 GeV/c.
The extracted direct photon spectrum from WA98 shows a clear excess
over the background for photon transverse momenta between 1.5 and
3.5 GeV/c.

In order to learn from these experimental results about the hadrons
and their interactions, one has to compare them to theoretical
pre- and postdictions. In $pp$ and $pA$ collisions the cross sections
for prompt photon production have been calculated within perturbative QCD.
These cross sections follow from folding the basic partonic processes,
e.g. $q\bar q \rightarrow g\gamma $, with the parton distributions
of the hadrons.
% Using next-to-leading or even next-to-next-to-leading
% order corrections plus a optimization procedure for fixing the parameters
% (renormalization and factorization scales), a reasonable description
% of most of the data can be obtained although their are some clear
% discrepancies in particular at low collision energies.
Using next-to-leading or even next-to-next-to-leading
order corrections plus an optimization procedure for fixing the parameters
(renormalization and factorization scales), a reasonable description
of most of the data can be obtained although there are some clear
discrepancies in particular at low collision energies.
Assuming an intrinsic
momentum broadening in the parton distributions, a better agreement with the
various data sets can be obtained, of course at the expense of introducing a
new parameter. Indications for an additional nuclear broadening (Cronin effect)
have also been found in $pA$ collisions.

In a spatially and temporally extended system like the fireball in
a nucleus-nucleus collision, it is not sufficient to compute the cross
sections or rates for photon production. Rather, one has to convolute
the rates with the space-time evolution of the fireball, for which usually
hydrodynamical models are adopted. Furthermore, one has to consider the
photon production from different stages, i.e. prompt photons from initial hard
collisions and thermal photons from the QGP and the HHG. In this way,
photon spectra are obtained which can be compared to experimental
spectra. In order to draw a conclusion on the possible presence
of a QGP phase, one has to compare predictions for spectra with and without
a QGP phase to experimental results.

The calculation of the photon production rate from an equilibrated
(or chemically non-equilibrated)
QGP is based on perturbative QCD at finite temperature. To lowest order the
basic processes are quark-antiquark annihilation and Compton scattering.
Owing to infrared singularities the HTL resummation technique has to
be employed. Assuming the weak coupling limit, the leading logarithm
contribution can be extracted in this way. However, beyond the leading
logarithm infinitely many diagrams, corresponding for example to
bremsstrahlung, contribute to the same order. Hence, the photon production
rate from the QGP beyond the leading logarithm approximation
may not be perturbatively describable. In the most recent calculations of
photon spectra, the 2-loop rate within the HTL improved perturbation theory
has been used as an educated guess. The photon production rate from
the HHG, on the other hand, is based on effective models for hadronic
interactions. The most important contributions to the rate come from
interactions between $\pi$- and $\rho$-mesons. In particular, the $a_1$
resonance plays an important role for the photon production. Also the
HHG rate suffers from a number of uncertainties such as assumptions
about the effective Lagrangians, medium effects, etc.

For deriving the spectra from the rates,
various hydrodynamical models, describing the expansion of
the fireball in 1, 2 or 3 spatial dimensions, fixing the
initial conditions in different ways, using different EOS,
and including a chemically non-equilibrated QGP have been
employed. However, a systematic and comprehensive treatment is still missing.
In other words, theoretical predictions are subject to uncertainties in the
rates as well as the hydrodynamical description.

The present calculations
do not allow to infer about the existence of a QGP phase in
central $Pb+Pb$ collisions at a beam energy of 158 A GeV. However, the
data are consistent with a thermal source, either QGP or HHG,
for photons with $p_T<2.5$ GeV/c and with 
enhanced prompt photons for $p_T>2.5$ GeV/c.

The situation will change drastically at RHIC and LHC. The RHIC
experiments including PHENIX with its excellent photon detection have
started to take data at $\sqrt{s_{NN}}=200$ GeV. They will measure
photon spectra with very high statistics. The ALICE experiment planned
for the CERN LHC is gearing up to measure photons in heavy-ion
reactions at still higher beam energy. A much larger
temperature and life-time of the fireball in these collisions
will cause a copious production of thermal photons.
Most estimates of the photon spectra at RHIC and LHC
predict a window around $p_T=2$ GeV, where the QGP contribution should
dominate. However, in order to confirm this prediction, new developments
in the calculation of the rates from QCD as well as a consistent description
of the space-time evolution would be essential. This conclusion
applies also to all the other signatures for the QGP, as quantitative
predictions of non-perturbative QGP properties are required, taking into
account the dynamical evolution of the fireball at the same time.
Unfortunately, at the moment there is no non-perturbative, dynamical
approach to QCD available. Hence, the prospects and problems for discovering
the QGP from the direct photon production are similar as for other
signatures. However, direct photons from relativistic heavy-ion
collisions will definitely help to reveal and
understand important and interesting properties of
strongly interacting matter.

\bigskip

\leftline{\large \bf Acknowledgements:}

We would like to thank P. Aurenche, T.C. Awes, A. Dumitru, K. Gallmeister,
F. Gelis, K. Haglin, P. Huovinen, B. K\"ampfer, P. Levai, M. Mustafa,
V. Ruuskanen, D. Srivastava, and H. Zaraket for helpful comments and
discussions. We are also grateful to K. Reygers and F. Steffen for 
their careful reading of the manuscript. M.T. has been supported by
the Heisenberg program of the Deutsche Forschungsgemeinschaft. M.T. would
also like to thank CERN for its hospitality, where a large part of this 
work has been performed.

\section{Appendix A: Perturbative Calculation of the Photon Production 
Rate}\label{sec:appa}

%\begin{figure}[hbt]
%       \centerline{\includegraphics{fig41.eps}}
%        \caption{}
%        \protect\label{fig4.1}
%\end{figure}

In this Appendix the photon production rate of energetic photons ($E\gg T$)
is derived to lowest order perturbation theory from the diagrams of 
Fig. \ref{fig2.1},
corresponding to quark annihilation and Compton scattering. Starting from
the definition Eq. (\ref{rate1}) of the rate, using the matrix elements for
these processes, Kapusta et al. \cite{Kapusta91} calculated the rate, assuming
Maxwell-Boltzmann distributions for the incoming partons, i.e. 
$n_i\sim \exp (-E_i/T)$ for $i=1,2$ in Eq. (\ref{rate1}). This facilitates 
the evaluation of the momentum integrals over $p_1$ and $p_2$ in Eq. 
(\ref{rate1}),
which otherwise can be done only numerically. The use of the Boltzmann
distribution instead of the exact quantum statistical distributions for
quarks and gluons is justified in the case of energetic photons, $E\gg T$.
Owing to energy conservation, the sum of the incoming 
parton energies is much larger than the temperature, $E_1+E_2 = E\gg T$, 
and the phase space
for small $E_1$ or $E_2$ is unfavorable \cite{Kapusta91}. Comparing to the 
numerical calculation using Fermi and Bose distributions, 
one observes that the error introduced by the Boltzmann 
approximation is less than a few percent \cite{Neubert89}.

But even assuming Boltzmann distributions for the incoming particles,
the momentum integrations are still rather involved, although they can be 
performed analytically \cite{Staadt86}. However, the calculation can
be facilitated further on by computing the inverse process, photon absorption,
and using the principle of detailed balance \cite{Shuryak78,Thoma95a}.
Accordingly, the photon production rate is related to the photon damping
or absorption rate $\gamma $ by \cite{Thoma95a}
\beq
\frac{dN}{d^4xd^3p}=\frac{4}{(2\pi)^3}\> e^{-E/T}\> \gamma.
\label{balance}
\eeq
The damping rate is defined as the imaginary part of the dispersion 
relation $\omega (p)$ 
of a real photon in the QGP, $\gamma = -{\rm Im}\, \omega(p)$,
i.e. it follows from 
\beq
\omega^2-p^2-\Pi_T(\omega ,p)=0,
\label{dispersion}
\eeq
where 
\beq
\Pi_T(p_0,p)=\frac{1}{2}\> \left (\delta_{ij}-\frac{p_ip_j}{p^2}\right )\>
\Pi_{ij}(p_0,p)
\label{polar}
\eeq
is the transverse polarization tensor at finite temperature. ($\Pi_{ij}$
denotes the spatial components of the polarization tensor.) Assuming no
overdamping, i.e. $\gamma \ll p$, we find
\beq
\gamma=-\frac{1}{2p} {\rm Im}\, \Pi_T(p_0=p,p).
\label{damping}
\eeq

Using cutting rules \cite{Weldon83}, the damping rate can be calculated
alternatively from the matrix elements. In the case of Compton scattering,
i.e. the inverse process of the one in Fig. \ref{fig2.1}, 
namely photon absorption and
gluon emission, the damping rate is given as
\bea
\gamma_{\rm comp}&=&\frac{1}{4E}\int \frac{d^3p_3}{(2\pi)^32E_3}\> n_F(E_3)\>
\int \frac{d^3E_2}{(2\pi)^32E_2}\> [1+n_B(E_2)]\>
\int \frac{d^3E_1}{(2\pi)^32E_1}\nonumber \\
&& [1-n_F(E_2)]\>
(2\pi )^4\> \delta^4(P+P_3-P_2-P_1)\> \sum_{i} \langle |{\mathcal M}|^2
\rangle _{\rm comp}.
\label{compton1}
\eea
In the case of pair creation, i.e. the inverse process of quark-antiquark
annihilation in Fig. \ref{fig2.1}, we have
\bea
\gamma_{\rm pair}&=&\frac{1}{4E}\int \frac{d^3p_3}{(2\pi)^32E_3}\> n_B(E_3)\>
\int \frac{d^3E_2}{(2\pi)^32E_2}\> [1-n_F(E_2)]\>
\int \frac{d^3E_1}{(2\pi)^32E_1}\nonumber \\
&& [1-n_F(E_1)]\>
(2\pi )^4\> \delta^4(P+P_3-P_2-P_1)\> \sum_{i} \langle |{\mathcal M}|^2
\rangle _{\rm pair}.
\label{pair1}
\eea
In Eqs. (\ref{compton1}) and (\ref{pair1}) the factor $1/(4E)$ instead of the usual 
$1/(2E)$ comes from the definition of the damping rate as the imaginary 
photon energy. The sums in front of the matrix elements indicate sums
over the initial states of the incoming parton, since all possible states
of the partons interacting with the photon have to be counted in the rate.
The matrix elements summed over all initial and final parton states
are given by \cite{Halzen84}
\bea
\sum_{i} \langle |{\mathcal M}|^2 \rangle _{\rm comp} &=& -32\> \frac{5}{9}\>
e^2\> g^2 \> \left (\frac{u}{s} + \frac{s}{u}\right ), \nonumber \\
\sum_{i} \langle |{\mathcal M}|^2 \rangle _{\rm pair} &=&  16\> \frac{5}{9}\>
e^2\> g^2 \> \left (\frac{u}{t} + \frac{t}{u}\right ), 
\label{matrix}
\eea
where we neglected the quark masses as the bare up and down quark masses
are much smaller than the temperature. The Mandelstam variables are 
$s=(P+P_3)^2$, $t=(P-P_2)^2$, and $u=-s-t$. The factor $5/9$ comes 
from the sum over the square of the electric charges of the up and down 
quarks. 

The Boltzmann approximation for the incoming particles in the production rate
corresponds to using Boltzmann distributions for the outgoing particles
in the damping rate. This means that we simply neglect the distribution
functions in the Pauli blocking or Bose enhancement factors in Eqs. (\ref{compton1})
and (\ref{pair1}) since they do not exist in classical statistics. 
This can also be seen by writing, for example, $1+n_B(E_2)=\exp (E_2/T)\,
n_B(E_2)
\simeq 1$. The advantage compared to the direct calculation of the
production rate is now that there are
no distributions functions in the momentum integrations over $p_1$ and $p_2$.
Hence we can easily evaluate these integrals by transforming to the center
of mass system using the Lorentz invariant phase space factor \cite{Halzen84}
\beq
dL=(2\pi)^4\> \delta^4(P+P_3-P_2-P_1)\> \frac{d^3E_2}{(2\pi)^32E_2}
\frac{d^3E_1}{(2\pi)^32E_1}=\frac{dt}{8\pi s}.
\label{phase}
\eeq
Then the integration over $t$ from $-s-\Lambda ^2$ to $-\Lambda ^2$, where
$\Lambda $ is a cutoff for the logarithmic IR divergence of the $t$ and
$u$ channel, leads to
\bea
\int dL \> \sum_{i} \langle |{\mathcal M}|^2 \rangle _{\rm comp}
&=& \frac{20}{9\pi}\> e^2\> g^2\> \left (\ln \frac{s}{\Lambda^2}+\frac{1}{2}
\right ), \nonumber \\ 
\int dL \> \sum_{i} \langle |{\mathcal M}|^2 \rangle _{\rm pair}
&=& \frac{20}{9\pi}\> e^2\> g^2\> \left (\ln \frac{s}{\Lambda^2}-1
\right ).
\label{integration1}
\eea
The remaining integral over $p_3$ can be done using $s=2pp_3(1-\hat p\cdot 
\hat p_3)$ and 
\bea
\int_0^\infty dp_3\> p_3\> n_B(p_3) &=& \frac{\pi^2T^2}{6}, \nonumber \\
\int_0^\infty dp_3\> p_3\> n_F(p_3) &=& \frac{\pi^2T^2}{12}, 
\label{integrals1}
\eea
and
\bea
\int_0^\infty dp_3\> p_3\> \ln \frac{p_3}{\Lambda}\> n_B(p_3) &=& 
\frac{\pi^2T^2}{6}\> \left [\ln \frac{T}{\Lambda}+1-\gamma-\frac{\zeta'(2)}
{\zeta(2)}\right ],\nonumber \\
\int_0^\infty dp_3\> p_3\> \ln \frac{p_3}{\Lambda}\> n_F(p_3) &=& 
\frac{\pi^2T^2}{12}\> \left [\ln \frac{2T}{\Lambda}+1-\gamma-\frac{\zeta'(2)}
{\zeta(2)}\right ],
\label{integrals2}
\eea 
where $\gamma =0.57722$ is Euler's constant and $\zeta (z)$ is Riemann's
zeta function with $\zeta'(2)/\zeta(2)=-0.569996$. Using detailed 
balance, Eq. (\ref{balance}), we arrive at
\bea
\left (\frac{dN}{d^4xd^3p}\right )_{\rm comp}=\frac{5}{54\pi^2}\> \alpha \>
\alpha_s\> T^2\> \frac{e^{-E/T}}{E}\> \left [\ln \frac{8ET}{\Lambda^2}+
\frac{1}{2}-\gamma +\frac{\zeta'(2)}{\zeta(2)}\right ],\nonumber \\
\left (\frac{dN}{d^4xd^3p}\right )_{\rm pair}=\frac{5}{27\pi^2}\> \alpha \>
\alpha_s\> T^2\> \frac{e^{-E/T}}{E}\> \left [\ln \frac{4ET}{\Lambda^2}-1
-\gamma +\frac{\zeta'(2)}{\zeta(2)}\right ],
\label{comp_pair}
\eea
which has also been found by Kapusta et al. \cite{Kapusta91} in a direct 
calculation of the production rate. Adding the two
contributions above leads to Eq. (\ref{pre_htl}) if we replace the IR
cutoff $\Lambda $ by the bare quark mass $m_0$ and keep only
the leading logarithm assuming $ET\gg m_0^2$.

\section{Appendix B: Hard Thermal Loops and Photon Production}\label{sec:appb}

The HTL resummation technique has been invented in order to cure serious
problems of gauge theories at finite temperature 
\cite{Braaten90,Pisarski89,Pisarski89a,Braaten90d,Frenkel90,Frenkel92}. 
For a review of this method and its applications see 
\cite{LeBellac96,Thoma95,Thoma00,Blaizot01}. It consists out of three steps:
extraction of the HTLs, resummation of HTLs into effective Green functions,
and use of the resummed Green functions. Using these effective propagators
and vertices corresponds to an effective perturbation theory which yields
gauge invariant results \cite{Braaten90,Braaten90c} 
with an improved IR behavior.

\medskip

{\it 1. Step: Extraction of the HTLs.} The starting point for isolating
HTL diagrams is the distinction between the soft momentum scale, $gT$,
and the hard one, $T$, which is possible in the weak coupling limit,
$g\ll 1$. HTLs are 1-loop diagrams (self energies and
vertex corrections) containing a hard loop momentum
but exclusively soft external momenta. The HTL 
approximation is equivalent to
the high-temperature limit of these diagrams 
\cite{Klimov82,Weldon82,Weldon82a} 
and the semiclassical approximation 
\cite{Blaizot01,Silin60,Blaizot93,Kelly94}. In the HTL
limit, analytic and  gauge invariant expressions are obtained.

\begin{figure}[hbt]
\centerline{\resizebox{12cm}{!}{\includegraphics{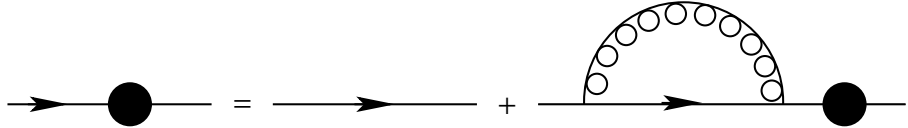}}}
\caption{Dyson-Schwinger equation defining the effective quark 
propagator.}
\protect\label{figB.1}
\end{figure}

As an example, we discuss the
quark self-energy, which is needed for the photon production rate as discussed
below.
The most general ansatz for the self-energy of a massless fermion
interacting with a heat bath at temperature $T$ is given by \cite{Weldon82}
\beq
\Sigma(P)=-a(p_0,p)P\sla - b(p_0,p)\gamma _0,
\label{self}
\eeq
where $a$ and $b$ are scalar functions of the fermion energy $p_0$
and the magnitude of the momentum $p=|{\bf p}|$. Due to the choice
of the heat bath as rest frame, the self-energy depends on
$p$ and $p_0$ separately 
and has a term proportional to $\gamma _0$. It should be
noted that the ansatz Eq. (\ref{self}) respects chiral symmetry. The functions
$a$ and $b$ are related to traces of the self energy
\bea
a(p_0,p) & = & \frac {1}{4p^2}\> \left [{\rm tr}(P\sla \> \Sigma)
- p_0\> {\rm tr}(\gamma _0 \> \Sigma)\right ],\nonumber\\
b(p_0,p) & = & \frac {1}{4p^2}\> \left [P^2\> {\rm tr}(\gamma _0\> \Sigma)
- p_0\> {\rm tr} (P\sla \> \Sigma)\right ].
\label{ab}
\eea
The HTL quark self-energy follows from the 1-loop diagram of 
Fig. \ref{figB.1}, where the 
internal hard momentum is much larger than the soft external. Using the
imaginary or real time formalism within thermal field theory
\cite{Kapusta89,Chou85,Landsmann87}, one finds in the HTL approximation
\cite{Thoma95}
\bea
{\rm tr}(P\sla \> \Sigma) & = & 4\> m_q^2\; ,\nonumber \\
{\rm tr}(\gamma _0\> \Sigma) & = & 2\> m_q^2\> \frac {1}{p}\> \ln \frac
{p_0+p}{p_0-p},
\label{traces}
\eea
where $m_q^2=g^2T^2/6$ is the effective quark mass. This result 
has been derived first in the high-temperature limit 
\cite{Klimov82,Weldon82}. Despite the appearance of a gluon propagator,
the HTL quark self-energy is gauge invariant. Furthermore, the effective
quark mass does not violate chiral symmetry as Eq. (\ref{self}) is chirally
invariant. For details of the computation of HTL self-energies 
see e.g. \cite{Thoma00}. 

\medskip

{\it 2. Step: Resummed Green Functions.} After having extracted the HTLs,
we will construct effective Green functions from them. E.g.,
the effective quark propagator is obtained by resumming the HTL quark
self-energy within the Dyson-Schwinger equation of Fig. \ref{figB.1}, 
resulting in
\beq
S(P)^{-1}=P\sla -\Sigma(P).
\label{prop1}
\eeq
For massless fermions it is convenient to use the helicity 
representation \cite{Braaten90b}
\beq
S(P)=\frac{\gamma _0-{\bf \hat p} \cdot \mbox{\boldmath$\gamma$}}{2D_+(p_0,p)}
+\frac{\gamma _0+{\bf \hat p} \cdot \mbox{\boldmath$\gamma$}}{2D_-(p_0,p)},
\label{prop2}
\eeq
where
\beq
D_\pm(p_0,p)=(-p_0\pm p)\> [1+a(p_0,p)]-b(p_0,p).
\label{Dpm}
\eeq

\begin{figure}[hbt]
\vspace*{-13cm}
\hspace*{5cm}
\centerline{\resizebox{10cm}{!}{\includegraphics{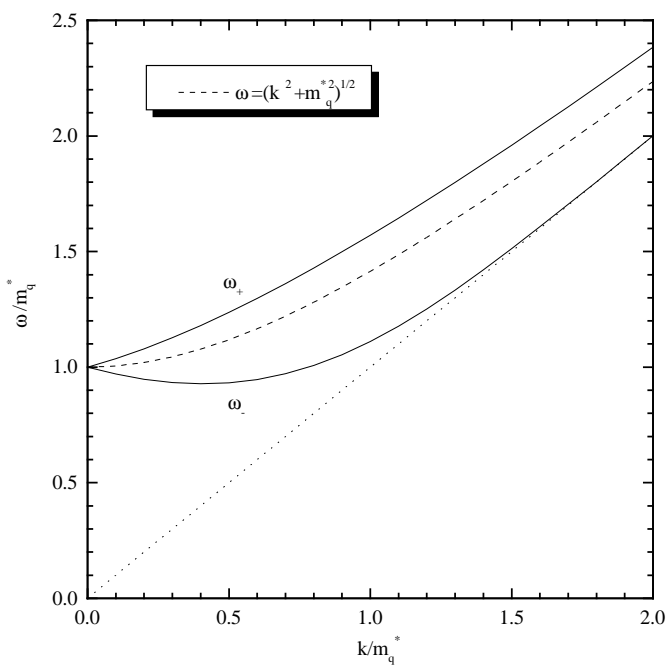}}}
\vspace*{6.5cm}
\caption{Quark dispersion relation in the QGP.}
\protect\label{figB.2}
\end{figure}

The bare quark propagator in the helicity representation follows from 
Eqs. (\ref{prop2}) and (\ref{Dpm}) for $a=b=0$. The HTL resummed propagator
is given by substituting Eq. (\ref{ab}) together with Eq. (\ref{traces})
into Eqs. (\ref{prop2}) and (\ref{Dpm}). It describes
the propagation of collective quark modes in the QGP. The poles of 
the effective quark propagator determine the in-medium dispersion 
relations shown 
in Fig. \ref{figB.2}. This dispersion relation exhibits two branches, where 
the lower one, $\omega_-(p)$,  
coming from $D_-(p_0,p)=0$, has a negative ratio of helicity
to chirality. Such a mode, called plasmino \cite{Braaten92}, does not exist 
in vacuum, but appears in a medium, similar as longitudinal photons 
(plasmons). For large momenta ($p\gg gT$) the spectral strength of the plasmino
is exponentially suppressed. The upper branch, $\omega_+(p)$, coming from 
$D_+(p_0,p)=0$, on the other hand, reduces to the vacuum mode with
$\omega (p)=p$ for large momenta. At zero momentum both branches agree
with $\omega_\pm (0)=m_q$. The minimum in the plasmino branch
has interesting consequences, leading to Van Hove singularities in
the low-mass dilepton production rate \cite{Braaten90b} and in the
spectral functions of hadronic correlators \cite{Karsch01a}. It can be shown
that the minimum in the plasmino dispersion relation is a general property
of massless fermions at finite temperature, independent of the approximation
for the effective quark propagator \cite{Peshier00}. Therefore, Van Hove
singularities in the low-mass dilepton production might provide a unique 
signature for the presence of deconfined, collective quarks in 
relativistic heavy-ion collisions \cite{Peshier00}.
Within the HTL approximation the in-medium quarks (quasiparticles) are 
undamped. However, the HTL quark self-energy Eqs. (\ref{self}) to (\ref{traces})
exhibits an imaginary part below the light cone
($p_0^2<p^2$) corresponding to Landau damping,  which
describes the collisionless energy transfer from a collective quark 
to the heat bath \cite{Lifshitz81}. Hence virtual, time-like 
in-medium quarks are damped in the HTL approximation. 

Besides the HTL resummed quark propagator, also a HTL gluon propagator and
HTL vertices exist in QCD. The latter are a consequence of the
Ward or Slavnov-Taylor
identities which relate propagators to vertices \cite{Braaten90d}, e.g.
\beq
ie\> [\Sigma(P_1)-\Sigma (P_2)]=(P_1+P_2)_\mu \Gamma^\mu (P_1,P_2),
\label{ward}
\eeq
where $\Gamma^\mu$ is the quark-gluon vertex. In the following, however,
we will not need HTL vertices.

\medskip

{\it 3. Step: Effective Perturbation Theory.} Now we can use the HTL resummed 
propagators and vertices as in ordinary perturbation theory. However, they
are only necessary if all energies and momenta of the external
legs of the Green functions under consideration are soft, i.e. of order $gT$.
Otherwise bare Green functions are sufficient. This can be seen, for instance,
in the case of the quark propagator Eq. (\ref{prop1}). If $p_0$ and $p$ are soft,
the HTL self-energy in Eq. (\ref{prop1}) is of the same order $gT$ according to
Eqs. (\ref{self}), (\ref{ab}), and (\ref{traces}). Hence it 
cannot be neglected in the propagator. 
In other words, the HTL resummed propagator contributes to the same order
as the bare propagator if its energy and momentum are soft. 
On the other hand, if $p_0$ or $p$
are hard, i.e. of order $T$ or larger, the bare propagator is sufficient
\cite{Braaten90}\footnote{This cannot be seen from Eqs. (\ref{self}) to 
(\ref{traces}) since Eq. (\ref{traces}) has been derived under the assumption
$p_0\sim p\sim gT$.}.

At the same time, 
the use of HTL self-energies takes into account important medium effects of 
the QGP (see e.g. \cite{Thoma00}). 
For example, the HTL gluon self-energy contains Debye screening,
which improves the IR behavior of IR divergent diagrams and quantities, in
which a gluon is exchanged. On the other hand, there is no static magnetic 
screening in the HTL gluon polarization tensor, which requires a 
non-perturbative treatment \cite{Linde80,Gross81}. Therefore certain 
quantities, e.g. 
the damping rates of a hard quark or gluon, which are quadratically IR 
divergent using a bare gluon propagator, are (even to leading order)
still logarithmically IR divergent if the HTL gluon propagator is taken.

After all, the HTL resummation technique means a very important progress for
finite temperature gauge theories since it leads to gauge invariant results,
in which, compared to naive perturbation theory, diagrams of the same order 
are included. In many cases the HTL method allows (at least to leading order)
IR finite results in contrast to naive perturbation theory. An important
example for this is the photon production rate discussed below.

The HTL resummation technique has been extended to a finite chemical potential
\cite{Vija95,Manuel96}, which is important for treating a finite baryon density
in heavy-ion collisions, in particular at SPS, and in quark matter, 
which might exist in the interior of neutron stars \cite{Schertler97}. 
In particular, the HTL method has been used to describe color 
superconductivity in dense quark matter \cite{Pisarski00}.
It has also been extended to a chemically
non-equilibrated QGP \cite{Baier97,Carrington99}, as it is expected at
RHIC and LHC \cite{Biro93}.

The HTL resummation technique has been adopted for calculating important 
properties of the QGP, such as parton damping rates, transport coefficients,
the energy loss of energetic partons, and dilepton and photon production
rates\footnote{For a review and references on these applications see e.g. 
\cite{Thoma95,Thoma98}.}. Here we want to discuss the calculation of
the production rate of energetic photons in this way. For this purpose, we 
start from the definition of the photon rate using the polarization
tensor Eq. (\ref{rate2}). To lowest order in the HTL improved perturbation
theory we have to consider the diagram shown in Fig. \ref{fig2.3}. Here we have
replaced one bare quark propagator by a HTL resummed one. Due to 
energy-momentum conservation, only one quark momentum can be soft in the 
case of energetic photons ($E\gg T$). Therefore, we do not need to dress 
both quark propagators at the same time. Also it is not necessary to use an 
effective quark-photon propagator. This is different for soft photons and
dileptons, where two HTL propagators and vertices have to be considered
\cite{Braaten90b,Baier91}. Note that the polarization tensor of 
Fig. \ref{fig2.3} has
an imaginary part also for on-shell photons, in contrast to a bare quark
loop since the HTL self-energy contained in the effective quark propagator
has an imaginary part. Therefore, the physical process leading to
the photon production from Fig. \ref{fig2.3} 
is related to Landau damping, i.e. the
interaction of a soft quark with thermal gluons. Note also that the
diagram of Fig. \ref{fig2.3} contains infinitely many quark-gluon loops as the
HTL quark self-energy is resummed in the effective propagator in 
Fig. \ref{figB.1}.
The imaginary part of Fig. \ref{fig2.3} corresponds therefore to the scattering 
diagrams of Fig. \ref{fig2.1} 
(quark-antiquark annihilation, Compton scattering),
where the exchanged bare quark is replaced by an in-medium quark.

Since the diagram of Fig. \ref{fig2.3} has to be considered only if the 
dressed 
quark line is soft, we introduce a separation scale $q_c$.
For soft quark momenta, we calculate the photon production rate from  
Fig. \ref{fig2.3}, whereas for hard momenta we adopt a bare quark propagator as
in Fig. \ref{fig2.1} and 
Fig. \ref{fig2.2}. Hence, the hard contribution for the rate follows from
the result of Appendix A, Eq. (\ref{comp_pair}), 
where the IR cutoff $\Lambda $
is replaced by the separation scale $q_c$. Assuming $gT\ll q_c\ll T$, which
is possible in the weak coupling limit, the arbitrary separation scale 
cancels once the hard and the soft contributions are added, as it should be 
the case for a consistent leading-order calculation \cite{Braaten91}.

Following Kapusta et al. \cite{Kapusta91}, the imaginary part of the
polarization tensor of Fig.~\ref{fig2.3}, 
entering the soft photon production rate
according to Eq. (\ref{rate2}), can be written as
\bea
Im\, \Pi_\mu^\mu (E) &=& \frac{5e^2}{12\pi}\> \int_0^\infty
dk \int_{-k}^k d\omega \> [(k-\omega) \rho_+(\omega ,k)\nonumber \\
&&+ (k+\omega) 
\rho_-(\omega ,k)]\> \theta (q_c^2-k^2+\omega^2),
\label{soft_polar}
\eea
where we have assumed two massless quark flavors and $E\gg T$. Furthermore,
we have chosen a covariant separation scale, i.e. $\omega^2-k^2<q_c^2$,
in accordance with the covariant cutoff $\Lambda^2$, introduced in 
Eq. (\ref{integration1}), in the hard part.
Here $\rho_\pm $ are the spectral functions of the effective quark 
propagator Eq. (\ref{prop2}) defined as 
\beq
\hspace*{-0.5cm}
\rho_\pm(\omega ,k)=\frac{1}{\pi }\> Im\, \frac{1}{D_\pm (\omega ,k)}.
\label{spectral}
\eeq
In the HTL approximation these spectral functions are given by 
\cite{Braaten90b}
\beq
\rho_{\pm} (k_0, k) = {k_0^2 - k^2 \over 2 m_q^2} 
[\delta (k_0 - \omega_{\pm}) + \delta (k_0 + \omega_{\mp})]+
\beta_{\pm} (k_0, k) \theta(k^2 -k_0^2).
\label{rho_htl}
\eeq
The first part of Eq. (\ref{rho_htl}) corresponds to the pole contribution 
of the HTL propagator. The second part, corresponding to the cut contribution 
from the imaginary part of the HTL quark self energy, reads 
\bea
\hspace*{-0.5cm}\beta_{\pm} (k_0, k) &=& 
-{m_q^2 \over 2}\> (\pm k_0 - k)\> \Biggl \{\biggl[k(-k_0 \pm k) + m_q^2 
\biggl( \pm 1 - {\pm k_0 -k \over 2k}
\ln{k+k_0 \over k-k_0} \biggr)\biggr]^2\nonumber \\
&& +\biggl[ {\pi \over 2}
m_q^2 {\pm k_0 -k \over k} \biggr]^2 \Biggr \}^{-1}.
\label{beta}
\eea
For real photons, only the cut contribution Eq. (\ref{beta}) has to be 
considered, since $\omega^2<k^2$ according to Eq. (\ref{soft_polar}), i.e.
the exchanged quark is time-like. For virtual photons decaying
into dileptons, on the other hand, also the pole part of 
Eq. (\ref{rho_htl}) contributes \cite{Thoma97}.

Due to the complex momentum dependence of Eq. (\ref{beta}), the integrations in
Eq. (\ref{soft_polar}) cannot be done analytically \cite{Kapusta91,Baier92}.
However, using generalized Kramers-Kronig relations, the so-called Leontovich
relations \cite{Thoma00a},
it can be shown, that only the high energy limit of the HTL quark
propagator is needed. Then the imaginary part of the
photon polarization tensor Eq. (\ref{soft_polar}) reduces to \cite{Thoma00a}
\beq
Im\, \Pi_\mu^\mu (E) = -\frac{5e^2}{12\pi}\> \int_0^{q_c} dq\> q\> 
\frac{2m_q^2}{q^2+2m_q^2},
\label{leon}
\eeq
where $2m_q^2=g^2T^2/3$ is the high energy limit of the effective
quark mass, $\omega_+^2(p\rightarrow \infty)=p^2+2m_q^2$. As a matter of fact,
the approach based on the Leontovich relation allows in principle
a more general
evaluation of the photon production rate and other quantities beyond the
HTL approximation if the high energy limit of the full quark propagator
is known \cite{Thoma00a}. The integral in Eq. (\ref{leon}) can easily be
done yielding
\beq
Im\, \Pi_\mu^\mu (E) = -\frac{5e^2}{12\pi}\> m_q^2 \ln \frac{q_c^2}{2m_q^2},
\label{imag_polar}
\eeq
where we assumed $q_c\gg m_q$ in accordance with $gT\ll q_c\ll T$ . 
This result was also found
independently by Kapusta et al. \cite{Kapusta91} and Baier et al.
\cite{Baier92}, where the factor $1/2$ under the 
logarithm could be derived only numerically using the full spectral functions
Eq. (\ref{rho_htl}). 

Combining the soft part with the hard part Eq. (\ref{comp_pair}), 
where the IR cutoff 
$\Lambda $ is replaced by the separation scale $q_c$, we obtain the
final result Eq. (\ref{1-loop}) for the production rate of energetic photons to 
leading logarithm
\beq 
\frac{dN}{d^4xd^3p}
=\frac{5}{18\pi^2}\> \alpha \alpha_s\> e^{-E/T}\> \frac{T^2}{E}
\> \ln \frac{0.2317 E}{\alpha_s T}.
\label{result}
\eeq
Here the separation scale $q_c$, serving as an IR cutoff for the hard 
part, drops out as expected \cite{Braaten91} since the hard 
part and the soft part have the same factors in front of the logarithm.
Using the Boltzmann approximation for the hard part (see Appendix A) and the 
Leontovich relation for the soft part, the photon production rate 
to lowest order in the HTL approximation Eq. (\ref{result}) could be derived 
analytically. At finite chemical potential \cite{Traxler95} and in 
non-equilibrium
\cite{Baier97} the Boltzmann approximation cannot be used since there is no
cancellation of the hard and the soft parts in this case. Using the correct
quantum statistical distributions, the separation scale drops out. However, 
the rates have to be calculated numerically 
in these cases.

In the leading logarithm approximation, the photon rate Eq. (\ref{result})
agrees with the 
result obtained from the diagrams of Fig. \ref{fig2.1} 
in naive perturbation theory 
if the thermal quark mass is used as IR cutoff. The HTL method 
allows to compute also the coefficient under the logarithm, i.e. the next term
beyond the leading logarithm. However, as discussed in Section
\ref{subsubsec:trq}, this
term is not complete but there are higher order contributions within the HTL 
improved perturbation theory to the same order.

\end{document}